\documentclass[11pt]{article}

\usepackage{amsmath,amsthm,amssymb,color,multirow,algorithmic,url,bm}
\usepackage[top=1in,bottom=1.1in,left=1.05in,right=1.05in]{geometry}
\usepackage[boxed]{algorithm}
\usepackage{longtable,color}
\usepackage[normalem]{ulem}

\usepackage{hyperref}

\usepackage[sort&compress]{natbib}
\makeatletter 
\renewcommand\@biblabel[1]{#1.} 
\makeatother %

\usepackage{ifpdf}
\ifpdf
\usepackage[pdftex]{graphicx}
\else
\usepackage[dvips]{graphicx}
\fi

\newcommand{\blue}{\textcolor{black}}
\newcommand{\rev}{\textcolor{black}}

\newcommand\E{\ensuremath{\mathbb{E}}}

\newcommand\R{\ensuremath{\mathbb{R}}}

\newcommand\eps{\epsilon}

\theoremstyle{remark}
\newtheorem*{remark}{Remark}

\theoremstyle{definition}
\newtheorem{prop}{Proposition}

\makeatletter
\renewcommand\@biblabel[1]{#1.}
\makeatother
\usepackage[explicit]{titlesec}
\titleformat{\section}
  {\normalfont\bfseries\scshape\fontsize{12}{0}}{\thesection}{1em}{\MakeUppercase{#1}}
\titleformat{\subsection}
  {\normalfont\bfseries\fontsize{12}{0}}{\thesubsection}{1em}{#1}
\titleformat{\subsubsection}
  {\normalfont\itshape\fontsize{10}{0}}{\thesubsubsection}{1em}{#1}

\graphicspath{{.}{appendixFiles/japmale/}{appendixFiles/japfemale/}{appendixFiles/ukmale/}{appendixFiles/ukfemale/}{appendixFiles/usfemale/}}

\begin{document}
\title{Gaussian Process Models for Mortality Rates and Improvement Factors}

\author{Mike Ludkovski \and Jimmy Risk \and Howard Zail \footnote{Ludkovski is with the Department of Statistics \& Applied Probability, University of California, Santa Barbara CA 93106-3110; Risk is with the Department of Mathematical Sciences, Cal Poly Pomona. H. Zail is an Actuary and Founder of Elucidor, LLC {ludkovski@pstat.ucsb.edu,risk@cpp.edu,hzail@elucidor.com}}}

\date{\today}

\maketitle

\begin{abstract}
We develop a Gaussian process (``GP'') framework for modeling mortality rates and mortality improvement factors. GP regression is a nonparametric, data-driven approach for determining the spatial dependence in mortality rates and jointly smoothing raw rates across dimensions, such as calendar year and age. The GP model quantifies uncertainty associated with smoothed historical experience and generates full stochastic trajectories for out-of-sample forecasts.  Our framework is well suited for updating projections when newly available data arrives, and for dealing with ``edge'' issues where credibility is lower. We present a detailed analysis of Gaussian process model performance for US mortality experience based on the CDC (Center for Disease Control) datasets. We investigate the interaction between mean and residual modeling, Bayesian and non-Bayesian GP methodologies, accuracy of in-sample and out-of-sample forecasting, and stability of model parameters. We also document the general decline, along with strong age-dependency, in mortality improvement factors over the past few years, contrasting our findings with the Society of Actuaries (``SOA") MP-2014 and -2015 models that do not fully reflect these recent trends.
\end{abstract}



\section{Introduction}
\blue{Publishing of pension mortality tables and mortality improvement factors for use by actuarial professionals and researchers in longevity risk management is a major endeavor of the actuarial professional organizations. In the US, the Society of Actuaries (SOA) runs the Retirement Plans Executive Committee (RPEC); its most recent publication is known as the RP-2014 mortality tables and the MP-2015 improvement scales \citep{SOA2014,MP2014}. In the UK, annual tables are released in the form of the Continuous Mortality Investigation reports \citep{CMI2015}. Being official proposals of the actuarial Societies, such tables enjoy wide use and are also heavily used in the valuation of pension and post-retirement medical liabilities. For example, in the US the SOA tables have been included by the Internal Revenue Service for the purposes of the Pension Protection Act of 2005, or by the Congressional Budget Office for long-term forecasts.}


 The basic aim in constructing the tables is to convert the raw mortality data 
 into a graduated table of yearly mortality rates and improvement factors, broken down by age and gender. Since the goal is to forecast future mortality from retrospective experience, the process involves two fundamental steps: \emph{smoothing} raw data to remove random fluctuations resulting from finite data sizes; and \emph{extrapolating} future rates.  To maximize actuarial credibility of the tables, graduation techniques are applied, in particular for estimating mortality improvement trends based on past experience and then projecting those trends into future years. For example, see the RPEC reports \cite{SOA2014,MP2014} for the full description of constructing the US tables/scales, as well as more general SOA longevity studies in \cite{Purushotham11,Rosner13}.

In the present article, we propose a new methodology to graduate mortality rates and generate mortality improvement scales within a single statistical model. More precisely, we advocate the use of Gaussian process regression, a type of Bayesian nonparametric statistical model. Our aim is to provide a data-driven procedure that produces an alternative to existing methods while enjoying a number of important advantages:

\begin{itemize}
\setlength{\itemsep}{0pt}
 \item The GP framework is Bayesian, offering rich uncertainty quantification. The model produces mortality curves smoothed over multiple dimensions, as well as credible intervals which quantify the uncertainty of these curves. This is generated for in-sample smoothing and out-of-sample forecasts. In their basic form, the latter forecasts are Gaussian, allowing for a simple interpretation of the uncertainty by the actuary. Moreover, the GP model is able to generate stochastic \emph{trajectories} of future mortality experience. We demonstrate this projection over both age and calendar year, but the GP model can be consistently applied over higher dimensional data as well. From this, full predictive distributions for annuity values, life expectancies, and other life contingent cash-flows can be produced. Such analyses can provide core components of stress testing and risk management of mortality and longevity exposures.

 \item Using GPs leads to unified modeling of mortality rates and mortality improvement; one may analytically differentiate the mortality surface to obtain mortality trends (and corresponding credible bands) over calendar years. This structure offers a coherent approach to both tables, jointly quantifying uncertainty in rates and improvement factors.

  \item Standard graduation techniques are sensitive to edge issues, i.e.~the experience in the latest few years. For example, to achieve a better prediction, the MP-2015 method extrapolates rates from 2011 onwards, effectively excluding the last several years of data (as of this writing, CDC data go up to 2014). In contrast, our GP approach intrinsically handles the specific shape of the data and is well suited to incorporating missing  data. Therefore, dropping the ``edge years" is not necessary with GP, with its self-adjusting credible bands.

  \item The GP approach provides natural ``updating'' of mortality tables in terms of incorporating the latest mortality experience. The end users can  easily update the tables, no longer requiring reliance on official  updates.
\end{itemize}

To recapitulate, the main contribution of this article is to propose the use of Gaussian process regression for constructing mortality  tables and improvement factors. While being a relatively new ``machine learning'' paradigm, the underlying statistical methodology and most crucially the software implementation has matured significantly in the past decade. To wit, all of the case studies below have been implemented straightforwardly using publicly-available, free, well-documented software, and required only basic programming skills. As a companion to the manuscript, we provide an \texttt{R} markdown (.rmd) notebook file containing a concise version of the tables and figures produced in this document, along with the \texttt{R} code used to produce them; please see \href{https://github.com/jimmyrisk/GPmortalityNotebook}{github.com/jimmyrisk/GPmortalityNotebook}.  \rev{Our main focus is on the GP methodology, and accordingly we concentrate on \emph{describing} the important components of this framework in the context of mortality modeling. We illustrate its application over several mortality datasets and show that GPs are competitive with existing methods in terms of their performance and predictive power.}


\blue{From the empirical direction, our data-driven analysis sheds light on the question of latest mortality experience, whereby mortality improvements appear to have significantly moderated from past trends. Specifically, after implementing the above framework on the latest US mortality experience, we document that as of 2015, mortality improvement factors are (statistically) zero, and possibly \emph{negative}  for ages 55--70 since as early as 2012.  These estimates diverge significantly from SOA projections embedded in MP-2015 that continue to bake in past improvements. Lower mortality improvement rates would have a material impact across the pension industry. This paper offers statistical support to the anecdotal demographic evidence of declining US longevity and calls into question traditional backward-looking methods for constructing mortality improvement factors.}


\subsection{Comparison to Other Approaches}\label{sec:comparison}

Mortality experience is summarized by a mortality surface, indexed by Age (rows $i$) and calendar Year (columns $j$). Typical data consists of two matrices $\bm{D}$ and $\bm{E}$ (or $\bm{L}$), listing the number of deaths $D$,  exposed-to-risk $E$, or the mid-year population $L$, respectively.
In the first step, one postulates a relationship between the individual elements of these matrices, ${D}_{ij}$ and ${E}_{ij}$, in terms of the latent (logarithmic) mortality state $\mu_{ij}$. In the second step, one estimates $\mu_{ij}$ through a statistical fitting approach.  We may identify two classes of estimation: (i) data-driven models that infer $\mu_{ij}$ by statistical smoothing techniques; (ii) factor models that express $\mu_{ij}$ in terms of several one-dimensional indices. For example, in Age-Period-Cohort (``APC'') models those factors are additive and correspond to Age, Year and Cohort effects; in Lee-Carter \citep{lee1992modeling} models they correspond to Age, Year, and an Age-Year interaction term. A common distinction is to assume a non-smooth evolution of the mortality surface in time, coupled with a smooth Age effect. The latter Age-modulating terms are then fitted non-parametrically by maximum likelihood, or given a fixed functional form, such as linear or quadratic in Age \citep{cairns2006two,HuntBlake14}. Imposing an underlying one-dimensional structure facilitates interpretation of the fitted output, but potentially leads to model risk. In contrast, the data-driven methods, dating back to the classical graduation technique of \cite{Whittaker22}, maintain an agnostic view of mortality experience, and solely focus on removing random fluctuations in observed deaths. Modern frameworks typically work with various types of splines, extending the seminal work by \cite{Currie04} (see also a modern software implementation in \cite{MortalitySmooth}). Here, the main challenge is appropriate smoothing across both Age and Year dimensions; some of the proposed solutions include constrained and weighted regression splines \citep{HyndmanUllah07}, extensions to handle  cohort effects that generate ``ridges'' \citep{HyndmanDokumentov14}, and a spatio-temporal kriging approach \citep{DebonMontes10}. \rev{After completing the first version of this work, we learned of the independently executed PhD thesis of \cite{wu2016gaussian} that among other things also considered a GP model for mortality.}
A mixed strategy of first smoothing the data non-parametrically, and then inferring underlying factor structure was proposed and investigated in \cite{HyndmanUllah07}. Finally, we also mention Bayesian approaches \citep{CzadoDenuit05,KingBook} that replace MLE-based point estimates with a posterior distribution of the mortality rate. To date, there is no consensus on which framework is the most appropriate.  For example, the influential study by \cite{Cairns09} considered eight different mortality models.  Another recent study by \cite{Currie16} looked at 32 models, nesting the former eight.

A further reason for the large number of models is the use of different link functions (log-Poisson, logit-Poisson, logit-Binomial, etc.), that connect the logarithmic mortality state to deaths and exposures. \rev{These choices correspond to using different generalized linear models (GLM) and affect the optimization procedure (usually some variant of maximum likelihood) for model calibration.} The Binomial model is defined as $D_{ij} \sim Bin( E_{ij}, e^{\mu_{ij}})$ \citep{HyndmanUllah07}; the Poisson model $D_{ij} \sim Poisson( L_{ij} e^{\mu_{ij}})$ \rev{\citep{BrouhnsDenuit02,renshaw1996modelling,sithole2000investigation}}; and the Gaussian model $\frac{D_{ij}}{E_{ij}} \sim \mathcal{N}( e^{\mu_{ij}}, \sigma^2 E_{ij})$ \citep{KingBook}. A related issue is regularization of the estimated factors that can be achieved via penalization, see  \cite{delwarde2007smoothing,Currie13}.

In terms of forecasting future mortality, a popular strategy is to differentiate the treatment of the Age index, which is incorporated directly into the mortality state and smoothed appropriately, vis-a-vis the Year index, whose impact is estimated statistically using time-series techniques. This is the basic idea of Lee-Carter models, which construct a time-series process for the Year factor(s) to extrapolate mortality trends and assess forecast uncertainty. More generally, this can be viewed as a principal component approach, expressing the Age-effect as a smooth mortality curve $\mu_t(x_{ag})$ \rev{in the age dimension $x_{ag}$}, fitted via functional regression or singular value decomposition techniques, and then describing the evolution of this curve over time \citep{RenshawHaberman03,HyndmanUllah07} as a multivariate time-series. In contrast, in the pure smoothing methods, all covariates are given equal footing, and forecasting is done by extrapolating the fitted \emph{surface} to new input locations.

Precise methods for constructing mortality tables are not without controversy, especially when it comes to extreme age longevity or future forecasts.  Ideally one ought to just let the ``data speak for itself''. However, this is in fact a very challenging issue, not least because the question of predictive forecasting must acknowledge that any given fixed forecast is only a \emph{point estimate}, and that there is always an element of uncertainty around the prediction. A common paradigm is to specify a stochastic model for mortality which directly prescribes future uncertainty. This is especially relevant for risk management or pricing applications, where the actuary wishes to incorporate (and hopefully manage) mortality risks.  However, most stochastic mortality frameworks are ``reduced-form'' in the sense of specifying a low-dimensional stochastic system with just a few parameters/degrees-of-freedom. For implementation, one ``calibrates'' the model to data by minimizing e.g.~the mean-squared error.  In contrast, the RP-2014 mortality table is bottom-up, aiming to directly specify the full mortality experience with minimal a priori specifications. Relative to these two basic strategies, the approach proposed in this article views uncertainty in forecasts as intrinsic to the statistical model, so that all credible bands are obtained simultaneously both in-sample and out-of-sample.

\rev{Our proposed approach recasts mortality surface calibration as a  \emph{spatial regression} task. Originating in environmental/spatial statistics, Gaussian process regression (also known as kriging) takes a functional nonparametric approach to learning the latent response surface $f$~\citep{cressieBook}. Inference of $f$ is viewed as conditioning on the observed data and fitting the GP concentrates on estimating the spatial dependence.  Recently, GP models have gained currency as a machine learning tool for spatio-temporal forecasting thanks to their ability to capture complicated nonlinear dynamics with a high degree of analytic tractability
and a minimum of tunable hyperparameters. An introduction to the vast GP landscape can be found in the monograph by~\cite{WilliamsRasmussenBook}. }

\subsection{Mortality Dataset}\label{sec:mortalityDataset}

Our study is US-centric and originated from discussions of the SOA's MP-2014 and successor tables. There has been some controversy that the scale excluded more recent trends, specifically a slowing of mortality improvement that was not fully reflected in the MP-2014 tables.  Indeed, a year later, the SOA updated the MP-2014 tables to the MP-2015 tables to include two additional years of mortality experience, and the new tables did in fact reflect a material drop in mortality improvement. In the interim, the CDC has also released new data showing a continued decline in mortality improvement levels. 

The mortality data we use comes from  Centers for Disease Control (CDC). The CDC data covers ages 0--84 and goes up to 2014 as of the time of writing.  For each cell of the table, the CDC data specifies the raw mortality rate for the exposed population. The mid-year exposures $L_{ij}$ are based on inter-censal estimates interpolated based on the 2000 and 2010 census counts. Thus, $e^{\mu_{ij}}$ corresponds to central death rates. Table \ref{table:2011mortality} provides a snapshot of the latest year of CDC data (2014).   The rapid decrease in sample size causes large variability in reported mortality rates at extreme ages. For a visual representation, two representative years of raw CDC data for Males aged 60--70  are plotted as the solid lines in Figure \ref{fig:GP-smooth-multiyear} in Section \ref{sec:retro-analysis}.  The figure shows the (super-) exponential increase in mortality with respect to age, along with a clear need for data smoothing.

\begin{table}
\centering
\begin{tabular}{cc|cc|cc} \hline
 \multicolumn{2}{c|}{Inputs $x^n$}  &  \multicolumn{2}{c|}{Log Mortality Rate $y^n$} & \multicolumn{2}{c}{Mortality Rate $\exp(y^n)$}\\ \hline
Age $(x^n_{ag})$ & Year $(x_{yr}^n)$ & \multicolumn{1}{c}{Male} & \multicolumn{1}{c|}{Female} & \multicolumn{1}{c}{Male} & \multicolumn{1}{c}{Female}\\
50 & 2011 & -4.931 & -5.437 & 0.00722 & 0.00435\\
64 & 2011 & -4.264 & -4.707 & 0.01406 & 0.00901\\
74 & 2011 & -3.435 & -3.821 & 0.03222 & 0.02191\\
84 & 2011 & -2.408 & -2.714 & 0.08999 & 0.06625\\
\end{tabular}
\caption{Excerpt of CDC mortality data to compare exposures and mortality rates over Ages and gender for calendar year 2011.  \emph{Mortality} is the observed proportion $D^n/L^n$ of the deceased during the Year relative to the mid-year population.\label{table:2011mortality}}
\end{table}


As our training dataset, we used the CDC database covering ages 50--84 in years 1999--2014. \blue{(Another data source is provided by Social Security Administration (SSA) and was utilized by RPEC.)} \rev{Since our main aim is to obtain the \emph{present} mortality rates and to forecast short-term calendar trend (through estimating mortality improvement factors) most relevant for actuarial applications, we only consider older ages and recent years. Our analysis targets the early retired group; additional challenges related to handling very young (e.g.~infant mortality) and very old $85+$ ages are discussed in Section \ref{sec:inhomogeneousGPmodels}. Furthermore, since we approach mortality as a non-stationary surface evolving in Age and Year, we discard most of $20^{th}$ century data, as
 distant mortality experience is less influential for our analysis.   To understand the impact of excluding some data, we also considered several subsets listed in Table~\ref{table:MortData}.}

In comparison to our dataset, the most recent MP-2015 scales incorporate actual smoothed rates up to 2010 with projections thereafter.  However, the CDC already provides actual mortality experience up to 2014.  Further results are provided in the Appendix based on matching US Female data. Additionally, in an online supplement \href{https://github.com/jimmyrisk/UKJapanResults}{github.com/jimmyrisk/UKJapanResults} we present results for Japan and UK males and females based on Human Mortality Database (HMD) \citep{wilmoth2010human} datasets. To keep the analysis consistent across countries, in all cases we worked with the equivalent of All Data, i.e.~ages 50--84 and years 1999--2014 (HMD contains additional years and ages if desired).



\begin{table}
\centering
\begin{tabular}{l|l|l} \hline
Set Name & Training Set & Test Set\\ \hline
All Data & 1999--2014, ages 50--84 & N/A: In-Sample\\
Subset I & 1999--2010, ages 50--84 & 2011--2014, ages 50--84\\
Subset II & 1999--2010, ages 50--84 \& 2011--2014, ages 50--70 & 2011--2014, ages 71--84\\
Subset III & 1999--2010, ages 50--70 & 2011--2014, ages 71--84\\ \hline
\end{tabular}
\caption{Data sets used in analysis.  Mortality data is taken from CDC as described in Section \ref{sec:mortalityDataset}.}\label{table:MortData}
\end{table}

\section{Gaussian Process Regression for Mortality Tables}
In this paper, we focus on analyzing mortality rates over a two-dimensional input space, namely Age and Year. The mortality data is viewed as a table of $N$ ``cells'' (see rows of Table \ref{table:2011mortality}), represented by inputs $x^n$ and outputs or responses $y^n$, $n=1,\ldots, N$. In our case, $x^n$ is in fact a tuple and represents the pair $(x^n_{ag}, x^n_{yr})$. For example, $x^n = (78, 2016)$ is the input for ``78-year old in 2016'' cell. We use the logarithmic central mortality rate for $y^n$, namely $y^n = \log (D^n/L^n)$ where $D^n$ and $L^n$ represent the annual deaths and midyear count of lives, respectively, for the $n$-th cell.  The overall inputs $\mathbf{x}=x^{1:N}$ and  observations $\mathbf{y}=y^{1:N}$ are denoted by boldface and aggregated into the mortality dataset $\mathcal{D} = (\mathbf{x},\mathbf{y})$. Superscripts identify individual inputs/outputs, subscripts distinguish coordinates, e.g.~$x^n_{ag}$.


\begin{remark}
This point of view treats calendar Year as simply another covariate and is easily extendible to further input dimensions, such as Select Period, et cetera. Also the format easily allows for missing cells,
which, for example, is a common issue for dealing with extreme ages (95+).
\end{remark}

\subsection{Basics of Gaussian Processes}
\blue{In traditional mortality regression, a parametric function, $f$, is postulated which maps the inputs $\mathbf{x}$ to the noisy measurements of the log-mortality rate, $\mathbf{y}$.} A cell is modeled as
\begin{equation}\label{eq:YequalsFpluserror}
y^i = f(x^i)+\eps^i,
\end{equation}
where $\eps^i$ is the error term. With a GP, the function $f$ is deemed to be latent and is modeled as a random variable. Consequently, a GP is defined as a set of random variables $\{f(x) | x \in  \mathbb{R}^d\}$ where any finite subset has a multivariate Gaussian distribution with  mean $\bm{m}(\cdot)$ and covariance $\bm{C}(\cdot, \cdot)$.  That is for any $n$-tuple $\mathbf{x}=x^{1:n}$:
$$
f(x^1), \ldots, f(x^n) \sim \mathcal{N} \left(mean= \bm{m}(\mathbf{x}), covariance= \bm{C}(\mathbf{x},\mathbf{x}) \right).
$$
 In shorthand, we write $f(\mathbf{x}) \sim GP(\bm{m}(\mathbf{x}), \bm{C}(\mathbf{x},\mathbf{x}))$.  An important concept of a GP is that each mortality rate is correlated with every other mortality rate: above, $\bm{C}$ is a $n \times n$ matrix with entry $C(x^i, x^j)$ representing the covariance between the $i$-th and $j$-th cells.

 \begin{remark}
 \rev{We emphasize that the assumption that $f(\cdot)$ forms a GP is solely a statistical representation of the mortality surface, similar for example to assuming that $\mu$ can be described in terms of splines. Practically, the assumption is about the \emph{shape} of $\mu(\cdot)$; for typical kernels the corresponding functional space is dense in the class of continuous functions. It can be compared to the APC models that decompose $\mu$ as a sum of \emph{one-dimensional} factors. A more relevant question concerns the mapping from the latent surface to the observed mortality; \eqref{eq:YequalsFpluserror} assumes an additive noise structure like in the classical least-squares framework. Because mortality experience comes from count data, a generalized linear model could be viewed as better suited, see Section~\ref{sec:obs-model}.}
 \end{remark}

Once we collect data $\mathcal{D}$
, the next step is to determine the posterior distribution for $f$, namely $p(f|\mathcal{D})$. That is, we want to know the distribution of mortality rates, given the experience data. Using Bayes' rule, we have
\begin{align*}
  p(f|\mathcal{D}) \propto p(\mathbf{y}|f,\mathbf{x})p(f) = \{ likelihood \} \cdot \{ prior \}
\end{align*}
where $p(\mathbf{y}|f,\mathbf{x})$ is the ``likelihood'' and $p(f)$ the ``prior''. To complete the definition of the GP, we therefore need to define the ``prior'',  $p(f)$.  This is equivalent to setting the initial assumptions for mean function $\mathbf{m}$ and covariance function $\mathbf{C}$.

\textbf{The Prior Mean Function}:  \blue{the prior mean $m(x)$ stands in for our belief about mortality rate at input $x$ in the absence of any historic data.} We might, for example, define $m(\cdot)$ as a Gompertz or Makeham curve in the age coordinate $x_{ag}$.  However, we will show that the choice of $m(\cdot)$ has little impact on the output of the GP model for purposes of \emph{in-sample} smoothing.  Even if we set $m(x)=0$ or $m(x)=\beta_0$ for some constant $\beta_0$ and for all $x$, the results will be largely unaffected\blue{, since the posterior mean is largely dominated by the impact of the data}.  
  However, for purposes of \emph{out-of-sample} projections, we will conversely show that a more realistic choice of $m(\cdot)$ is required for long term mortality projections.

\textbf{The Covariance Function:}  A core concept of a GP is that for any cells $i,j$, if $x^i$ and $x^j$ are deemed to be ``close'', then we would expect the outputs, $y^i$ and $y^j$, to be ``close'' too. For example, the mortality rate for a 60 year old in 2016 $(x^i = (60, 2016))$ will be closer to that of a 61 year old in 2017 $(x^j = (61, 2017))$, than that of a 20 year old in 1990 $(x^j = (20,1990))$. This idea is mathematically encapsulated in  $C$: the closer $x^i$ is to $x^j$, the larger the covariance $C(x^i, x^j)$.  It follows, that if $x^i$ and $x^j$ are very close, knowledge of $y^j$ will greatly affect our expectations of $y^i$.  Conversely, if $x^i$ is far from  $x^j$, then knowledge of $y^j$ will have little influence on our expectations of $y^i$.

\textbf{The Posterior Function:}  To project mortality, we evaluate the GP function on new Age and/or Year inputs $\mathbf{x}_*$, i.e.~evaluate $\bm{f}_* = f(\mathbf{x}_*)|\mathcal{D}$. We show in the next subsection that when $m$ is a constant and the likelihood function is Gaussian, then the posterior distribution for $\bm{f}_*$ can be determined analytically. In fact, this posterior itself is a new GP $\bm{f}_*(\mathbf{x}_*)| \mathcal{D} \sim GP( \bm{m}_*(\mathbf{x}_*), \bm{C}_*(\mathbf{x}_*,\mathbf{x}_*))$  with an updated mean and covariance functions, specified in \eqref{eq:GP-mean-vector}. The posterior mean $\bm{m}_*(\mathbf{x}_*)$ is interpreted as the model prediction for inputs $\mathbf{x}_*$, and the posterior covariance $\bm{C}_*(\mathbf{x}_*,\mathbf{x}_*)$ gives a goodness-of-fit measure for this prediction.

The posterior function can be used for both projecting mortality, as well as producing in-sample smoothed mortality curves.  For the latter, all we need to do is set $\mathbf{x}_*= \mathbf{x}$, namely the training set inputs.  In this case, the mean $\bm{m}_*(\mathbf{x})$ of the posterior will produce a smooth set of mortality rates, and the posterior variance $\bm{C}_*(\mathbf{x},\mathbf{x})$ quantifies the uncertainty around $\bm{m}_*(\mathbf{x})$.  Alternatively, if $\mathbf{x}_*$ represents inputs of future calendar years, then the posterior will produce an out-of-sample projection of the mortality curves.  By fitting a GP, and then analyzing the posterior we are able to achieve the following:
\begin{itemize}\setlength{\itemsep}{0pt}
\item Estimate the historic smoothed mortality curves by calendar year ($\bm{m}_*(\mathbf{x})$ above);
\item	Estimate a credible interval around such curves (use the posterior covariance $\bm{C}_*(\mathbf{x},\mathbf{x})$);
\item	Project the curves forward ($\bm{m}_*(\mathbf{x}_*)$ for future inputs $\mathbf{x}_*$);
\item	Estimate the credible intervals for such projections ($\bm{C}_*(\mathbf{x}_*,\mathbf{x}_*)$);
\item   Generate stochastic future forecasts (sample from the random vector $\mathbf{f}_*(\mathbf{x}_*)$ as a future mortality scenario);
\item	Smooth curves over all dimensions, using automatically determined tuning parameters.
\end{itemize}

Note that the above projections are about $\bm{f}_*$. Depending on the context, an actuary might also wish to project future mortality experience $\mathbf{y}_*$ whose marginal credible intervals are necessarily wider. When the noise $\epsilon$ is additive and has a Gaussian distribution, $\mathbf{y}_*$ in fact remains a GP with same mean as $\bm{f}_*$, and a modified variance due to the variance of $\epsilon$. Practically, forecasting realized mortality (for example, in connection with realized annuity payouts) requires also predicting future exposures $E$.

\begin{remark}
In a Lee-Carter framework one first postulates a parametric form for the mortality experience, such as
\begin{align}\label{eq:lee-carter}
\mu_{ag,yr} = \alpha_{ag} +\beta_{ag} \kappa_{yr} + \eps_{ag,yr}
\end{align}
where $\bm{\alpha}$ is the Age shape, $\bm{\beta}$ is the age-specific pattern of mortality change and $\bm{\kappa}$ is the Year trend. In the second step, after fitting $\bm{\alpha},\bm{\beta}$ by maximum likelihood, one then postulates a time-series model for the $\kappa$ factor. Relative to a pure regression model such as ours, the Lee-Carter method treats Age and Year dimensions completely differently; moreover the fit for the Age/Period factors is done globally (i.e.~from the full dataset used), so that even spatially distant data \blue{directly influences all predictions}. Finally, Lee-Carter has no mechanisms for (i) smoothing in-sample experience (beyond model calibration), and (ii) incorporating the uncertainty of the Age/Period factors in out-of-sample forecasts; its forecasts are stochastic only insofar as the time-trend is uncertain.
\end{remark}

\subsection{Mathematical Details}\label{sec:mathematical-details}

GP regression takes a response surface approach, postulating an unknown, nonparametric functional dependence between covariates (inputs) $\mathbf{x}$ and outputs $\mathbf{y}$,
\begin{equation}\label{eq:kriging}
\mathbf{y} = \mathbf{f}(\mathbf{x})+\eps,
\end{equation}
where $f$ is the \emph{response surface} (or regression map) and $\eps$ is the mean-zero noise term \blue{with observation variance $\sigma^2(x)$}, independent across $x$'s. \blue{The meaning of the noise term are the statistical fluctuations that lead to deviations between observed raw mortality rates and the latent ``true'' rates that are being modeled. The strength of these fluctuations $\sigma(x)$ is interpreted as the credibility of the corresponding mortality cell exposure.} We remind the reader that throughout the paper, $y^n$ represents log-mortality\blue{, and  $x^n=(x^n_{ag},x^n_{yr})$ is a two--dimensional age--year pair}. In Gaussian process regression, the map $f$ is assumed to be a realization of a Gaussian process with covariance kernel $C$ that controls the spatial smoothness of the response surface. The GP model starts with a prior on $f$'s over the function space $\mathcal{M}$ and then computes its posterior distribution conditional on the data $\mathcal{D}$
. The function space specifying potential $f$'s is a reproducing kernel Hilbert space based on the kernel $C$. The GP assumption that $f$ is generated by a Gaussian process implies that the posterior distributions are also Gaussian. Hence at any fixed input $x$, the marginal posterior is $f_*(x) \sim \mathcal{N}( m_*(x), C_*(x,x)),$ where $m_*$ is the predictive mean (also the posterior mode, hence maximum a posteriori (MAP) estimator), and $C_*(x,x)$ is the posterior uncertainty of $m_*$. $C_*(x,x)$ offers a principled empirical estimate of model accuracy, serving as a proxy for the mean-squared error of $m_*$ at $x$.

A GP model $GP( \bm{m}(\mathbf{x}), \bm{C}(\mathbf{x},\mathbf{x}))$ is specified through its mean function $m(x^i)=\E[f(x^i)]$ and covariance \blue{$C(x^i,x^j) = \E[ \left(f(x^i)-m(x^i)\right) \left(f(x^j)-m(x^j)\right)]$}. Specifically, the prior of $\bm{f}(\mathbf{x})$ is  $p(\bm{f} | \mathbf{x} ) = \mathcal{N}( \mathbf{m}, \bm{C})$, where $\mathbf{m} = \blue{\left(m(x^i)\right)_{1 \leq i \leq N}}$ and  $\bm{C} = (C(x^i, x^j))_{i,j}$. In the standard case, it is further assumed that the noisy observations vector $\mathbf{y}$ has a Gaussian relationship to the latent $\bm{f}$, i.e.~$\eps^i \sim\mathcal{N}\left(0,\sigma^2(x^i) \right)$, so that
 \begin{align}\label{eq:y-likelihood}
 p(\mathbf{y} | \bm{f}) = \mathcal{N}( \mathbf{y} | \bm{f}, \bm{\Sigma}),
  \end{align}
  where 
  $\bm{\Sigma} = diag( \sigma(x^i)^2)$ is the $N\times N$ noise variance matrix. \blue{Certainly, assuming $\eps$ to be Gaussian with a prescribed variance is not realistic for mortality modeling, but as we show this has minimal statistical effect; we return to this point later.} Equation \eqref{eq:y-likelihood} implies that if $\bm{f} \sim GP(\mathbf{m}(\mathbf{x}), \bm{C}(\mathbf{x},\mathbf{x}))$  then $\mathbf{y} \sim GP( \mathbf{m}(\mathbf{x}), \bm{C}(\mathbf{x},\mathbf{x}) + \bm{\Sigma})$.

Thanks to the Gaussian assumption, determining the posterior distribution $p(\bm{f} | \mathbf{y})$ reduces to computing the predictive mean $\bm{m}_*$ and  covariance $\bm{C}_*$.  Combining the above likelihoods and denoting by $\Theta$ the hyper-parameters of the GP model, the log-likelihood
 is
\begin{align}\label{eq:log-lik}
\log p(\mathbf{y} | \mathbf{x}, \Theta) = -\frac{1}{2} \mathbf{y}^T (\bm{C} + \bm{\Sigma})^{-1} \mathbf{y} - \frac{1}{2} \log | \bm{C} + \bm{\Sigma} | - \frac{N}{2} \log (2\pi),
\end{align}
\blue{where $\mathbf{y}^T$ denotes vector transpose.}

The basic GP model treats the prior mean function $m$ as given (i.e.~known and fixed). \blue{In Section \ref{sec:UnivKriging} we discuss the more relevant case where we simultaneously infer a parametric prior mean $m(x)$ and the kernel hyperparameters, which is known as Universal Kriging. For now,} by de-trending via $\mathbf{f}-\bm{m}(\mathbf{x}),$ we may assume without loss of generality that $\bm{f}$ is centered at zero and $m \equiv 0$. The resulting 
posterior distribution $\bm{f}_*(\mathbf{x}_*)$ at a vector of inputs $\mathbf{x}_*$ is multivariate Gaussian \citep{kmPackage-R} with mean/covariance:
\begin{align}
\bm{f}_*(\mathbf{x}_* | \mathbf{x},\mathbf{y}) \sim GP \Bigl( mean &= \bm{C}(\mathbf{x},\mathbf{x}_*)^T (\bm{C}+\bm{\Sigma} )^{-1} \mathbf{y},  \\ covariance &= \bm{C}(\mathbf{x}_*,\mathbf{x}_*) - \bm{C}(\mathbf{x},\mathbf{x}_*)^T (\bm{C}+\bm{\Sigma} )^{-1} \bm{C}(\mathbf{x}_*,\mathbf{x}) \Bigr),\label{eq:GP-mean-vector}
\end{align}
where $\bm{C}^T$ is the transpose of $\bm{C}$.

\blue{The effect of \eqref{eq:GP-mean-vector} is that if we have new inputs $\mathbf{x}_*$, then draws from the posterior distribution of $\bm{f}_*$ at $\mathbf{x}_*$ will be primarily influenced by historic data that have inputs close to $\mathbf{x}_*$. Marginally at a single cell $x_*$, and similar to kernel regression, the predicted value $m_*(x_*)$ is a linear combination of observed $y^i$'s, capturing the idea of the GP model nonparametrically smoothing the raw mortality data.  The covariance kernel $C$ quantifies the relative contribution of different $y^i$'s in terms of the distance of their $x^i$'s to $x_*$,  see Section \ref{sec:kernel-families} below.}
\subsubsection{Observation Model}\label{sec:obs-model}

The observation noise matrix $\bm{\Sigma}$ represents the credibility of the corresponding observations $y$'s and is used by the GP to automatically
%
determine how much of interpolation versus smoothing to carry out; in the limiting case $\sigma=0$, \blue{the posterior mean}  exactly interpolates the observation $y^i$ at $x^i$: $m_*(x^i) = y^i$.

\rev{In reality, the credibility of mortality experience is non-constant because of the different number of exposed-to-risk in different age brackets. Indeed, in the existing literature it is common to replace the additive noise structure of \eqref{eq:y-likelihood} with a GLM approach to  match the fact that observed mortality is based on the counts $D_x, L_x$. A popular choice is a (log-link) Poisson GLM model that replaces \eqref{eq:y-likelihood} (equivalent to $D_x = e^{f(x) + \eps(x)} L_x$) with
\begin{align}\label{eq:poisson-obs}
D_x \sim Poi( e^{f(x)} \cdot L_x)
\end{align}
and constructs a linear model
\begin{align}\label{eq:poisson-glm}
f(x) = \sum_i \beta_i B_i(x),
\end{align}
see \citep{BrouhnsDenuit02,CzadoDenuit05}.  Because mortality data tend to exhibit \emph{over-dispersion}, other approaches like Negative binomial GLM have been proposed instead of \eqref{eq:poisson-obs}. In \cite{Currie04} a Poisson GLM model was adjusted by fitting an age-dependent overdispersion factor with a spline.}

\rev{Conceptually, it is straightforward to combine a GLM link function with a GP model: one simply adjusts the log-likelihood function in \eqref{eq:log-lik} and proceeds to fit the GP hyperparameters. This is equivalent to working with \eqref{eq:poisson-obs} where $D_x, L_x$ are observations and $f(\mathbf{x}) \sim GP(\bm{m}(\mathbf{x}), \bm{C}(\mathbf{x},\mathbf{x}))$ is the latent Gaussian process.
 A small caveat is that the non-Gaussian observations \eqref{eq:poisson-obs} ruins the Bayesian conjugacy, so that the posterior $f_*$ is no longer Gaussian. The typical solution is a Laplace approximation which constructs a Gaussian distribution for $f_*$ around the posterior mode. Such details are gracefully handled by the software packages and do not pose practical difficulties. Because the exposed counts $E_x$ are very large (on the order of $10^5$ or more), the Gaussian likelihood approximation to $D_x/E_x$ is very close, see Table~\ref{table:Poisson-gp}.}

\rev{To capture the non-uniform credibility of the different cells one may take the noise level $\sigma(x^i)$ to be state-dependent.
Specifically, the mortality table structure can be used to estimate $\sigma(x^n)$: $L^n \cdot \exp\{y^n\} $ is expected to be binomially distributed with parameters $p^{(n)} := D^n/E^n$, and  size $E^n \simeq L^n + D^n/2$.  We then have $\text{Var}\left(\exp\left\{y^n \right\}\right) = p^{(n)} (1-p^{(n)})/E^n,$ and large population  $L^n$ implies the delta-method estimate
\begin{equation}\label{eq:deltamethod}
\sigma^2(x^n)=\text{Var}\left(y^n\right) \simeq \frac{(1-p^{(n)})}{p^{(n)} E^n}.
\end{equation}
We find however that \eqref{eq:deltamethod} does not perform well, partly due to the mentioned over-dispersion effect and  partly because the computed $p^{(n)}$  is not the true mortality rate. In fact, our experience has been that a precise estimate of $\sigma(x^n)$ is not important for GP performance, because $\sigma$ is only used for smoothing.}
%
\blue{Specifically, in our main analysis we take  $\sigma^2(x^n) \equiv \sigma^2$ to be an unknown constant, estimated as part of fitting the model.} We return to this issue in Section \ref{sec:results}.

\subsection{Covariance Kernels and Parameter Estimation}\label{sec:kernel-families}
Given the covariance kernel $C$,
\eqref{eq:GP-mean-vector} fully specifies the posterior distribution $\bm{f}_*(\mathbf{x}_*) | \mathcal{D}$ conditional on the dataset $\mathcal{D}$. GP inference is thus reduced to simply applying the above formulas, akin to the ordinary least-squares (OLS) equations that specify the coefficients of a linear regression model. Of course in practice the kernel $C$ is not known and must be inferred itself. This corresponds to fitting the hyperparameters $\Theta$.

Our examples use the separable, spatially-stationary kernel of the squared-exponential family, which written out explicitly takes
\begin{align}\label{eq:sqExp}
{C(x^i,x^j)  =  \eta^2 \exp \left( -\frac{(x^i_{ag}-x^j_{ag})^2}{2\theta_{ag}^2} - \frac{(x^i_{yr}-x^j_{yr})^2}{2\theta_{yr}^2}\right)}.
\end{align}
In \eqref{eq:sqExp}, covariance between $y^i$ and $y^j$ is determined by the distance between inputs of the respective cells, measured through the (squared) difference in Ages and  Years between $x^i, x^j$, and modulated by the $\theta$'s. \blue{This use of spatial dependence can be straightforwardly extended to incorporate other dimensions, such as year-of-birth cohorts to conduct an APC allocation, or to include duration, to create a select and ultimate mortality table in the context of life insurance mortality analysis.}

\blue{The hyper-parameters ${\theta}_\ell$ are called characteristic length-scales and their effect on the model is quite subtle. Informally, }
larger $\theta$'s result in smoother mortality curves, i.e.~correlation dissipates slower. Smaller lengthscales reduce smoothing and lead to ``rougher'' curves. (\blue{The form of \eqref{eq:sqExp} implies that the mortality curves are infinitely differentiable both in Age and Year dimensions.})
Note
that the two lengthscale parameters $\theta_\ell$ for Age and  Year are different, so that the covariance
kernel is anisotropic.
The lengthscales also determine the speed at which the latent process reverts back to its prior outside the dataset. \blue{ For example, considering the Year coordinate and the question of projecting mortality rates into the future,  the GP prediction will automatically blend smoothed mortality rates derived from the experience data and the specified Year trend. Indeed, $\bm{m}_*(\mathbf{x}_*)$ is a weighted average of observed experience $\mathbf{y}$, and $\mathbf{m}(\mathbf{x}_*)$, with the weights determined by the lengthscale parameters $\theta_{yr}$ and $\theta_{ag}$. We contrast this to APC-type models where such blending is \emph{ad hoc} based on user-defined parameters.}

Two further GP parameters are the process variance $\eta^2$ which controls the natural amplitude of $f$ and the observation noise $\sigma^2$ in \eqref{eq:YequalsFpluserror} which is viewed as a constant to be estimated. Thus, the overall hyperparameter set is $\Theta \doteq (\theta_{ag}, \theta_{yr}, \eta^2, \sigma^2)$.

The classical method for inferring $\Theta$ is obtained by optimizing the marginal likelihood  $p(\mathbf{y} | \mathbf{x}, \Theta) = \int p( \mathbf{y} | \bm{f}, \bm{\Theta}) p(\bm{f} | \mathbf{x}, \bm{\Theta}) d\bm{f}$ which can be written out explicitly since all the integrands are Gaussian. This leads to a nonlinear optimization problem of simultaneously fitting $\theta_\ell$'s and variance terms $\eta^2, \sigma^2$.  Details on this procedure can be found in Section 3.2 of \cite{picheny2013nonstationary}. Alternatively, it is possible to directly specify $C$, for example from expert knowledge regarding the expected correlation in mortality rates. Given $\theta$'s, the MLEs for $\eta$ and $\sigma$ can be analytically inferred \citep{picheny2013nonstationary}. This approach increases interpretability of the final smoothing/prediction and makes the GP model less of a black-box.

\subsubsection{\blue{Fitting the Mean Function}} \label{sec:UnivKriging}

A generalized version of  \eqref{eq:kriging} incorporates a parametric prior mean  of the form $m(x) = \beta_0  + \sum_{j=1}^p \beta_j h_j(x),$ where $\beta_j$ are constants to be estimated, and $h_j(\cdot)$ are given basis functions. The coefficient vector $\bm{\beta} = (\beta_1, \dots, \beta_p)^T$ is obtained in parallel with computing $\mathbf{m}_*, \bm{\Sigma}$.
Letting $\bm{h}(x) \doteq \left(h_1(x), \ldots, h_p(x)\right)$ and $\bm{H} \doteq \left(\bm{h}(x^{1}), \ldots, \bm{h}(x^{N})\right),$  the posterior mean and variance at cell $x$ are \citep{kmPackage-R}
\begin{align} \label{eq:uk-equations}
\left\{ \begin{aligned}
\hat{\bm{\beta}} & =  \left(\bm{H}^T (\bm{C} + \bm{\Sigma})^{-1}\bm{H}\right)^{-1}\bm{H}^T (\bm{C} +  \bm{\Sigma})^{-1}\mathbf{y}; \\
m_{*}(x_*) &= \bm{h}(x_*)\hat{\bm{\beta}} + \bm{c}(x_*)^T (\bm{C} +  \bm{\Sigma})^{-1}(\mathbf{y}-\bm{H}\hat{\bm{\beta}});\\
s^2_{*}(x_*) &= C(x_*,x_*) + \left(\bm{h}(x_*)^T - \bm{c}(x_*)^T (\bm{C} +  \bm{\Sigma})^{-1}\bm{H}\right)^T\left(\bm{H}^T (\bm{C} + \bm{\Sigma})^{-1} \bm{H}\right)^{-1} \cdot \\
 & \qquad\qquad   \cdot\left(\bm{h}(x_*)^T - \bm{c}(x_*)^T (\bm{C} + \bm{\Sigma})^{-1} \bm{H}\right),
\end{aligned}\right.
\end{align}
where $\bm{c}(x_*) = \left(C(x_*,x^{i})\right)_{1 \leq i \leq N}$.
\blue{Note that \eqref{eq:uk-equations} reduces to \eqref{eq:GP-mean-vector} when $\bm{h} \equiv \bm{0}$.  We also see that}  the fitted coefficients $\bm{\beta}$ are in analogue to the classical least-squares linear model. 
\blue{A non-constant mean function is important for imposing structural constraints about the shape of the mortality curve, as well as the long-term improvement trends in mortality rates.} Appropriate choices for parameterizing $m$ are needed to be able to give reasonable out-of-sample projections, which corresponds to extrapolating in Age, or in calendar Year.


Use of a mean function for the GP via \eqref{eq:uk-equations} combines the idea of parametrically de-trending the raw data through a fitted Age shape, and then modeling the residual fluctuations into a single step. We note that as $m$ is assigned more and more structure, the residuals necessarily decrease and becomes less correlated. This calls to attention the typical over-fitting concern. Standard techniques, such as cross-validation or information criteria could be applied as safeguards, but their precise performance within the GP framework is not yet fully analyzed. We therefore confine ourselves to a qualitative comparison regarding the impact of the prior mean $m(\cdot)$ on the GP model output.

\subsubsection{Bayesian GP and Markov Chain Monte Carlo}\label{sec:bayesian}
One can also consider a fully Bayesian GP model, where the mean and/or covariance parameters have a prior distribution, see \citet{WilliamsRasmussenBook}. Bayesian GP implies that there is additional, intrinsic uncertainty about $C$ which is propagated through to the predictive distributions $f_{*}$.
 Starting from the hyper-prior $p(\Theta)$, the posterior distribution of the hyperparameters is obtained via $p(\Theta | \mathcal{D}) \propto p(\Theta) p( \mathbf{y} | \mathbf{x}, \Theta)$.  This hierarchical posterior distribution is typically not a GP itself.  Practically this means that one  draws realizations $\Theta^m$, $m=1,2,\ldots$ from the posterior hyperparameters and then applies \eqref{eq:GP-mean-vector} to each draw to compute $m_*(\mathbf{x}_* | \Theta^m), C_*(\mathbf{x}_*, \mathbf{x}_*) | \Theta^m)$.

 In general, sampling from $p(\Theta | \mathcal{D})$ requires approximate techniques such as Markov Chain Monte Carlo (MCMC). The output of MCMC  is a sequence $\Theta^1, \Theta^2, \ldots, \Theta^M$ of $\Theta$ values which can be used as an empirical approximation for the marginal distribution of $\Theta$, namely $p(\Theta | \mathbf{y},\mathbf{x})$.  From this sequence, it possible to calculate means and modes of the model parameters or use the $\Theta$ sequence directly to conduct posterior predictive inference.  A hybrid approach first specifies hyperparameter priors but then simply uses the MAP estimates of $\Theta$ for prediction (thus bypassing the computationally intensive MCMC steps).
 This idea is motivated by the observation that
 under a vague prior $p(\Theta) \propto 1$, the posterior of $\Theta$ is proportional to the likelihood, so that the MAP estimator $\hat{\Theta}$ which optimizes $p(\Theta | \mathbf{y},\mathbf{x})$ becomes identical to the MLE maximizer above.

 We note that standard MCMC techniques are not well suited for GP as the components of $\Theta$ tend to be highly correlated resulting in slow convergence of the MCMC chains. One solution is to use Hamiltonian Monte Carlo (HMC)~\citep{MCMChandbook} which is better equipped for managing correlated parameters.

\subsubsection{Setting Priors for the Bayesian Model}
To improve the efficiency of the MCMC routines, we first standardize the input covariates, for example $x_{ag,std}^i := (x_{ag}^i- mean(\mathbf{x}_{ag}))/sd(\mathbf{x}_{ag})$. We then set priors relative to this standardized data  model. Note that for comparative purposes with non-Bayesian models, the resulting posteriors of $\bm{\beta}$ and $\Theta$ then need to be transformed back to the original scale.

Priors are taken to be weakly informative, accounting for the specifics of each hyperparameter. For the lengthscale, $\theta_\ell$ should be below the scale of the input $x_\ell$, otherwise the resultant model will be essentially linear in the $\ell^{th}$ input dimension \citep{Stan}. Thus a prior that curtails values much beyond the data scale is appropriate.  After standardization, we found that $ \log \theta_\ell \sim \mathcal{N}(0,1)$ is reasonable. The $\eta$ parameter plays a role similar to that of the prior variance for linear model weights in a standard linear regression, and we found $\log \eta^2 \sim \mathcal{N}(0,1)$ prior to be reasonable for the linear and quadratic-mean models. The prior for $\sigma$ should reflect the noise in the data. For the CDC data, we set the prior $\sigma^2 \sim \mathcal{N}_+(0,0.2)$, restricted to be positive. When including trend, priors for the $\beta$ parameters are also required.  These are set similarly to standard regression coefficients.  In our analysis, we tested both Cauchy priors of Cauchy$(0, 5)$ or Gaussian priors of $\mathcal{N}(0,5)$ and found both to be reasonable.  For the intercept coefficient we chose $\beta_0 \sim \mathcal{N}(-4,5)$ to reflect log-mortality, whereby $\exp(y) \simeq 2\% = \exp(-3.9)$.

\begin{remark}
The Bayesian hierarchical approach for determining the parameters of the covariance matrix is also coined ``automatic relevance determination''. The Bayesian model will automatically select the values of $\theta_\ell$ and $\eta$ without the need for using cross-validation or other approaches to set the parameter levels.  Smaller values of $\theta$ amplify the effect of the difference calculation in the covariance matrix,  hence  determining the relevance of an input dimension.  Thus the Bayesian approach automatically sets the level of covariance among the $y$-values.
\end{remark}

\subsection{Software}\label{sec:software}

There are several software suites that implement Gaussian process modeling and can be used for our application. The software is complementary in terms of its capabilities and approaches, in particular for inferring the covariance kernel $C$ and for handling extensions of GPs discussed in Section \ref{sec:extensions} below.

To implement Bayesian GP models, we built models in Stan \citep{Stan}. Stan is a probabilistic programming language and is a descendant of other Bayesian programming languages such as BUGS and JAGS. In its default setting,  Stan's engine utilizes Markov chain Monte Carlo techniques, and in particular a version of Hamiltonian Monte Carlo (HMC) \citep{MCMChandbook}. Stan also allows the option of working with the MAP estimate $
\hat{\Theta}$ or the incorporation of non-conjugate priors, and implementation of idiosyncratic features within a model. Stan automatically infers the GP hyperparameters, specifically the lengthscales $\theta$'s, that determine the smoothness of the mortality curves. This allows for a more data-driven approach compared to traditional graduation that a priori imposes the degree of smoothing to apply to raw data.

Within the \texttt{R} environment, we utilized the package ``DiceKriging'' \citep{kmPackage-R}. \texttt{DiceKriging} can fit both standard and parametric trend \eqref{eq:uk-equations} models, and works with \rev{several} different kernel families (Gaussian, exponential, Mat\'ern). Moreover, \texttt{DiceKriging} can handle non-constant observation noise and has multiple options regarding the underlying nonlinear optimization setup. It estimates hyper-parameters through maximum likelihood (but does not do MCMC).

\section{Results}\label{sec:results}

We implemented a GP model for CDC mortality rates using a squared-exponential \eqref{eq:sqExp} covariance structure. To analyze and compare the different choices available within the GP framework we have experimented with:
\begin{enumerate}
  \item Other covariance kernel families, in particular Matern-5/2;
  \item MLE and Bayesian approaches to inference of hyperparameters $\Theta$;
  \item A variety of mean function specifications;
  \item Choice of inhomogeneous noise variance $\sigma(x).$
\end{enumerate}

For easier reading, the Figures and Tables below show the results for the Males; in the Appendix we report the corresponding Figures and Tables for Females. Most of the conclusions are identical for both genders; where appropriate we make further remarks.

We tested both the \texttt{DiceKriging} and Stan models as described in Section \ref{sec:software}. Table \ref{table:hyperparameters} reports the MLE and MAP hyperparameter estimates for the intercept-only models fitted with All data (Males aged 50--84, years 1999--2014, see Table \ref{table:MortData}).  All of the MLEs are quite close to the MAP estimates and both fall in the 80\% credible intervals for the MCMC runs. Closer analysis of the Stan output
revealed that the hyper-parameter posteriors are reasonably uncorrelated, justifying the use of the MAP estimates and corresponding marginal credible intervals.

Comparing both methods showed the resulting posterior distributions for the GP to be near identical, with the posterior means $m_*$ on average within 0.3\% (relative error) of each other, and the credible bands within 1.2\% of each other. This indicates stability of the GP estimates given slightly different hyper-parameters.

Consequently, the rest of the analysis in this paper is done using the simpler \texttt{DiceKriging} model which is quicker to fit and produces a convenient Gaussian posterior for the log-mortality (the fully-Bayesian model built in Stan can be viewed as a mixture-of-Gaussians). Similarly, there was no major difference in prediction and smoothing when picking different covariance kernels.  In general, picking a kernel is like picking a basis family for linear regression; basic caveats apply, but it is mostly a secondary effect.  Below we focus on the squared-exponential kernel. One benefit of this choice is that the resulting scenarios $f_*$ are guaranteed to be infinitely differentiable, which enables analytic treatment of instantaneous mortality improvement $\partial_{yr} f_*$, see Section \ref{sec:forecastingmortalityimprovement}.

\begin{table}[ht]
\centering
\begin{tabular}{l|r|rrr}
&  \multicolumn{1}{|c|}{\texttt{DiceKriging}} & \multicolumn{3}{|c}{Stan}\\ \hline
&  \multicolumn{1}{|c|}{MLE} & \multicolumn{1}{c}{MAP} & \multicolumn{1}{c}{MCMC Mean} & \multicolumn{1}{c}{MCMC 80\% Posterior CI}\\ \hline
$\theta_{ag}$ & 15.8384 & 14.7988 & 11.0401 & (6.3369, 17.0395) \\
$\theta_{yr}$ & 15.5308 & 15.7910 & 25.8306 & (14.6287, 39.4763) \\
$\eta^2$ & 1.8468 & 1.2365 & 1.9920 &  (0.8744, 3.3930) \\
$\sigma^2$ & 2.808e-04 & 2.753e-04 & 2.760e-04 & (2.536e-04, 2.998e-04) \\
$\beta_0$ & -3.8710 & -3.8003 & -3.8302 &  (-4.7305, -2.9350)\\ \hline
\end{tabular}
\caption{ Hyperparameter estimates based on maximum likelihood (\texttt{DiceKriging}) and maximum a posteriori probability (Stan), along with MCMC summary statistics.  The GP is fitted to all data and uses squared-exponential covariance kernel \eqref{eq:sqExp} with prior mean $m(x) = \beta_0$. Stan hyper-priors (on standardized data) were $\log \theta_{ag}, \log \theta_{yr}, \log \eta^2 \sim \mathcal{N}(0,1)$ i.i.d., $\sigma^2 \sim \mathcal{N}_+(0,0.2)$, $\beta_0 \sim \mathcal{N}(-4,5)$.\label{table:hyperparameters}}
\end{table}\



For the observation noise, estimating a constant noise variance led to MLE of $\hat{\sigma}^2 = 2.808\cdot 10^{-4}$. \blue{Figure \ref{fig:residualanalysis} in Appendix B gives a descriptive analysis of the resulting residuals; we observe that both the Gaussian assumption and the i.i.d~assumptions are statistically plausible.
As a further check, we tried to work with a non-constant $\sigma^2(x)$ by plugging-in the delta method estimate in \eqref{eq:deltamethod}. However, this  led to credible bands that are too narrow in terms of coverage ratios due to the aforementioned over-dispersion effect. Manual calibration found that $\check\sigma^2(x) =4\cdot(1-p_x)/(p_x E_x)$, i.e.~an overdispersion factor of 2, works fine. The resulting estimated $\check{\sigma}^2$ values  ranged over $[1.066\cdot 10^{-4}, 1.304\cdot 10^{-3}]$ with a mean of $4.36\cdot 10^{-4}$. This is close to the constant-$\sigma^2$ MLE estimate and the respective projections were very close, confirming that with a GP model the whole question of capturing observation errors is a ``higher order'' concern.} For ease of interpretation, we thus used a constant $\sigma^2$, estimated via MLE, for the remainder of the analysis.

\rev{As an alternative, we also implemented a Poisson GP model in Stan. The resulting parameter estimates are reported in Table \ref{table:Poisson-gp}. As can be seen, the resulting GP covariance structure is very similar to Table \ref{table:hyperparameters}.
We find that the actual outputs of the two models are also essentially identical. Namely, the mean percent error between the Gaussian and Poisson GP was -8.14e-05, indicating no systematic discrepancy between the two models and the root mean squared error was 4.99e-4 which is not material for the ultimate actuarial use. Our conclusion is that there is no statistical difference between using additive Gaussian noise or a Poisson link function.}

%

\begin{table}[ht]
\centering
\begin{tabular}{l|rr}
 & \multicolumn{1}{c}{MCMC Mean} & \multicolumn{1}{c}{MCMC 80\% Posterior CI}\\ \hline
$\theta_{ag}$  & 10.5955 &  (5.8652, 16.4272) \\
$\theta_{yr}$  & 26.1259 &  (14.4609, 40.1992) \\
$\eta^2$ &  1.8643  &  (0.7938, 3.2357) \\
$\sigma^2$ & 2.170e-04  & (1.945e-04, 2.400e-04) \\
$\beta_0$ & -3.7912 &   (-4.6851, -2.9317)\\ \hline
\end{tabular}
\caption{ Posterior Hyperparameter estimates from MCMC runs in Stan for the Poisson GP model. The GP is fitted to all data and uses squared-exponential covariance kernel \eqref{eq:sqExp} with prior mean $m(x) = \beta_0$. Stan hyper-priors (on standardized data) were $\log \theta_{ag}, \log \theta_{yr}, \log \eta^2 \sim \mathcal{N}(0,1)$ i.i.d., $\sigma^2 \sim \mathcal{N}_+(0,0.2)$, $\beta_0 \sim \mathcal{N}(-4,5)$.\label{table:Poisson-gp}}
\end{table}\


\subsection{Retrospective Analysis}\label{sec:retro-analysis}
\begin{figure}[ht]
  \centering
  \includegraphics[width=0.47\textwidth,height=2.2in,trim=0.15in 0.15in 0.15in 0.15in]{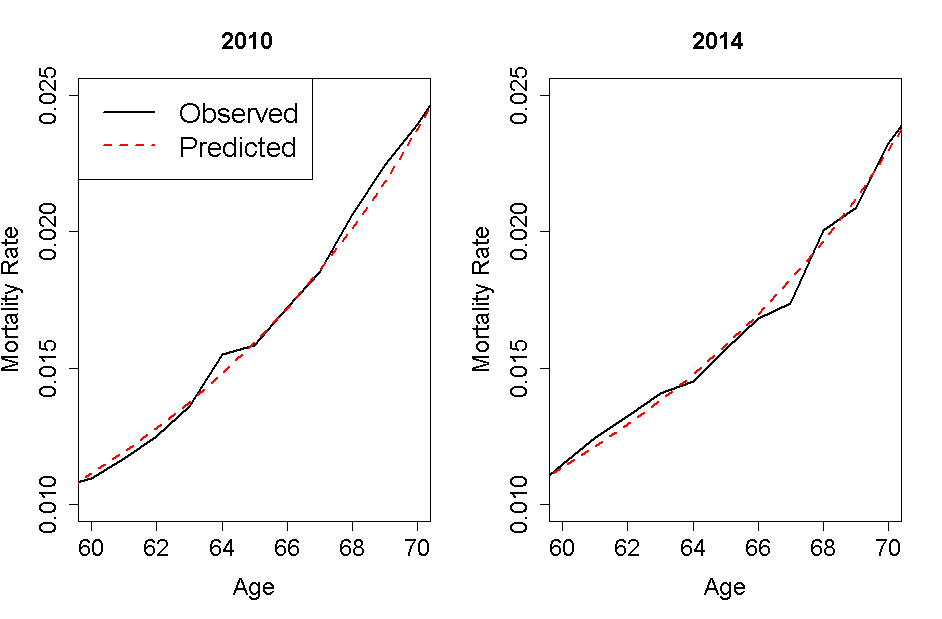}
 \includegraphics[width=0.47\textwidth,height=2.2in,trim=0.15in 0.15in 0.15in 0.15in]{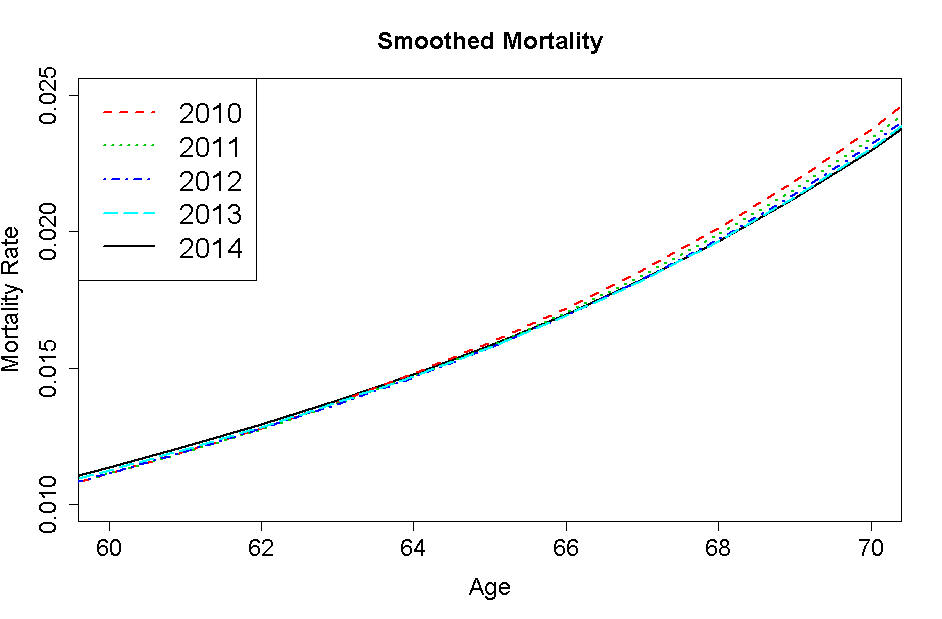}
  \caption{Mortality rates for US Males aged 60--70 during the years 2010--2014. Raw (solid) vs.~smoothed (dashed) mortality curves. Models are fit to 1999--2014 CDC data for Ages 50--84 (All data). Mean function $m(x)$ is intercept-only, $m(x)=\beta_0$.}\label{fig:GP-smooth-multiyear}
\end{figure}

\begin{figure}[ht]
  \centering
  \includegraphics[scale=0.2]{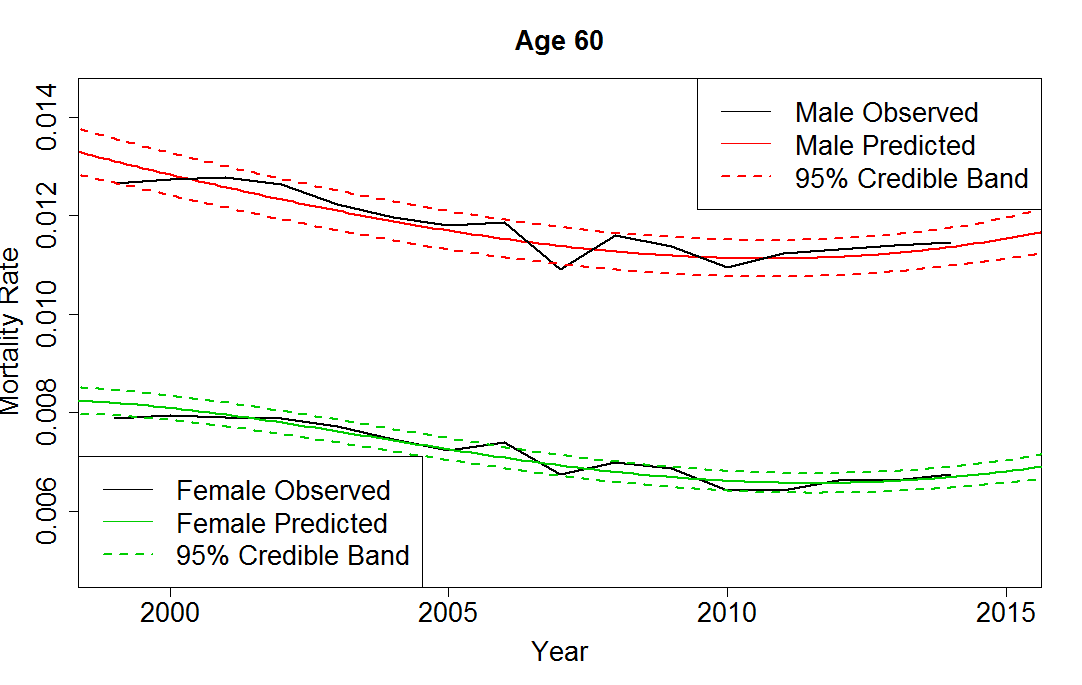}
  \includegraphics[scale=0.2]{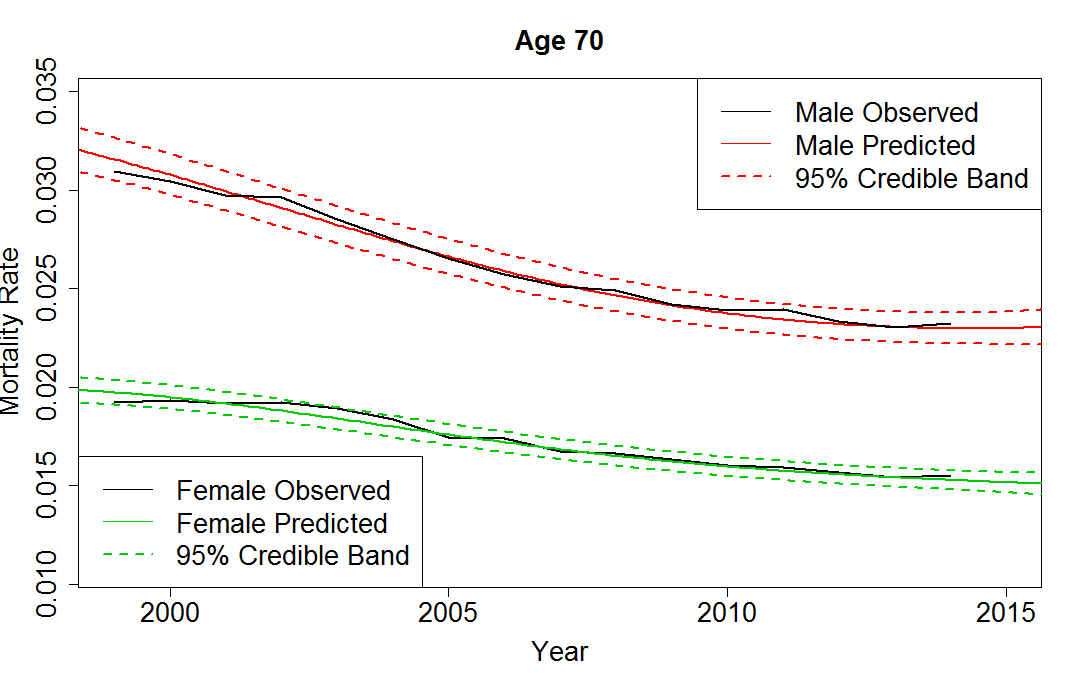}
  \includegraphics[scale=0.2]{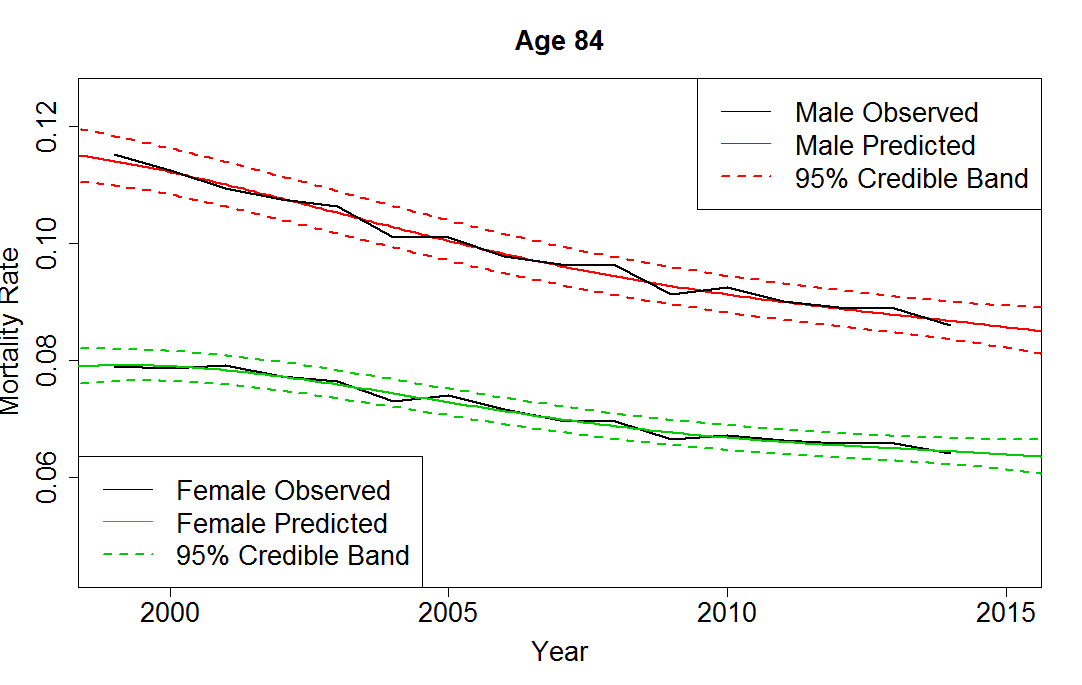}
  \caption{Mortality rates for US Males (top) and US Females (bottom) aged 60, 70 and 84 over time. The plots show raw mortality rates (solid black) for years 1999--2014, as well as predicted mean of the smoothed mortality surface (solid red) and its 95\% credible band, for 1999--2016. Models are fit to the 1999--2014 CDC data for Ages 50--84 (All data).  Mean function  is intercept-only, $m(x)=\beta_0$.\label{fig:MortalityVSTime} }
\end{figure}

We begin with a retrospective look at smoothed mortality experience over the recent past. \blue{Traditionally, this is done using actuarial graduation techniques; for the GP framework smoothing is simply the in-sample prediction $m_*(\mathbf{x})$}. Specifically, we fit a model using all the data, and investigate the mortality during the last 5 years of the period.
Figure \ref{fig:GP-smooth-multiyear} shows the estimated mortality rates as a function of age, specifically Males aged 60--70. The left panel compares the raw and GP-smoothed rates for 2010 and 2014, while the right panel shows the overall yearly trend for years 2010--2014.  As a complement to above, Figure \ref{fig:MortalityVSTime} provides a preliminary analysis of mortality improvement by plotting mortality rates against time. We show the observed and smoothed mortality rates against calendar years 1999--2014 for Males and Females aged 60, 70, and 84, along with the forecasted rates up to 2016. From the figure, we clearly observe the decrease of mortality at older ages which is, however, slowing down in the last few years.

\begin{figure}[ht]
  \centering
  \includegraphics[width=0.9\textwidth,height=2.5in,trim=0.15in 0.15in 0.15in 0.15in]{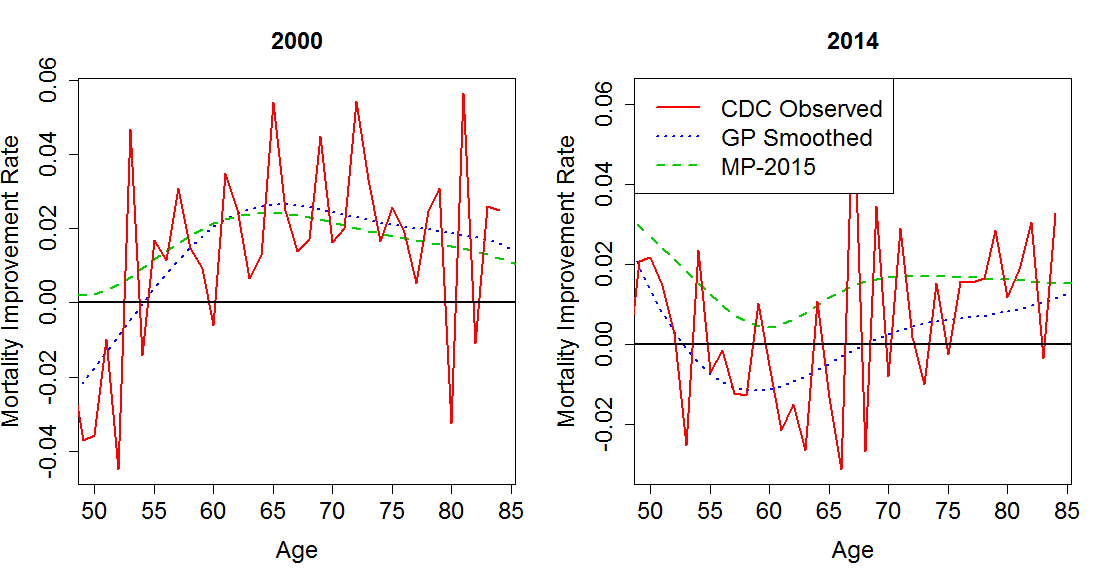}
  \caption{Mortality improvement factors for US Males using All Data. Solid lines indicate the empirical mortality experience $MI_{back}^{obs}(\cdot; yr)$ for years $yr \in \{2000, 2014\}$, the dotted and dashed lines are $\partial m_{back}^{GP}(\cdot; yr)$ from \eqref{eq:MIback}, and the MP-2015 improvement  scale $MI_{back}^{MP}(\cdot; yr)$, respectively. \label{fig:MP-multiyear} }
\end{figure}

\begin{figure}[ht]
  \centering
  \includegraphics[width=0.44\textwidth,height=2.2in,trim=0.15in 0.15in 0.15in 0.15in]{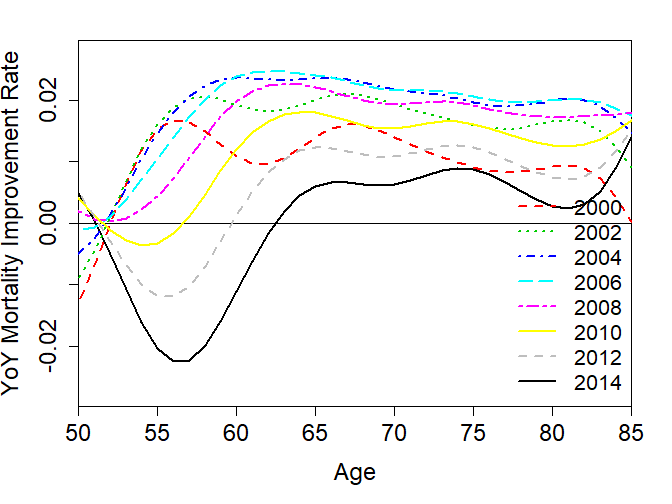}
  \includegraphics[width=0.48\textwidth,height=2.2in,trim=0in 0.15in 0.15in 0.15in]{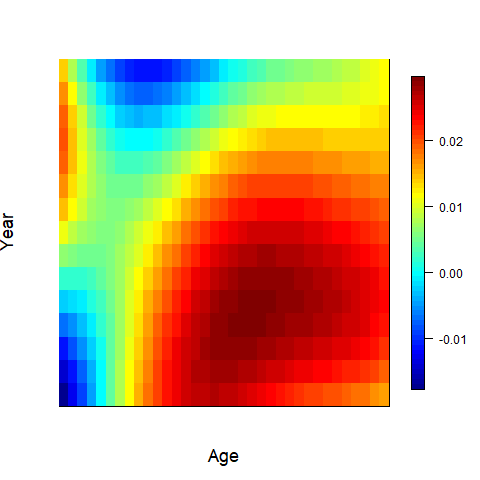}
  \caption{ \rev{Smoothed yearly mortality improvement factors $\partial m_{back}^{GP}(ag, yr)$ from \eqref{eq:MIback} for US Males using All data. Left: age factors for $yr=2000, \ldots, 2014$. The curves for 2000 and 2014 are the same as in Figure \ref{fig:MP-multiyear}. Right: Heatmap of the estimated YoY improvement factors}. \label{fig:mortTrend} }
\end{figure}

\blue{A key output of official tables are the mortality improvement scales, such as the MP-2015 rates $MI^{MP}_{back}(x_{ag}, yr)$, where we distinguish the common indexing by Age, keeping Year fixed. These are intuitively the smoothed version of the raw annual percentage mortality improvement which is empirically observed via}
\begin{align}\label{eq:MIback}
MI_{back}^{obs}\left(x_{ag}; yr\right) &\doteq 1-\frac{\exp\left(\mu(x_{ag},yr)\right)}{\exp\left(\mu(x_{ag},yr-1)\right)}
\end{align}
with $\mu(x_{ag},yr)$ the raw log-mortality rate for $(x_{ag},yr)$. In analogue to above, we can obtain the predicted mean improvement by replacing $\mu$'s by the GP model posteriors $f_*$'s and integrating over their posterior distributions:
\begin{equation}\label{eq:MIgp}
\partial m_{back}^{GP}\left(x_{ag},yr\right) := \E\left[MI_{back}^{GP}\left(x_{ag},yr\right)\right] \doteq \E\left[ 1-\frac{\exp\left(f_*(x_{ag},yr)\right)}{\exp\left(f_*(x_{ag},yr-1)\right)} \right] .
\end{equation}
\blue{Figure~\ref{fig:MP-multiyear} shows these different improvement scales for ages 50--85 and two sample years, 2000 and 2014; the MP-2015 curves are from the published SOA reports \citep{SOA-MP-2015}.}  We observe that the raw mortality improvements $MI^{obs}$ are extremely noisy, which is not surprising since they are based on the relative difference of two very similar raw mortality rates. 
 Figure \ref{fig:MP-multiyear} also indicates that the MP-2015 estimates are significantly higher than either the actual experience (which has moderated a lot in the past decade) or our fit $\partial m_{back}^{GP}$, with differences of as much as 2\% p/a in improvement factors. Figure \ref{fig:mortTrend} emphasizes that there is a downward trend in mortality improvement, and moreover non-uniform behavior across ages.  This throws into question the MP-2015 concept of a sustained, age-uniform projected long-term mortality improvement trend.

\subsection{Mean Function Modeling}\label{sec:trendassumptions}
  We tested three models for the prior mean function: an intercept-only model $m(x^n) = \beta_0$, a linear model, $m(x^n) = \beta_0 + \beta_{1}^{ag} x^n_{ag} + \beta_1^{yr} x^n_{yr} $, and a quadratic age model, $m(x^n) = \beta_0 + \beta_{1}^{ag} x^n_{ag} + \beta_1^{yr} x^n_{yr} + \beta_2^{ag} (x^n_{ag})^2$. Thus, the linear model has the log mortality increasing linearly in age and decreasing linearly in calendar year. The quadratic model then adds a convexity component to the age axis.

The coefficients of these functions were estimated concurrently with fitting the GP models using \eqref{eq:uk-equations}. A summary of the models and the coefficient estimates is shown in Table \ref{table:trendfunctions}. \blue{One finding is that the fitted year-trend coefficient $\beta_1^{(yr)}$ is consistently estimated by both the linear and quadratic model and indicates} a linear improvement in log mortality rates of about 1.4\% per calendar year in both of these models regardless of assumptions on age shapes.  Since this model is fitted to ages 50--70, these results are consistent with the long-term trend of improving mortality. \blue{As expected, the table also indicates a strong Age effect; we note that the fitted coefficient $\beta_2^{(ag)}=1.459 \cdot 10^{-4}$ for the quadratic age component confirms a significant convexity of log-mortality in Age.}

\begin{table}[ht]
\centering
\begin{tabular}{l|rrrr|rrrr}
&  \multicolumn{4}{|c|}{Mean Function Parameter MLE's} & \multicolumn{4}{|c}{GP Hyperparameter MLE's}\\ \hline
&  \multicolumn{1}{c}{$\beta_0$} & \multicolumn{1}{c}{$\beta^{ag}_1$} &  \multicolumn{1}{c}{$\beta^{ag}_2$} &\multicolumn{1}{c}{$\beta^{yr}_1$} & \multicolumn{1}{c}{$\eta^2$} & \multicolumn{1}{c}{$\sigma^2$} & \multicolumn{1}{c}{$\theta_{ag}$} & \multicolumn{1}{c}{$\theta_{yr}$}\\ \hline
Intercept &  -4.526  &  -  &  -  &  -  &  6.213e-01  &  3.428e-04  &  8.384  &  12.746  \\
Linear &   18.737  &  0.081  &   -  & -1.397e-02  & 8.521e-04  &  1.761e-04  &  3.610  &  3.543  \\
Quadratic  & 19.641  &  0.064  &  1.459e-04  &  -1.417e-02  &  1.403e-03  &  2.998e-04  &  3.629  &  3.475  \\
\end{tabular}
\caption{Fitted mean function and covariance parameters using Subset III (ages 50--70 and years 1999--2009) for Males. The mean functions are $m(x^n) = \beta_0$ for Intercept, $m(x^n) =\beta_0 + \beta_1^{ag} x^n_{ag} + \beta_1^{yr}x^n_{yr}$ for Linear, and $m(x^n) = \beta_0 + \beta_{1}^{ag} x^n_{ag} + \beta_1^{yr} x^n_{yr} + \beta_2^{ag} (x^n_{ag})^2$ for Quadratic.\label{table:trendfunctions}}
\end{table}

Intuitively, the mean function provides a fundamental explanation of mortality rates by age and year, while the covariance structure captures deviations from this postulated relationship based on nearby observed experience (with the influence depending on the lengthscale). Consequently, the choice of the mean function affects the covariance structure; a stronger trend/shape lowers the spatial dependence of the residuals. We observe this effect in Table \ref{table:trendfunctions}, where the intercept-only model has length-scales of $\theta_{ag} \approx 8.5, \theta_{yr} \approx 12.5$, while for the linear and quadratic models
the range of the length-scales is much smaller $\theta_\ell \approx 3.5.$  \blue{Another effect of the mean function is on the hyperparameter $\eta$ which can be viewed as the variance of the model residuals. If the mean function fits well then we expect smaller $\eta$. In turn, smaller $\eta$ translates into tighter credible intervals around in-sample smoothing and out-of-sample forecasts. Table \ref{table:trendfunctions} shows that the values for $\eta$ and $\sigma$ are similar across linear and quadratic models while the intercept-only model has uniformly larger values across parameters.}

\begin{figure}[ht]
\centering
\includegraphics[scale=0.24]{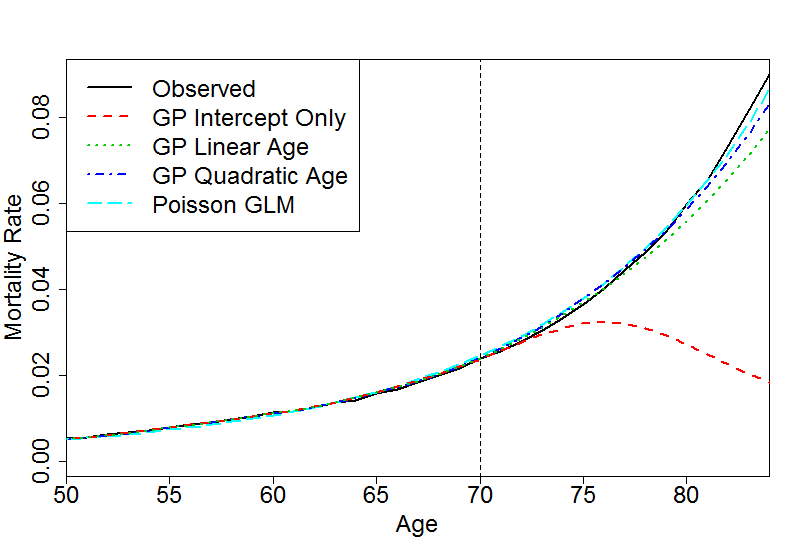}
\includegraphics[scale=0.24]{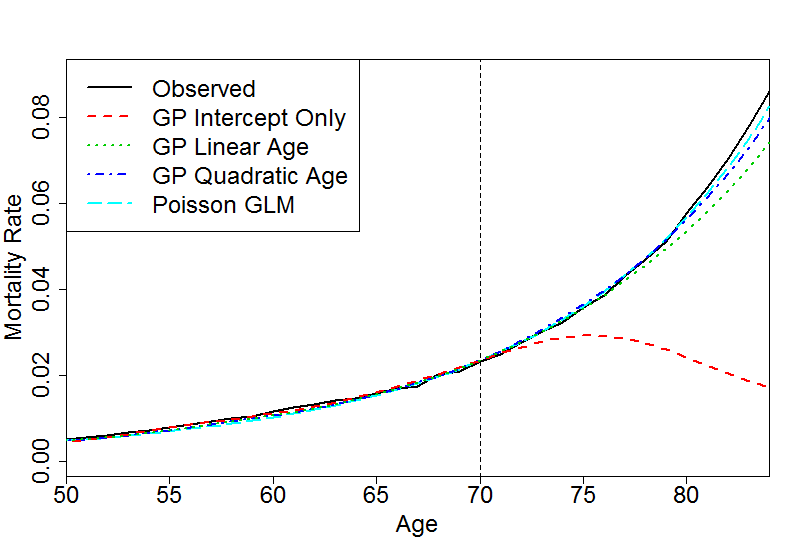}
\caption{ \label{fig:trendcomparison} Comparison of mean function choices in extrapolating mortality rates at old ages.  Models are fit to years {1999--2010} and ages 50--70 (Subset III) for US Males, with estimates made for Age 50--85 in {2011 and 2014}. The vertical line indicates the boundary of the training dataset in $x_{ag}$. The mean functions are given in Table \ref{table:trendfunctions}. \rev{Also shown is the fit based on a quadratic-age Poisson GLM model \eqref{eq:poisson-glm}, cf.~Table~\ref{tbl:poisson-glm}.}}
\end{figure}

Figure~\ref{fig:trendcomparison} illustrates these three models fit to Subset III which emulates deep out-of-sample extrapolation. \blue{As discussed, out-of-sample forecasting by the GP model can be viewed as blending the data-driven prediction with the estimated trend encapsulated by $m$. Specifically, as $x^n_{*,ag}$ moves beyond the age range of $\{50, 51, \ldots, 70\}$ in Subset III we have $m_*(x_*) \to m(x_*)$.} In the case of an intercept-only model, this implies that $m_*(x_*) \to \beta_0$, i.e.~the projected mortality is independent of either Age or Year. In Figure~\ref{fig:trendcomparison} the asymptotic projected rate was $\exp(\hat{\beta}_0) = 1.082\%$. A similar issue pertains to the linear-mean model whose long-range forecasts imply exponential Age dependence which is not appropriate for ages above 80. This discrepancy is successfully resolved by the quadratic $m(x_*)$ model. \blue{ The lengthscales $\theta$ control this transition; roughly speaking extrapolating more than $\theta$ distance away reduces to  $m_*(x_*) \simeq m(x_*)$. This can also be seen in Figure~\ref{fig:trendcomparison}: since the training data includes up to 2010, the forecast for 2011 is much more driven by past data compared to the one for 2014. As a result, for the intercept-only model with $\theta_{yr}=11.461$, the forecast is acceptable in 2011, but deteriorates dramatically for 2014. This effect is also present but less apparent in the trend models; due to smaller values of $\theta_{yr}$ the latter forecasts already rely more heavily on their mean functions for extrapolation.}

\rev{As an additional comparator, Figure~\ref{fig:trendcomparison} plots the fit of a Poisson GLM model using \eqref{eq:poisson-glm} and a quadratic age-trend. The respective fitted coefficients $\bm{\beta}$ are listed in Table~\ref{tbl:poisson-glm}. We observe that the estimated coefficients using a Poisson link function are very similar to those in Table~\ref{table:trendfunctions} (confirmed visually in the Figure), which is consistent with our earlier discussion that adding a link function does not materially modify the results. At the same time, the goodness-of-fit of the GP model is significantly better than a parametric GLM fit, confirming the complex spatial structure that cannot be captured in a simple GLM.}
\begin{table}[ht]
\centering
\begin{tabular}{l|rrrrr}
&  \multicolumn{5}{c}{Poisson GLM Mean Function Parameter MLE's}  \\ \hline
&  $\beta_0$ & $\beta^{ag}_1$ &  $\beta^{ag}_2$ & $\beta^{yr}_1$ 
\\ \hline
Intercept &  -4.442  &  -  &  -  &  - \\ 
Linear &  23.264  & 0.080 &  -  & -1.62e-02 \\ 
Quadratic & 24.218 & 0.0403 & 3.24e-04  & -1.608e-02  \\ 
\end{tabular}
\caption{\label{tbl:poisson-glm}\rev{Fitted Poisson GLM Parameters for \eqref{eq:poisson-glm} with three different age-year parameterizations (intercept-only $f(x^n) = \beta_0$, linear $f(x^n) = \beta_0 + \beta_{1}^{ag} x^n_{ag} + \beta_1^{yr} x^n_{yr} $, and quadratic-age $f(x^n) = \beta_0 + \beta_{1}^{ag} x^n_{ag} + \beta_1^{yr} x^n_{yr} + \beta_2^{ag} (x^n_{ag})^2$ based on US Male CDC Subset III data. }}
\end{table}

\blue{Another way for model comparison is to look at the widths of the respective credible intervals.}
For example, for year 2010 age 84, the observed mortality rate was $8.999\%$ and
 the intercept, linear and quadratic models generated 95\% credible intervals of $( 0.783\%, 10.254\%)$,$( 7.100\%, 8.562\%)$ and $( 6.379\%, 11.188\%)$ respectively. Certainly the first interval is too wide (partly due to the large $\eta$ and $\theta$'s of the intercept-only model), while the second interval is too narrow and does not even contain the raw data point (the linear model apparently underestimates $\eta$). On the other hand, for age 71 in year 2014, the raw rate was $2.489\%$ and the respective 95\% credible intervals were $( 2.258 \%, 2.927 \%)$, $( 2.378 \%, 2.703 \%)$ and $( 2.346 \%, 2.798 \%)$. While all models now contain the observed rate, the linear model again has the tightest credible interval, \blue{which might indicate poor goodness-of-fit}. 

\blue{ Returning to in-sample smoothing and looking again at  Figure \ref{fig:trendcomparison}, we observe that all three models generate very similar forecasts for ages 55--70 
This confirms that in-sample $m_*$ is data-driven and the choice of $m(\cdot)$ is secondary. To summarize, the most important criterion in including a mean function is whether the goal is to predict out-of-sample and if so, how far out-of-sample.  In general, mean modeling is crucial, but the precise choice of the mean function is not as clear. In Section \ref{sec:inhomogeneousGPmodels} we discuss one further method for mean-modeling based on Age-grouping.}

\subsection{Predictive Accuracy}

\blue{Figure~\ref{fig:trendcomparison} can also be viewed as a first glimpse into the predictive accuracy of a GP mortality model. Recall that in the Figure we fit to mortality data from 1999-2010 and then forecast 1 year out (2011) and 4 years out (2014). The Figure then compares these projections to the observed mortality experience in 2011 and 2014. As discussed, these projections are highly sensitive to the choice of $m(x)$, especially in terms of the Age-structure because the models are only given experience up to Age 70 and hence have zero information about how mortality behaves for higher ages.}

\blue{To provide a more ``fair'' comparison, Table~\ref{table:prediction} shows projections for other input datasets. Overall, we observe excellent predictive power for making projections 4-years out (fit using Subset I, forecast for 2014), confirming the competitive performance of the GP fitted models.}

\blue{Beyond the predictive mean $m_*(x_*)$, we also report the corresponding posterior marginal variance $s^2_*(x_*) := C_*(x_*,x_*)$ which is a proxy for the confidence the model assigns to its own prediction. This provides a valuable insight: for example if fitted to ages 50--70 (Subset III) and projecting for age 80 in year 2014: $\tilde{x} := (x_{ag},x_{yr}) = (80,2014)$, the intercept-only model reports minimal predictive power  which is reflected in the very large $s^{(III)}_*(\tilde{x})=0.4565$, in light of which the poor prediction $m^{(III)}_*(\tilde{x}) = -3.7177$  is simply a ``shot in the dark''. Indeed, the model predicts mortality rate of 2.43\% which is nowhere the realized 5.78\%, but is still within its 95\% credible interval of $(0.98\%, 6.03\%)$.
Including more ages (Subset I) gives a more reasonable and much more confident forecast of $m^{(I)}_*(\tilde{x})= -2.8416$ and $s^{(I)}_*(\tilde{x})=0.0463$, and including more years (which makes $\tilde{x}$ to be right at the edge of observed data) raises credibility even further, $m^{(All)}_*(\tilde{x})= -2.8579$ and $s^{(All)}_*(\tilde{x})=0.0170$.
Table~\ref{table:prediction} also quantifies the gains from using a more sophisticated $m$ -- the quadratic trend allows to shrink $s^{(I)}_*(\tilde{x})$ from $0.0463$ to $0.0333$, and brings the prediction $m^{(I)}_*(\tilde{x})$ closer to the eventually realized experience.}


\begin{table}[ht]
\centering
\begin{tabular}{l|cc|cc|cc|c} \hline
\multicolumn{8}{c}{Intercept-only $m(x) = \beta_0$} \\ \hline
& \multicolumn{2}{c|}{Fit to Subset III} & \multicolumn{2}{c|}{Fit to Subset I} & \multicolumn{2}{c|}{Fit to All Data} & Observed\\ \hline
$x_{ag}$ & \multicolumn{1}{c}{$m_*^{(III)}$} & \multicolumn{1}{c|}{($s_*^{(III)}$)} & \multicolumn{1}{c}{$m_*^{(I)}$} & \multicolumn{1}{c|}{($s_*^{(I)}$)} & \multicolumn{1}{c}{$m_*^{(All)}$} & \multicolumn{1}{c|}{($s_*^{(All)}$)} & $\mu$  \\ \hline
70 & -3.7520 & (0.0580) & -3.7380 & (0.0427) & -3.7702 & (0.0169) & -3.7630\\
80 & -3.7177 & (0.4565) & -2.8416 & (0.0463) & -2.8579 & (0.0170) & -2.8531\\ \hline \hline
\multicolumn{8}{c}{Quadratic $m(x) = \beta_0 + \beta_1^{ag} x_{ag} + \beta_1^{yr} x_{yr} + \beta_2^{ag} x^2_{ag}$} \\ \hline
& \multicolumn{2}{c|}{Fit to Subset III} & \multicolumn{2}{c|}{Fit to Subset I} & \multicolumn{2}{c|}{Fit to All Data} & Observed\\ \hline
$x_{ag}$ & \multicolumn{1}{c}{$m_*^{(III)}$} & \multicolumn{1}{c|}{($s_*^{(III)}$)} & \multicolumn{1}{c}{$m_*^{(I)}$} & \multicolumn{1}{c|}{($s_*^{(I)}$)} & \multicolumn{1}{c}{$m_*^{(All)}$} & \multicolumn{1}{c|}{($s_*^{(All)}$)} & $\mu$  \\ \hline
70 & -3.7507 & (0.0419) & -3.7711 & (0.0332) & -3.7671 & (0.0163) & -3.7630\\
80 & -2.8774 & (0.1046) & -2.8546 & (0.0333) & -2.8553 & (0.0164) & -2.8531\\ \hline
\end{tabular}
\caption{\blue{GP model predictions for US mortality in 2014 and Age 70/80 when fitted to various data subsets, cf.~Table \ref{table:MortData}, indicated by superscripts. We report the predictive mean $m_*(x)$ and the predictive standard deviation $s^2_*(x_*) = C_*(x_*, x_*)$.} \label{table:prediction}}
\end{table}

\begin{figure}[ht]
  \centering
  \begin{tabular}{cc} \hspace*{-20pt}
  \includegraphics[scale=0.18]{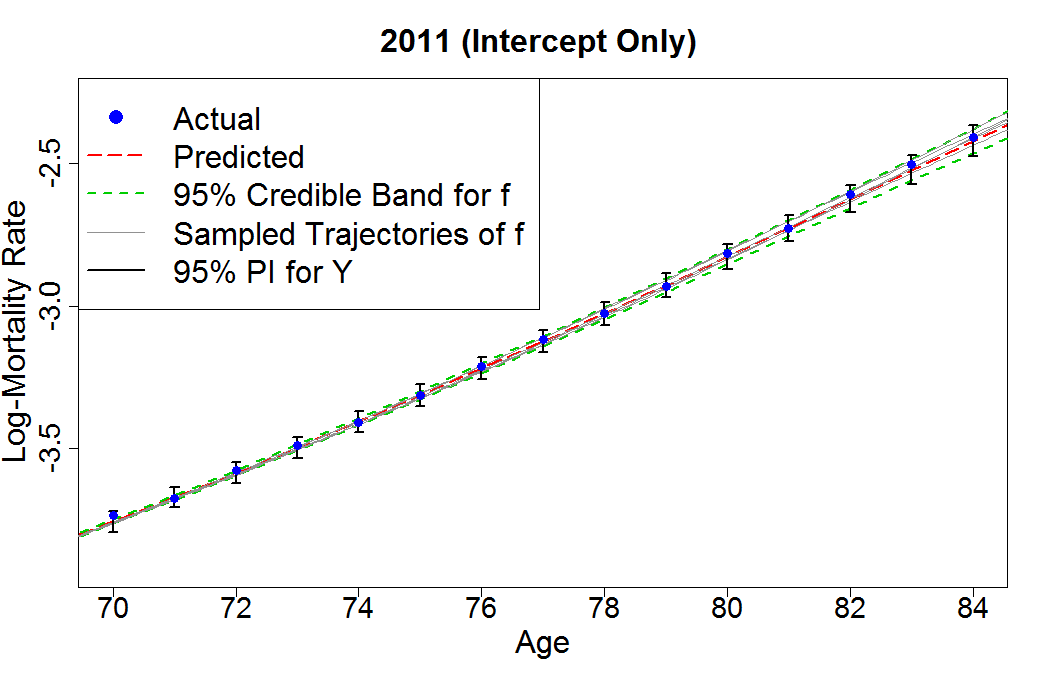} &
  \includegraphics[scale=0.18]{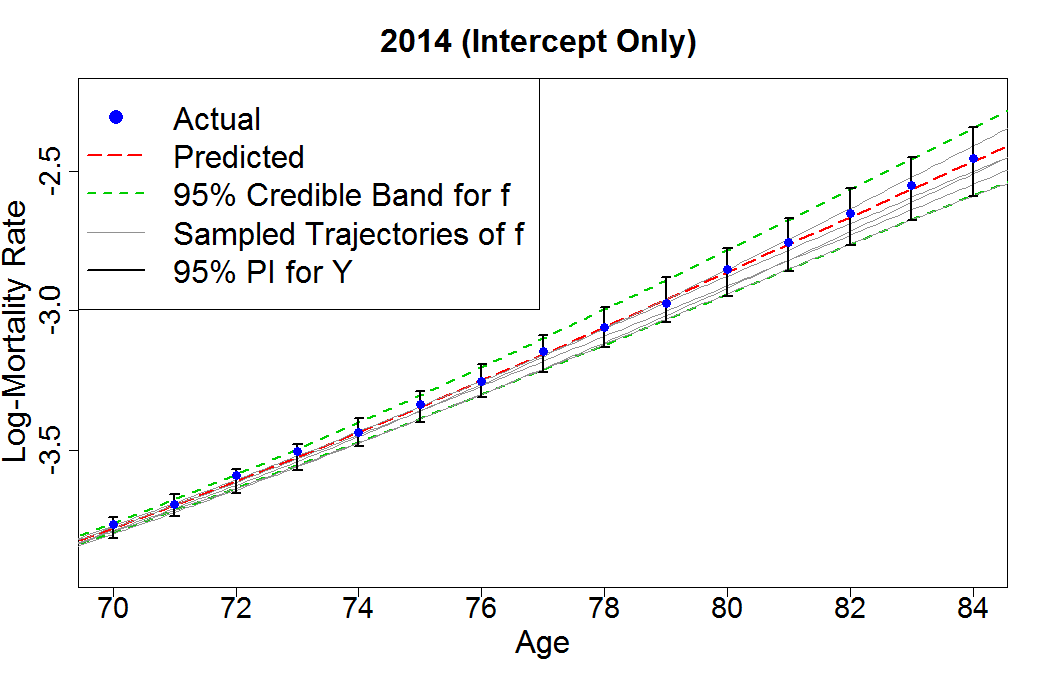} 
\end{tabular}
  \caption{ \label{fig:predict-2010} Mortality rate prediction for years 2011 and 2014 and ages 71--84.  Model is fit with Subset II data with intercept-only mean functions and squared-exponential kernel.  ``Simulated paths of $f$'' refers to simulated trajectories of the latent $\mathbf{f}_*$.  Credible bands are for the mortality surface $\mathbf{f}_*$; vertical intervals are for predicted observable mortality experience $\mathbf{y}_*$. }
\end{figure}

For another angle on forecasting with GP models,
Figure \ref{fig:predict-2010} shows that the intercept-only model still performs well when predicting only slightly out-of-sample. In this Figure, we fitted mortality curves using the ``notched'' Subset II: years 1999--2010 and ages 50--84, plus 2011--2014 with ages 50--70, and then predicted out-of-sample for mortality rates for 2011--2014 and ages 71--85.  This differs from the previous setup where the model had no prior information on ages 71--84. We observed that in this setup the uncertainty from the intercept-only model is only slightly worse (wider interval) relative to the quadratic trend model, confirming the reasonableness of using the simpler $m(x) = \beta_0$.

 Figure \ref{fig:predict-2010} also plots the marginal credible bands for $\mathbf{f}_*(\mathbf{x}_*)$ and intervals for future observations $\mathbf{y}_*$. As expected, the prediction uncertainty increases for the oldest ages and for later calendar years (compare credible intervals in Figure~\ref{fig:predict-2010} for 2014 vis-a-vis 2011). Also note that the intervals for $\mathbf{y}_*$ are always a fixed distance away from the pointwise bands of $\mathbf{f}_*$ regardless of Age/Year due to the assumed constant noise variance $\sigma^2$; this is much more noticeable when in-sample, where posterior variance $C_*(\mathbf{x}_*, \mathbf{x}_*)$ is negligible relative to $\sigma^2$.

As discussed, the GP model automatically generates credible intervals around any prediction, giving
a principled approach for assessing uncertainty in forecasts. Moreover, since GP considers the full covariance structure of mortality curves, one can analytically evaluate the joint predictive uncertainty of any number of mortality rates. \rev{In particular, one can use the posterior predictive distribution to generate conditional trajectories of mortality rates for any collection of cells. For instance, fixing a calendar year, we may sample from the multivariate normal distribution of $\mathbf{f}_*(\mathbf{x}_*)$ across ages $\mathbf{x}_*$ to obtain a stochastic scenario of the respective mortality age structure. This is  illustrated in Figure \ref{fig:predict-2010} that shows several such scenarios of log-mortality rates for calendar year 2014 (which is in-sample up to age 70 and extrapolating for ages 71-84), along with the overall credible band. Note that in contrast to factor models like Lee-Carter that force the log-mortality curve $\mu_{ \cdot, yr}$ to be confined to a low-dimensional space (e.g.~one degree of freedom in classical Lee-Carter), within a GP framework, the shape of $f_*(\cdot, yr)$ remains non-parametric and infinite-dimensional. Alternatively, we could sample potential evolutions of mortality at selected age into a desired future projection interval. Sampling such trajectories is crucial for quantifying aggregate mortality risk in a portfolio (say in a pension plan or life insurance context).}

\subsection{Comparison of GP and APC forecasts}
\blue{To provide a brief comparison of the popular stochastic mortality models, we fit a cohort extension of the Lee-Carter model in \eqref{eq:lee-carter}, introduced by \citet{renshaw2006cohort}, which is as follows:
\begin{align}\label{eq:lee-carter-cohort}
\mu_{ag,yr} = \alpha_{ag} +\frac{1}{n_a} \kappa_{yr}  + \frac{1}{n_a} \gamma_{yr-ag} + \eps_{ag,yr}
\end{align}
where $\gamma_{yr-ag}$ is the cohort effect and $n_a$ is the number of years in the data set.  Using the \texttt{StMoMo} software suite \citep{stmomo} on our data yielded a random walk with drift for $\kappa_{\cdot}$ and ARIMA(0,1,2) model for $\gamma$.  \citet{cairns2011mortality} showed that this model performed well in US male data analysis.}

\begin{figure}[ht]
  \centering
  \begin{tabular}{ccc}
  \includegraphics[scale=0.3]{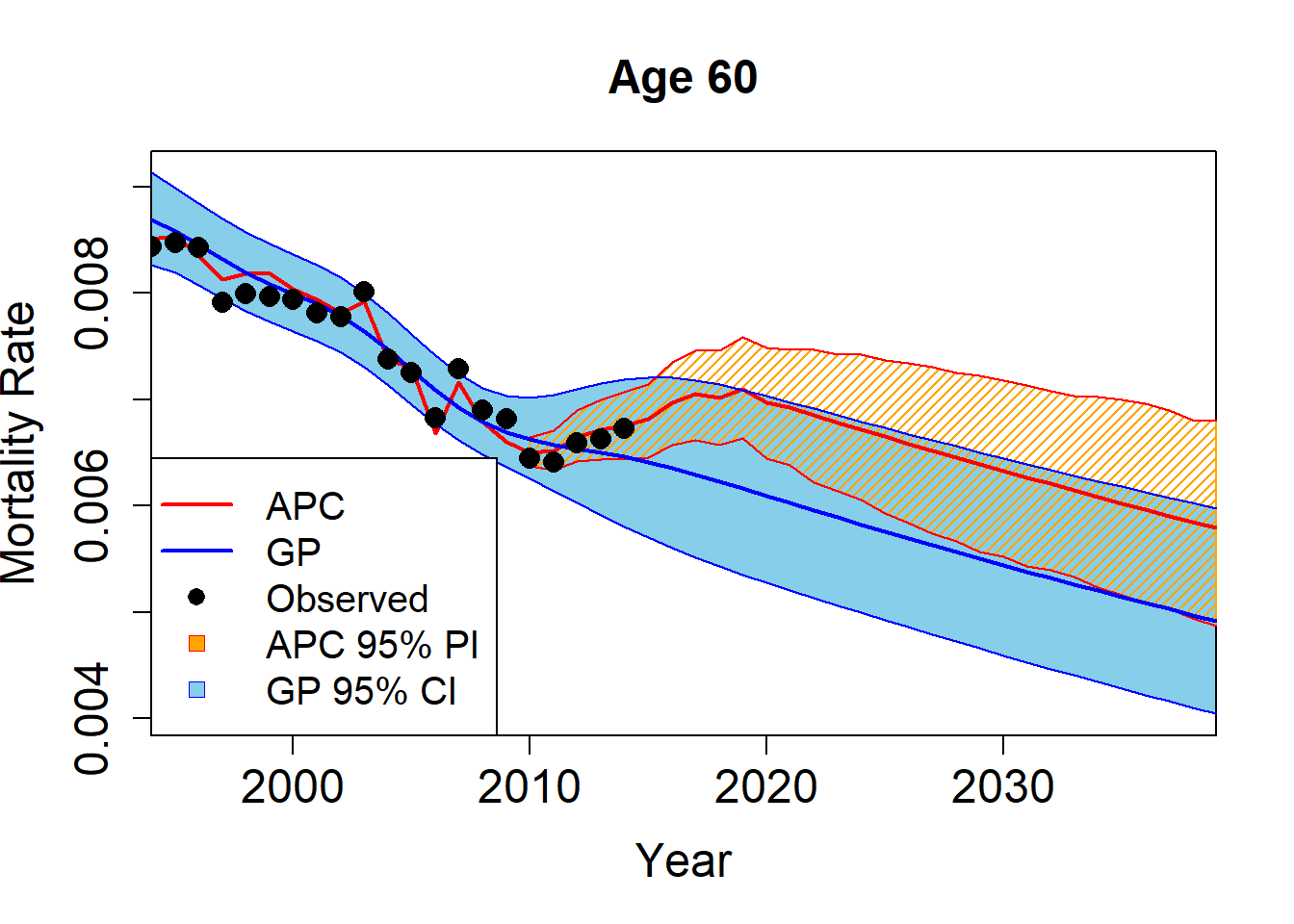} & 
  \includegraphics[scale=0.3]{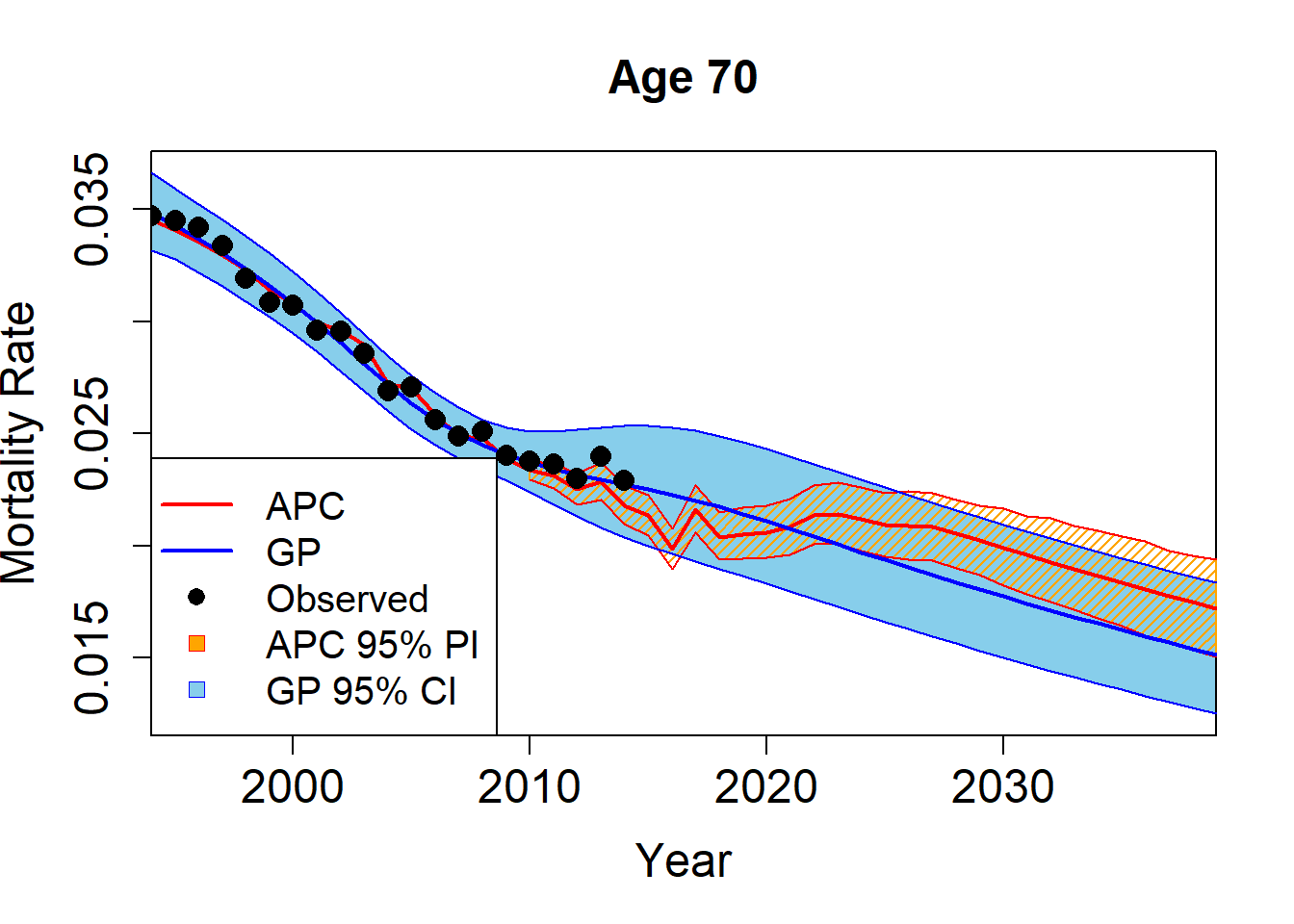} & 
  \includegraphics[scale=0.3]{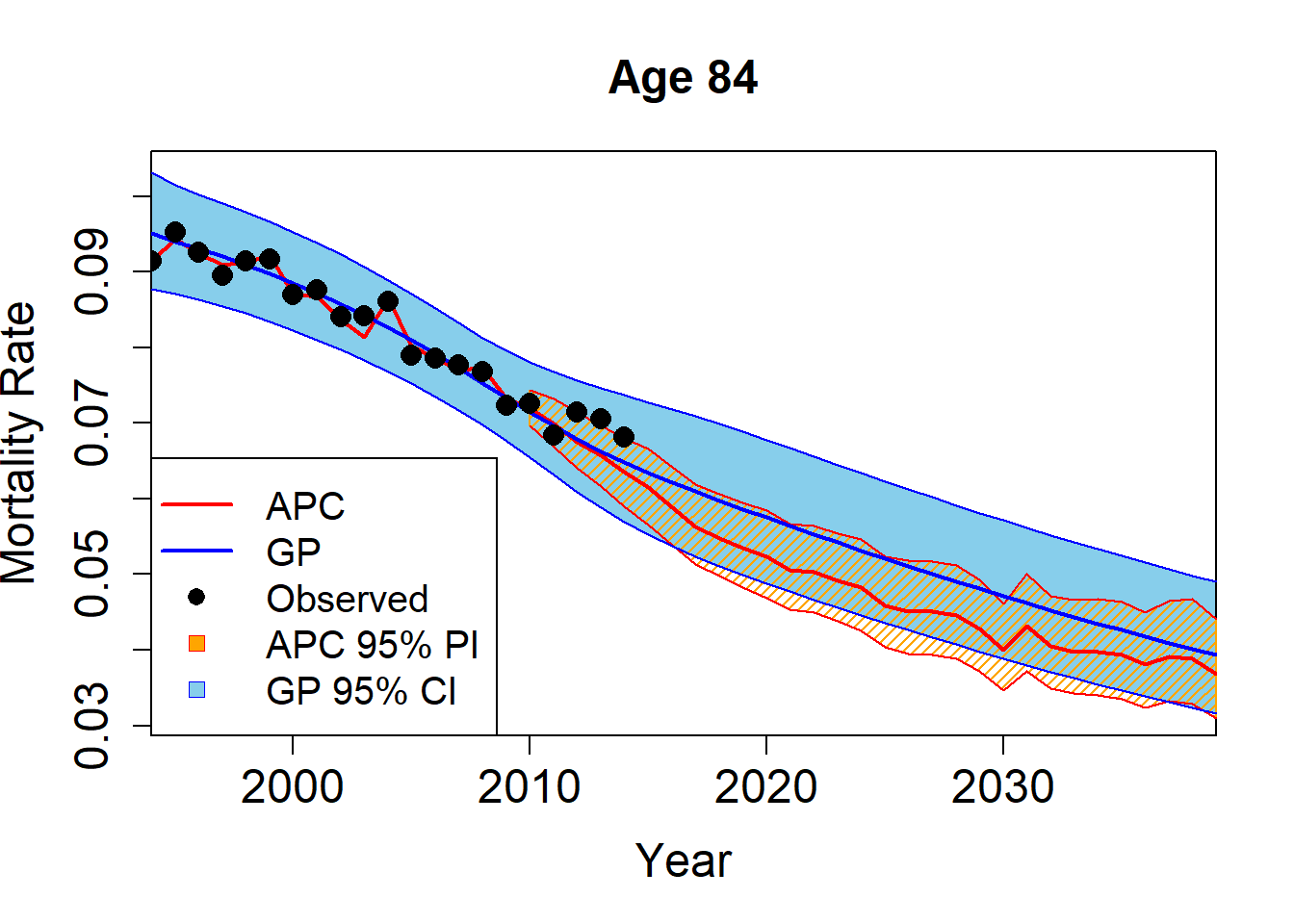} \\ 
  US Females & US Males & UK Females \end{tabular}
  \caption{Observed and predicted mortality rates for 1994-2040 for three representative datasets.  GP model uses quadratic mean function $m(x) = \beta_0 + \beta_1^{ag}x_{ag} + \beta_1^{yr}x_{yr} + \beta_2^{ag} x_{ag}^2$, and the APC model is as in Equation \eqref{eq:lee-carter-cohort}. Models fit to HMD data for 1994--2009 and ages 50--84.\label{fig:GPvsLC} }
\end{figure}

\rev{Figure \ref{fig:GPvsLC} compares the predictions from a GP model against those of an age-period-cohort (APC) model \eqref{eq:lee-carter-cohort} for three representative ages and across three different country/gender datasets. We use years 1994--2040, i.e.~both in-sample and up to 30 years into the future. We observe that relative to the GP model, the APC model generates both volatile in-sample projections (as it is not designed with smoothing in mind), and erratic short-term projections due to the underlying time series fitted to the $\kappa_{\cdot}$ and $\gamma_{\cdot}$ factors. Recall that the APC framework tries to average out trends via a parametric model which makes the projections dependent even on distant historical experience, while the GP effectively uses the history to learn the spatial dependence structure and then makes data-driven projections based on recent experience. Also, the APC has generally tighter predictive intervals (i.e.~it is more confident in its forecast) compared to the GP predictions, although the final uncertainty band in 2040 is about the same. We note that there is no general pattern ---sometimes the two models agree on the likely mortality trend, sometimes APC indicates larger improvements, sometimes GP does.}


\subsection{Forecasting Mortality Improvement}\label{sec:forecastingmortalityimprovement}

To focus more precisely on mortality \emph{improvement}, we proceed to analyze changes in $\mu(x_{ag}, \cdot)$ over time.
Section \ref{sec:retro-analysis} discussed already backward-looking annual (YoY) improvements $MI_{back}^{obs}$ and $\partial m_{back}^{GP}$ as defined in Equation \ref{eq:MIback}. For a more prospective analysis, one could consider a centered difference
\begin{equation}\label{eq:MIcentapprox}
1-\left(\frac{\exp\left(f_*(x_{ag},{yr}+h)\right)}{\exp\left(f_*(x_{ag},{yr}-h)\right)}\right)^{1/{2h}} \approx -\frac{f_*(x_{ag},{yr}+h) - f_*(x_{ag},{yr}-h)}{2h},
\end{equation}
which is possible to compute for any $h$ since the GP model for $f_*$ yields an an entire mortality surface spanning over all $(x_{ag},x_{yr}) \in \R^+ \times \R^+$. Note that since $f_*$ is a Gaussian process, the right hand side of \eqref{eq:MIcentapprox} remains Gaussian. We may also take the limit $h \rightarrow 0$ which gives the instantaneous rate of change of mortality in terms of calendar time. As an analogue to \eqref{eq:MIcentapprox}, we term the negative of the above differential as the instantaneous mortality improvement process
\begin{equation}
MI_{diff}^{GP}(x_{ag}; x_{yr}) \doteq -\frac{\partial f_*}{\partial x_{yr}}(x_{ag}, {yr}).
\end{equation}

A remarkable property of the Gaussian process is that $MI_{diff}^{GP}$ is once again a GP with explicitly computable mean and covariance functions \citep{WilliamsRasmussenBook}.

\begin{prop}\label{prop:GPderivative}
For the Gaussian Process $f_*$ with a twice differentiable covariance kernel $C,$ the limiting random variables
\begin{equation}\label{eq:GPderivative}
\frac{\partial f_*}{\partial x_{yr}}(x_{ag},{yr}) \doteq \lim_{h \rightarrow 0} \frac{f_*(x_{ag},{yr}+h) -f_*(x_{ag},{yr})}{h}
\end{equation}
exist in mean square and form a Gaussian process  $\frac{\partial f_*}{\partial x_{yr}}\sim GP(\partial m_{diff}, s_{diff})$.  Given the training set $\mathcal{D} = (\mathbf{x},\mathbf{y})$, the posterior distribution of $\frac{\partial f_*}{\partial x_{yr}} (x_*)$ has mean and variance
\begin{align}\label{eq:derivative-marginalE}
\partial m_{diff}(x_*) &= \E\left[\left.\frac{\partial f_*}{\partial x_{yr}} (x_*)\right|\mathbf{x},\mathbf{y}\right] = \frac{\partial C}{\partial x'_{yr}}(\mathbf{x},x_*) (\mathbf{C}+\mathbf{\Sigma})^{-1} \mathbf{y},\\ \label{eq:derivative-marginalvar}
s^2_{diff}(x_*) &= \text{Var}\left(\left.\frac{\partial f_*}{\partial x_{yr}} (x_*)\right|\mathbf{x},\mathbf{y}\right) =\frac{\partial^2 C}{\partial x_{yr}\partial x'_{yr}}(x_*,x_*) -\frac{\partial C}{\partial x'_{yr}}(\mathbf{x},x_*) (\mathbf{C}+\mathbf{\Sigma})^{-1}  \frac{\partial C}{\partial x_{yr}}(x_*,\mathbf{x}) ,
\end{align}
where $\frac{\partial C}{\partial x'_{yr}}(\mathbf{x},x_*) = \left[\frac{\partial C}{\partial x'_{yr}}(x^1,x_*), \ldots, \frac{\partial C}{\partial x'_{yr}}(x^N,x_*)\right]$ and each component is computed as the partial derivative of $C\left(x,x'\right).$
\end{prop}
See Theorem 2.2.2 in \citet{adler2010geometry} for more details. By analogy, Proposition \ref{prop:GPderivative} can also be extended to consider the differential of mortality to age or other covariates. Note that the squared exponential kernel in \eqref{eq:sqExp} is infinitely differentiable with derivatives
\begin{align}
\frac{\partial C}{\partial x_{yr}'}(x,x') &= -C(x,x')\frac{\eta^2}{\theta^2_{yr}}(x_{yr}-x'_{yr}) ,\\
\frac{\partial^2 C}{\partial x_{yr} \partial x'_{yr}}(x,x') &= C(x,x')\frac{\eta^2}{\theta^2_{yr}}\left(1-\frac{1}{\theta^2_{yr}}(x_{yr}-x_{yr}')^2\right).
\end{align}

Observe that the mean $\partial m_{diff}(x_*)$ mortality improvement is equal to the derivative of the predicted mortality surface, $\frac{\partial}{\partial x_{yr}} m(x_*)$, a desirable self-consistency property. However, Proposition \ref{prop:GPderivative} goes much further, providing also analytic credible bands around $\partial m_{diff}(x_*)$ and even the full predictive distribution of the mortality improvement process. Compare these features to a non-Bayesian smoothing model, such as P-splines, that only models $m(x_*)$ and therefore beyond direct differentiation provides no uncertainty quantification for $\frac{\partial f_*}{\partial x_{yr}}$.

 \begin{figure}[ht]
  \centering
  \includegraphics[width=0.35\textwidth,height=2in,trim=0.15in 0.15in 0.15in 0.15in]{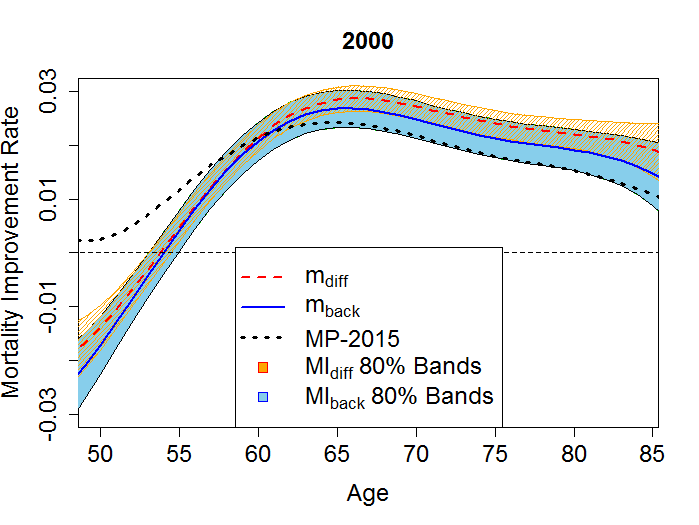}
  \includegraphics[width=0.35\textwidth,height=2in,trim=0.15in 0.15in 0.15in 0.15in]{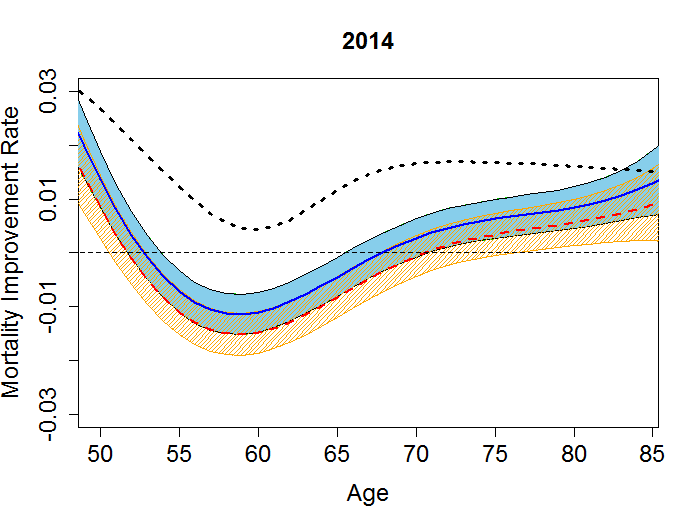}
  \caption{Estimated male mortality improvement using the differential GP model (instantaneous improvement) and the YoY improvement from the original GP model. We show the means and 80\% credible bands for $MI_{diff}^{GP}$ and $MI_{back}^{GP}$ for males aged 50--84 and years 2000 \& 2014.  Models used are fit to All Data with $m(x) = \beta_0$. \label{fig:improvbandplot}}
\end{figure}

\blue{To sum up the previous discussion, the GP framework yields a probabilistic estimate of the \emph{instantaneous} mortality improvement which is analytically consistent with the projected mortality rates.}
Figure \ref{fig:improvbandplot} shows mortality improvement estimates $\partial m^{GP}_{back}$, $\partial m^{GP}_{diff}$ and MP-2015 improvement factors for US ages 50--85 in years 2000 and 2014. The 80\% credible bands of $MI^{GP}_{back}$ and $MI^{GP}_{diff}$ are also shown. The bands for $MI^{GP}_{diff}$ were produced from \eqref{eq:derivative-marginalvar}, while for $MI^{GP}_{back}$ they were generated from empirical sampling from \eqref{eq:MIgp}.
While we observe similar overall structure (in terms of similar predicted values and similar predicted uncertainty), we also note that there are some differences which indicate the changing rate of mortality improvement. Thus, in 2000, mortality improvement was accelerating, leading to $\partial m_{diff}^{GP}(\cdot; 2000) > \partial m_{back}^{GP}(\cdot; 2000)$. In contrast, the fact that $\partial m_{diff}^{GP}(\cdot; 2014) < \partial m_{back}^{GP}(\cdot; 2014)$ suggests that mortality improvement continues to decelerate as of 2014, so that the gap with the level improvement scale embedded in MP-2015 is likely to grow. In our analysis, we find that this deceleration started around 2010, so that in the past 5-6 years mortality evolution over time has been convex, generating a growing wedge against the MP-2014/15 forecasts.

\begin{remark}
  In our analysis we concentrate on modeling the log mortality surface, obtaining the mortality improvement factors as a by-product. An alternative is to first directly calculate observed mortality improvement $MI_{back}$ and then model it with a GP. This would effectively replace the $\beta_1^{yr}$ component of the mean function with a richer structure.  This procedure is similar to that of \cite{mitchell2013modeling} where mortality improvement itself is modeled in a Lee-Carter framework.
\end{remark}


\subsection{Further Datasets}
\rev{As further empirical evidence, Appendix B presents analysis for US females, listing  the equivalents of Figures 1, 3-7 and Table 5. Additional results for four more datasets ---UK/Japan males/females---can be found in the online supplement \href{https://github.com/jimmyrisk/UKJapanResults}{github.com/jimmyrisk/UKJapanResults}. These were generated using the provided \texttt{R} notebook on the respective HMD datasets.}


Overall, the results are consistent, yielding similar covariance structure estimates. This suggests the possibility of building a hierarchical model that can improve credibility through cross-national borrowing of (statistical) information. The lengthscales in Age are all in the range $[8,16]$ and in Year are around 10. The estimated $\sigma^2$ values and $\eta^2$ are very similar throughout.
 The one exception is the UK males dataset which produced an exceptionally high $\theta_{ag}$ and an exceptionally low $\theta_{yr}$. 
 The low $\theta_{yr}$ causes out-of-sample forecasts to mean-revert too quickly, leading to poor prediction. This can be observed in the equivalent of Figures~\ref{fig:predict-2010}/\ref{fig:predict-2010-fem}, where the GP model trained on Subset II data under-estimates the mortality experience in 2014 for UK and Japan males, but does well for females. This is partly mitigated by the wider uncertainty bands for males, i.e.~the models are aware that its forecasts are less accurate.

Comparing male and female mortality, female mortality is always lower, but the Age-shapes are mostly the same. In US, we observe a smaller mortality deterioration for females around ages 50--65, and a slightly lower improvement $\partial m$ overall (compare Fig.~\ref{fig:mortTrend} and Fig.~\ref{fig:mortTrend-fem}). For smoothing, Figures \ref{fig:GP-smooth-multiyear} and \ref{fig:GP-smooth-multiyear-fem} are nearly identical in shape.  The curve in Figure~\ref{fig:MP-multiyear-fem} for 2014 is slightly different in shape compared to the male Figure \ref{fig:MP-multiyear} around ages 50--65 due to the observed mortality improvement declining in this range.

Comparing Table~\ref{table:trendfunctions-fem} with Table~\ref{table:trendfunctions} and Figure~\ref{fig:trendcomparison-fem} with Fig.~\ref{fig:trendcomparison}, we see that the trend model comparison results are near identical; the only noticeable differences are that the quadratic model is a much better fit on the test set for US females, and that the $\theta$ values for the intercept-only model are larger. We do notice differences across genders in the quadratic trend function parameters, see~Table~\ref{table:agesegmented-fem} vis-a-vis Table~\ref{table:agesegmented}. In particular, the intercept terms are different in magnitude, and some of the higher order terms differ in sign.  Thus the trend curves differ in shape between males and females in their respective age groups, which is unsurprising since the age group endpoints were chosen to match the male dataset.


\rev{
Some interesting features can be glimpsed by comparing the mortality improvement rate plots across datasets. 
First, raw YoY improvements are extremely noisy, even more than in US, so smoothing is essential for detecting trends. Second, relative to reported low and deteriorating mortality improvements in US, $MI$'s for Japan and UK are higher and are all positive, except around age 50 in UK in 2014. Japanese Males are experiencing the highest improvement rates, although their improvement pattern fluctuates a lot over the years. All populations indicate present mortality ``deceleration'' manifested in $\partial m_{\text{diff}} < \partial m_{\text{back}}$ in 2014, suggesting a trend of declining mortality improvement going forward as discussed in Section~\ref{sec:forecastingmortalityimprovement}.
Third, the age-shapes of $x_{ag} \mapsto \partial m( \cdot, x_{ag})$ are quite volatile over time and frequently ``rotate'' (perhaps due to cohort effects). Of interest is that the improvement in the 50-60  age range is very steady over the years for all females (1.5-2\% p/a), while it is all over the place for males or older ages. There is often a noticeable accelereration/deceleration of mortality, i.e.~a clear trend in $x_{yr} \mapsto \partial m(x_{yr}, ag)$ across years. For instance, decreasing improvements in Japan and UK females in their 60s, or increasing improvement in Japan males at ages 50-60.  These observations suggest diverging mortality experiences across different sub-groups. Finally, we clearly observe the difficulty in accurately learning mortality improvement rates: while the models are fairly confident about $MI$ back in year 2000, they are much less so for the ``edge'' year 2014, generating much wider relative uncertainty bands. }


\section{Extensions of GP Models}\label{sec:extensions}

\subsection{Inhomogeneous GP Models}\label{sec:inhomogeneousGPmodels}
Basic GP assumes a stationary covariance structure which may not be appropriate. If the spatial dependence in mortality experience is state-dependent, i.e.~$C(x^i,x^j)$ depends on $x^i,x^j$ (and not just $|x^i-x^j|$), this would introduce model misspecification and lead to poor model performance (i.e.~too much or too little smoothing).

To test for inhomogeneous correlation, we consider a GP model segmented by age. This means that we introduce a piecewise setup, fitting three different GP models depending on $x_{ag}$. The age grouping was done manually according to (younger) $x_{ag} \in \{50,\ldots,69\}$, (older) $x_{ag} \in \{70,\ldots,84\}$, as well as the full model $x_{ag} \in \{50, \ldots, 84\}$, and an extended model considering all ages $x_{ag} \in \{1, \ldots, 84\}$. Table \ref{table:agesegmented} presents the fitted trend and hyper-parameters for each group using a model fitted to all years 1999--2014 and quadratic mean function.

\begin{table}[ht]
\centering
\begin{tabular}{lrrrrrrrr} \hline
Ages Fit & \multicolumn{1}{c}{$\beta_0$} & \multicolumn{1}{c}{$\beta_{1}^{ag}$} & \multicolumn{1}{c}{$\beta_2^{ag}$} &  \multicolumn{1}{c}{$\beta_1^{yr}$}& \multicolumn{1}{c}{$\eta^2$} & \multicolumn{1}{c}{$\sigma^2$} & \multicolumn{1}{c}{$\theta_{ag}$} &\multicolumn{1}{c}{ $\theta_{yr}$} \\ \hline
Extended $[1,84]$ & -23.533  &  -0.005  &  8.402e-04  &  7.797e-03  &  1.904e-01  &  1.184e-03  &  3.966  &  12.795  \\
Younger $[50,69]$ & 10.521  &  0.084  &  -3.336e-05  &  -9.908e-03  &  2.633e-03  &  2.964e-04  &  4.501  &  4.196  \\
Older $[70,84]$ & 26.806  &  -0.016  &  7.113e-04  &  -1.635e-02  &  1.489e-03  &  1.517e-04  &  14.709  &  6.661  \\  \hline
All $[50,84]$ &  19.336  &  0.041  &  3.324e-04  &  -1.367e-02  &  1.760e-03  &  2.336e-04  &  4.543  &  3.825  \\   \hline
\end{tabular}
\caption{\label{table:agesegmented} GP models fitted by age groups. All models are fitted to years 1999--2014 and using a squared-exponential kernel with a quadratic mean function $m(x^n) = \beta_0 + \beta_1^{ag}x^n_{ag} + \beta_1^{yr}x^n_{yr} + \beta_2^{ag} (x^n_{ag})^2$. The reported hyper-parameter values are maximum likelihood estimates from \texttt{DiceKriging}.}
\end{table}

Table \ref{table:agesegmented} shows that the Extended age group trend/shape parameter estimates differ from the remaining groups, likely due to the fact that infant and adolescent mortality produce a non-quadratic mortality shape in age.  Furthermore, the respective positive coefficient of the Extended $\beta_1^{yr}$ parameter contradicts the idea of mortality improvement and possibly indicates poor goodness-of-fit.

Segmenting the older ages does generate some reasonable differences in fitted models: log-mortality is linear in the younger group, so that the $\beta_2^{ag}$ coefficient is negligible; it is larger in the older age group due to the rapid increase of mortality in age; combining the two as was done originally yields an average of the two estimates.  The estimates of $\beta^1_{yr}$ also support the claim of Older mortality improving faster than Younger mortality: log-mortality decreases annually at $1\%$ for for the Younger group and at $1.6\%$ for the Older group.
The $\theta_{yr}$ values are all similar across groups, except for the Extended group which needs to compensate for its poor trend fit.  The Younger and Extended fits share similar $\theta_{ag}$ values.  We attribute the larger $\theta_{ag}$ for Older ages to fitting issues due to a complicated age dependence and only 15 ages worth of data (it could also suggest that mortality rates of older ages are more correlated).  There is further evidence of this when comparing with Table \ref{table:agesegmented-fem} for females in the Appendix which also produces an unreasonably large value of $\theta^{Fem}_{ag}=44.118$ for Older ages.

\rev{In sum, this preliminary investigation suggests that a single model that includes all ages is inappropriate and both the mean and covariance structures have further age-dependence.  More detailed ``change-point'' analysis may be warranted to determine the best segmentation of data, and whether the lower cutoff at age 50 is appropriate. We remark that there exist hierarchical GP models \citep{tgpPackage} that attempt to automatically carry out such data splitting. See also \cite{LiOhare15} for a discussion about ``local'' versus ``global'' approaches to mortality.}


\subsection{Modeling Cause of Death Scales}
The raw CDC data are classified by cause of death and hence it is in fact possible to build a comprehensive mortality improvement model that is broken down beyond the basic Male/Female distinction. Understanding the different trends in cause-of-death can be important as there has been uneven progress (and in some situations reversal) of longevity improvements by cause. For example, the large improvement in mortality from coronary artery disease has not been matched by improvements in mortality from cancer. Different causes of death affect different ages, creating multiple ``cross-currents'' that drive mortality, a fact which is important for long-term projections.

Thus, mortality improvement models can benefit from analyzing by-cause data.  Building such models would need to balance the risk of over-specification with the benefit of incorporating additional data. Key issues and concepts in building a by-cause model are:
\begin{itemize} \setlength{\itemsep}{0pt}
\item	The mean function, $m$, would need to be fit to each cause.
\item	The covariance function controlling spatial correlation would also likely differ by cause.
\item	This paper focuses on modeling the log mortality rate.  A by-cause model would benefit instead from modeling the force of mortality from each cause, as the total force of mortality is simply a sum of the underlying by-causes forces of mortality. However this additive structure does not match the log transformation applied in this paper.
\item	Bayesian models with informative priors for mean function and other coefficients would provide a degree of protection against overfitting the models.
\item	A hierarchical model which builds in a relationship between the by-cause trend coefficients could be tested.
\end{itemize}

Such analysis is left for further research.

\subsection{Model Updating}
The GP model is convenient for analysis when new data becomes available.  This is in contrast to methods, such as splines, which require a full model refit.  With GPs, once the correlation structure is fit (and assuming it did  not change), the Gaussian posterior $\mathbf{f}_*$ allows for an updated $\mathbf{m}_*$ and $\mathbf{C}_*,$ see \citet[Section 5.1]{ludkovski2015kriging}  for details.  These formulas showcase the explicit impact of additional data, both for smoothing past experience, or projecting forward in time.

To illustrate the effect of a new year of data, we compute the predicted mean $\mathbf{m}_*$ and standard deviation $\mathbf{C}_*$ for age 65 and years 1999, 2013 and 2016, first based on data for all ages and calendar years 1999--2013, and then updated with year-2014 data. The results are listed in Table \ref{table:updating}.

\begin{table}[ht]
\centering
\begin{tabular}{l|cc|cc} \hline
& \multicolumn{2}{c|}{Before Updating (1999--2013)} & \multicolumn{2}{c}{After Updating (1999--2014)}\\ \hline
$x_{yr}$ & \multicolumn{1}{c}{$\E[f(65,x_{yr})|\mathbf{x},\mathbf{y}]$} & \multicolumn{1}{c|}{$s_*(65,x_{yr})$} & \multicolumn{1}{c}{$\E[f(65,x_{yr})|\tilde{\mathbf{x}},\tilde{\mathbf{y}}]$} & \multicolumn{1}{c}{$\tilde{s}_*(65,x_{yr})$} \\ \hline
1999 & -3.8845 & 0.0174 & -3.8849 & 0.0173\\
2013 & -4.1497 & 0.0174 & -4.1502 & 0.0170\\
2016 & -4.1197 & 0.0266 & -4.1248 & 0.0208
\end{tabular}
\label{table:updating}
\caption{GP model updating: $\mathbf{x}, \mathbf{y}$ refers to observed mortality for ages 50--84, years 1999--2013;  $\tilde{\mathbf{x}}, \tilde{\mathbf{y}}$ is the same data augmented with year-2014 experience. The mean function is intercept-only, $m(x) = \beta_0$; $s_*$ is posterior standard deviation.}
\end{table}

The additional year of credibility decreases posterior standard deviations $s_*$. Unsurprisingly, the impact on 1999-prediction is negligible since it is so far in the past. The standard deviation for 2013 has a slight decrease after updating, while 2016 has a much larger reduction: the original model was initially predicting 3 years out-of-sample, while the updated one does for just 2 years out-of-sample. Similarly, the in-sample means change only slightly, while the out-of-sample 2016 has a larger adjustment. The overall decrease in updated posterior means is consistent with the fact that the observed log-mortality for age 65 in 2014 was $ -4.1543$, lower than the predicted $-4.1443$ using the 1999--2013 model.

\subsection{Other Extensions}

A standard assumption is that mortality curves are increasing in Age, i.e.~$x_{ag} \mapsto f(x_{ag},\cdot)$ is monotone. The basic GP framework does not impose any monotonicity restriction. Such structural constraints on $f$ can help in improving mortality projection in terms of $m_*$ (especially for long-range forecasts), as well as reduce predictive uncertainty measured by $s^2_*$. At the same time, constraints are at odds with the underlying Gaussian random field statistical paradigm, introducing additional complexity in fitting and making inference from the constrained posterior.

One promising recent solution was proposed in
\cite{RiihimakiVehtari10} who suggested incorporating monotonicity by adding virtual observation points $\tilde{x}_i, \tilde{m}_i$ for the derivative of ${f}(x_i)$. Because the derivative $\bm{f}'$ also forms a GP, one can explicitly write down the joint covariance structure of $(\bm{f}, \bm{f}')$ (for example the posterior mean of $\bm{f}'$ is the derivative of $m_*$). Monotonicity is then implied by requiring the derivative to be positive at the given $\tilde{x}_i$'s. As the size of the latter collection increases, the resulting estimate is more and more likely to be increasing \emph{everywhere} in the domain. This strategy circumvents the direct monotonicity restriction while maintaining computational tractability through linear constraints. \cite{RiihimakiVehtari10} give a recipe for adaptively placing such virtual derivative points by iteratively adding new $\tilde{x}_i$'s where the current $m_*$ violates monotonicity. Further constraints, such as expert opinions about mortality at extreme ages (100+) could be beneficially added.

An additional extension involves use of multiple data sets; there are many instances where mortality data from one source might be more up-to-date than from other sources, for example CDC data provides at least 3 more years of information than SSA data.  The use of co-kriging models or the use of CDC data as an input to a GP used to model SSA data is another avenue of possible future research.  Such co-kriging models might also be helpful when using population improvement data to supplement a GP analysis of a specific insurance company's or pension fund's mortality experience.

\section{Conclusion}
We have proposed and investigated the use of Gaussian Process models for smoothing and forecasting mortality surfaces. Our approach takes a unified view of the mortality experience as a statistical \emph{response surface} that is noisily reflected in realized mortality experience. A statistical procedure is then used to calibrate the spatial dependence among the latent log-mortality rates. The GP model provides a consistent, non-parametric framework for uncertainty quantification in \emph{both} the mortality surface itself, as well as mortality improvement, which corresponds to relationship between $f$ and $x_{yr}$. This quantification can be done in-sample, by retrospectively smoothing raw mortality counts, or out-of-sample, by building mean forecasts, uncertainty bands, and full scenarios for future mortality/mortality improvement evolution. In contrast, traditional actuarial techniques for graduating data commonly and currently in use (e.g.~the Whittaker-Henderson model used by RPEC) focus on smoothing noisy data but fail to provide measures of uncertainty about the fit.

We have focused on population data and smoothing over age and year.  The model can be easily extended to additional dimensions, such as duration and net worth in the context of life insurance, or year-of-birth cohort for pension mortality analysis. Adding covariates to the definition of the covariance kernel $C(x,x')$ is straightforward, with the main challenge lying in interpreting the resulting GP parameters which would reflect a modified concept of spatial distance.

Perhaps the most useful application of our model is for analyzing the latest mortality data, i.e.~at the ``edge'' of the mortality surface. Here we find and document the statistical evidence that US mortality improvements have materially moderated across a large swath of ages. In particular, for Ages 55--70, US mortality has been effectively flat, or possibly even increasing in the 2010's. This points to a large divergence from the MP-2015 improvement scales that continue to assume significant mortality gains for all ages and would seem to be overstated at least in the near-term. Moreover, by explicitly computing the differential mortality improvement $MI^{GP}_{diff}$, our model gives the most current, instantaneous forecast on mortality improvement, in contrast to the traditional year-over-year estimates.

On a related note, our analysis quantifies the apparent correlation in observed mortality experience across Age and calendar Year. Thus, the obtained estimates of length-scales $\theta_{yr}$, imply that studies with very long historical analysis (e.g.~going back to 1950 or even 1900) may not add much value to our understanding of current or future projected trends in mortality improvement. Similarly, long-term projections of future mortality improvement (e.g.~MP-2015 which is used for projecting mortality often 40 to 60 years into the future) contain a higher degree of uncertainty than is typically recognized in actuarial analyses. Indeed, our results suggest that projections more than a decade into the future are entirely based on the assumed prior calendar trend and hence have almost no credibility based on observed experience.

\blue{Our results show that even a ``vanilla'' implementation of a GP model already produces useful statistical description of the mortality experiences that is competitive with existing methods in terms of its probabilistic richness and accuracy. We therefore see an enormous potential for further works in this direction, in particular to resolve some further methodological challenges.}
 Mean function modeling which is typically not an important component of GP models in other contexts, is critical for actuaries when projecting out-of-sample. Also, constrained GP models that structurally enforce the age-shape of mortality could be promising in creating better future forecasts. \blue{Yet another challenge is better blending of the data-influenced prediction and the prior mean for extrapolation which can be achieved with other Gaussian field specifications or other techniques \citep{salemi2013generalized,lee2015single}.}
A different challenge consists in creating meaningful backtesting analyses which would test not just predictive accuracy of $m_*$, but also the quality of the generated credibility intervals (both for mortality rates and mortality improvements), and the assumption of Age- and Year-stationary covariance structure. On that point, it would be worthwhile to investigate data from other countries to infer commonalities in mortality correlations.

\bibliographystyle{humanbio}
\bibliography{GPmortality}

\appendix
\section*{Appendix}
\section{Supplementary Plots}

\subsection{GP Model Residuals}

\begin{figure}[htb]
  \centering
  \includegraphics[width=0.47\textwidth,height=2.2in,trim=0.15in 0.12in 0.15in 0.15in]{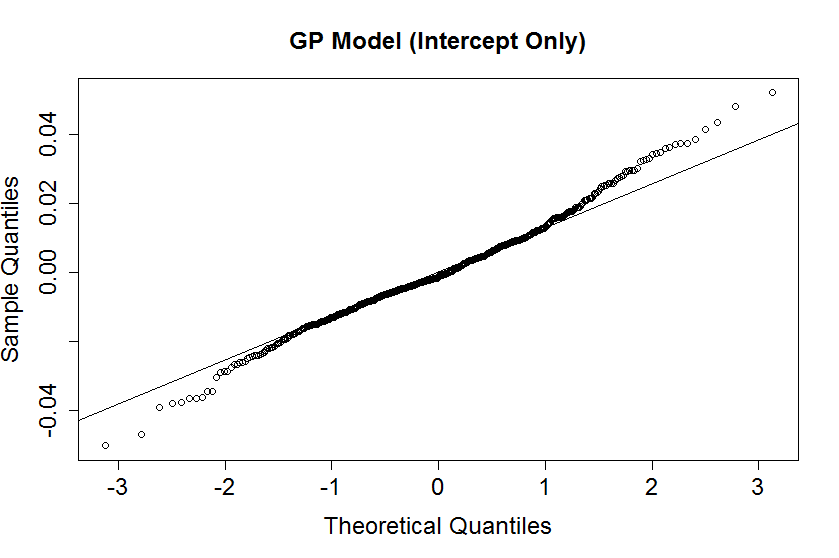}
 \includegraphics[width=0.47\textwidth,height=2.2in,trim=0.15in 0.12in 0.15in 0.15in]{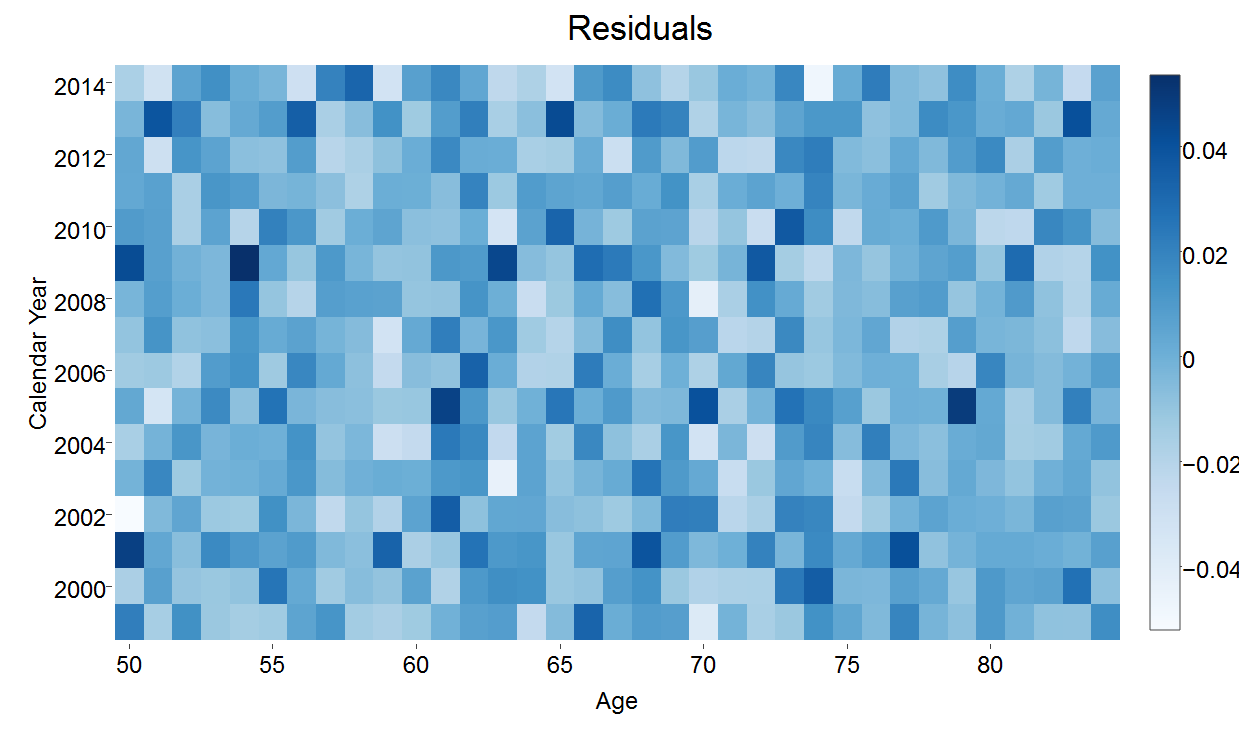}
  \caption{Left: Q-Q Plots for residuals of a fitted GP model with mean function $m(x)=\beta_0$. We use Male All Data to  test the normality assumption of $\eps$ in \eqref{eq:kriging}. We observe that the GP residuals are reasonably Gaussian with mildly heavy tails. Right: heatmap of $\eps(x)$ as a function of the two-dimensional input $x=(x_{ag}, x_{yr})$. We observe no apparent correlation in the fitted residuals.
  \label{fig:residualanalysis}}
\end{figure}

\clearpage

\section{Tables and Figures for US Female Data}

 \begin{figure}[ht]
  \centering
  \includegraphics[width=0.44\textwidth,height=2.2in,trim=0.15in 0.15in 0.15in 0.15in]{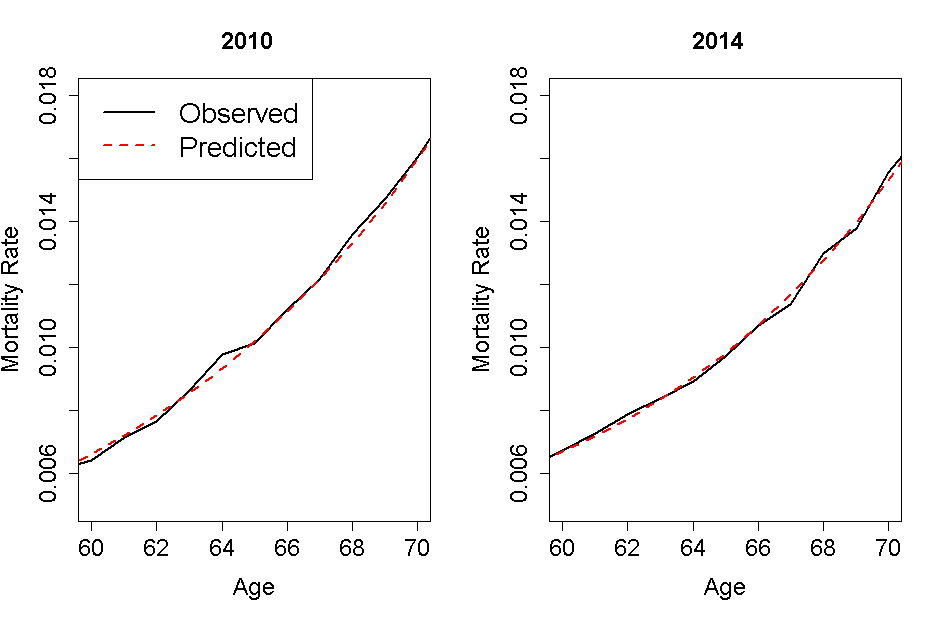}
 \includegraphics[width=0.44\textwidth,height=2.2in,trim=0.15in 0.15in 0.15in 0.15in]{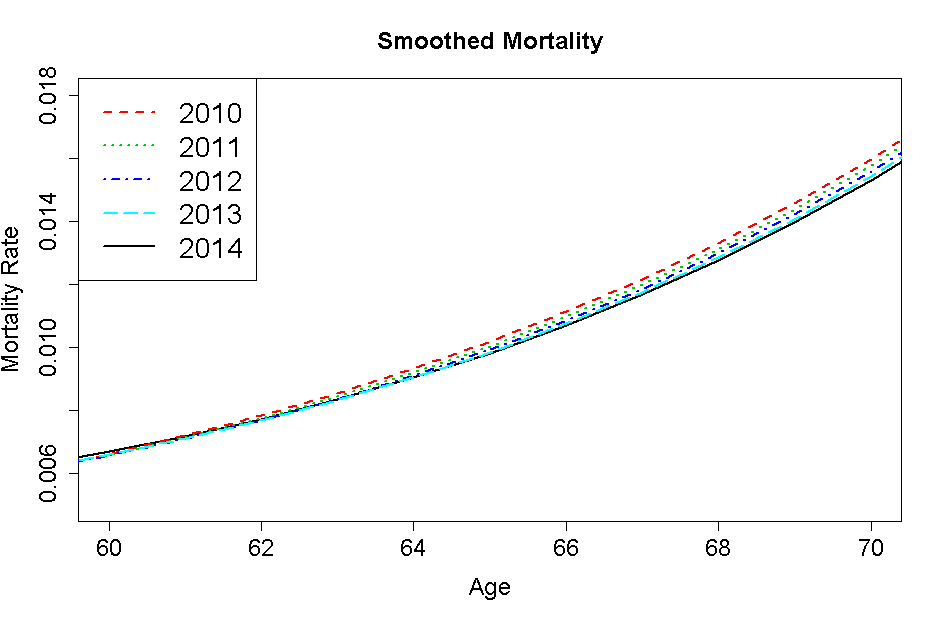}
  \caption{Mortality rates for US Females aged 60--70 during years 2010--2014. Raw vs.~estimated smoothed mortality curves.   {Models are fit to All Female data.}\label{fig:GP-smooth-multiyear-fem}   }
\end{figure}
 \begin{figure}[ht]
  \centering
  \includegraphics[width=0.45\textwidth,height=2.5in,trim=0.15in 0.15in 0.15in 0.15in]{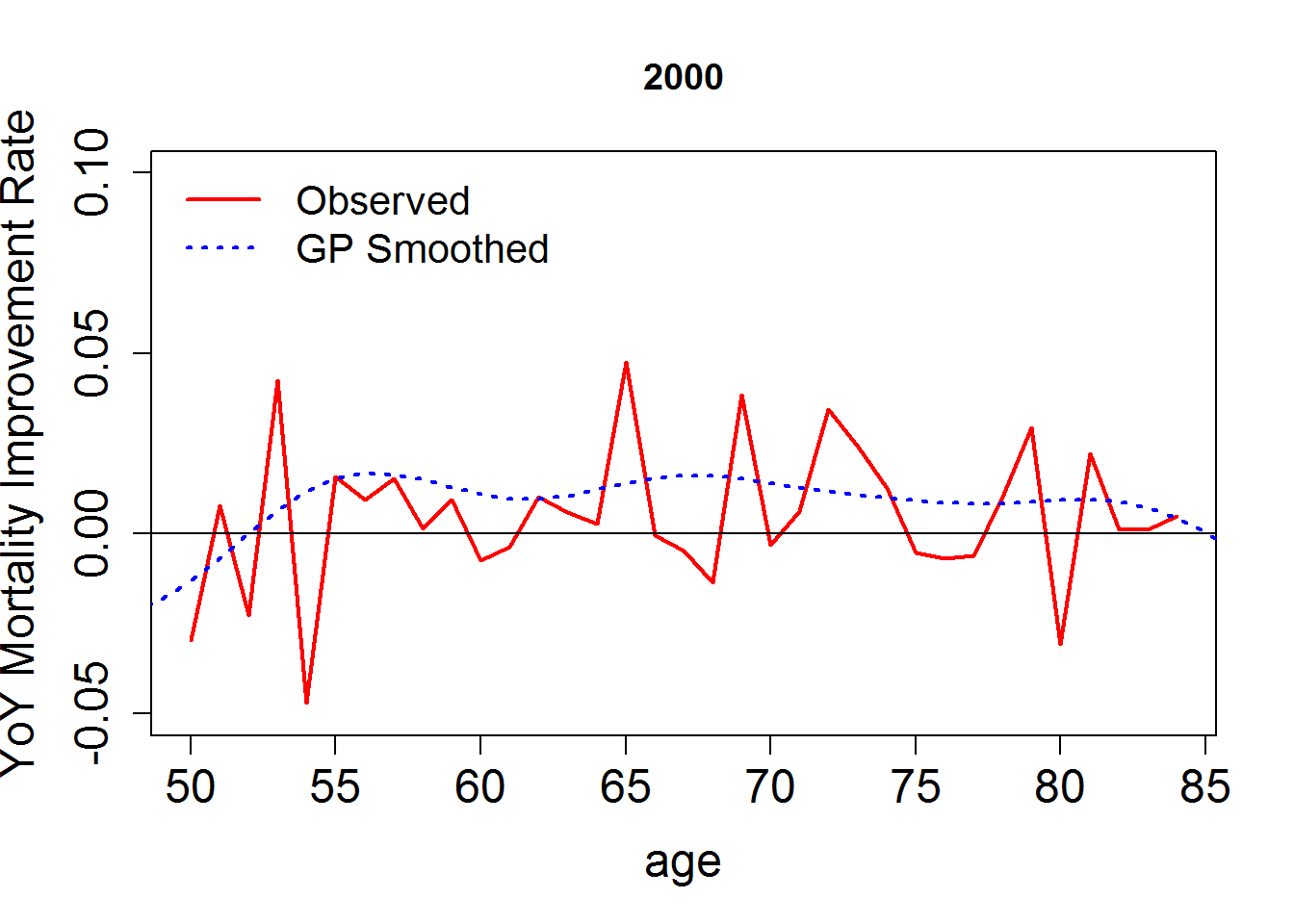}
  \includegraphics[width=0.45\textwidth,height=2.5in,trim=0.15in 0.15in 0.15in 0.15in]{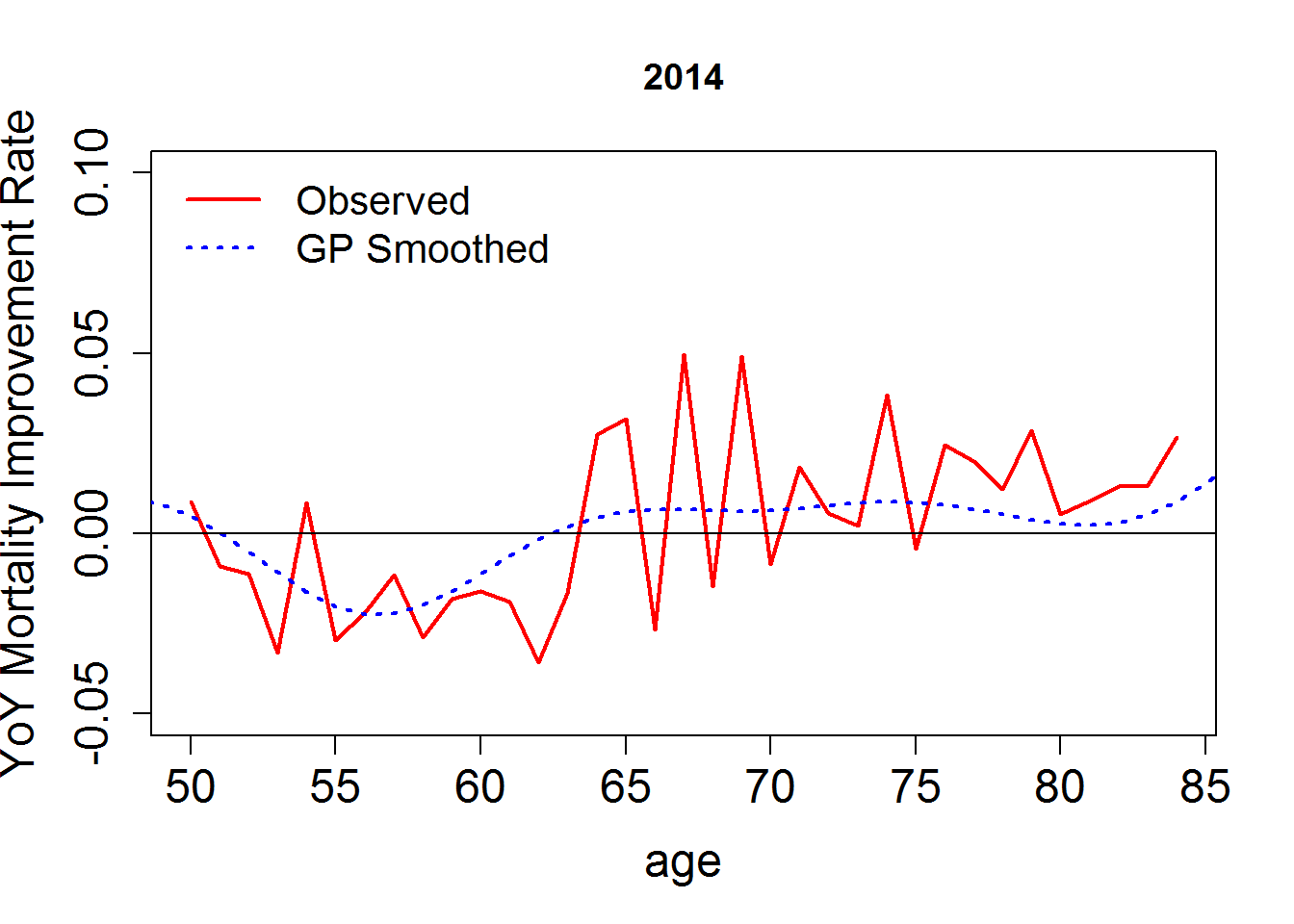}
  \caption{Mortality improvement factors for US Females using All data. Solid red lines indicate the empirical mortality experience; dotted blue lines are the smoothed estimates using a GP. \label{fig:MP-multiyear-fem} }
\end{figure}

 \begin{figure}[ht]
  \centering
  \includegraphics[width=0.5\textwidth,height=2.2in,trim=0.15in 0.15in 0.15in 0.15in]{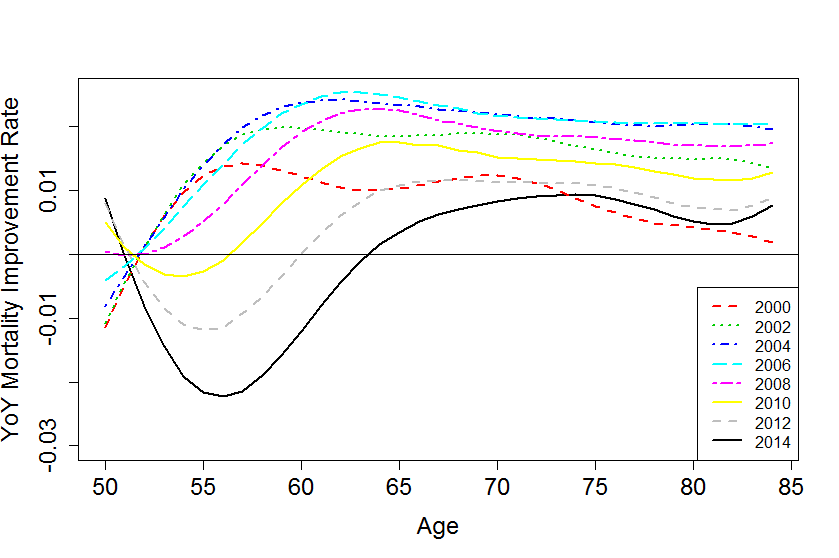}
  \caption{Comparison of yearly mortality improvement factors for US Females using All data. The curve for 2014 is the same as in Figure \ref{fig:MP-multiyear-fem}. \label{fig:mortTrend-fem} }
\end{figure}

\begin{table}[ht]
\centering
\begin{tabular}{l|rrrr|rrrr}
&  \multicolumn{4}{|c|}{Trend Parameter MLE's} & \multicolumn{4}{|c}{GP Hyperparameter MLE's}\\ \hline
&  \multicolumn{1}{c}{$\beta_0$} & \multicolumn{1}{c}{$\beta^{ag}_1$} &  \multicolumn{1}{c}{$\beta^{ag}_2$} &\multicolumn{1}{c}{$\beta^{yr}_1$} & \multicolumn{1}{c}{$\eta^2$} & \multicolumn{1}{c}{$\sigma^2$} & \multicolumn{1}{c}{$\theta_{ag}$} & \multicolumn{1}{c}{$\theta_{yr}$}\\ \hline
Intercept  &  -5.101  &  -  &  -  &  -  &  4.444e-01  &  2.968e-04  &  7.363  &  10.882  \\
Linear  & 4.484  &  0.083  &  - &  -7.167e-03   &  2.802e-03  &  3.682e-04  &  4.432  &  4.505  \\
Quadratic & 11.207  &  0.054  &  2.712e-04  &  -1.014e-02  &  2.053e-03  &  2.911e-04  &  4.464  &  4.384  \\
\end{tabular}
\caption{Mean functions and fitted covariance parameters using Set I US Female Data (ages 50--70 and years 1999--2010). The mean functions are $m(x) = \beta_0$ for Intercept, $m(x) =\beta_0 + \beta_1^{ag} x_{ag} + \beta_1^{yr}x_{yr}$ for Linear, and $m(x) = \beta_0 + \beta_{1}^{ag} x_{ag} + \beta_1^{yr} x_{yr} + \beta_2^{ag} x_{ag}^2$ for Quadratic.\label{table:trendfunctions-fem}}
\end{table}

\begin{figure}[ht]
\centering
\includegraphics[scale=0.22]{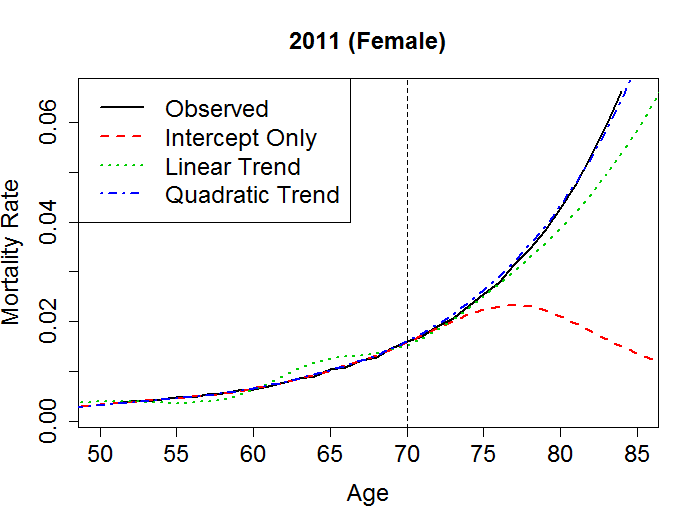}
\includegraphics[scale=0.22]{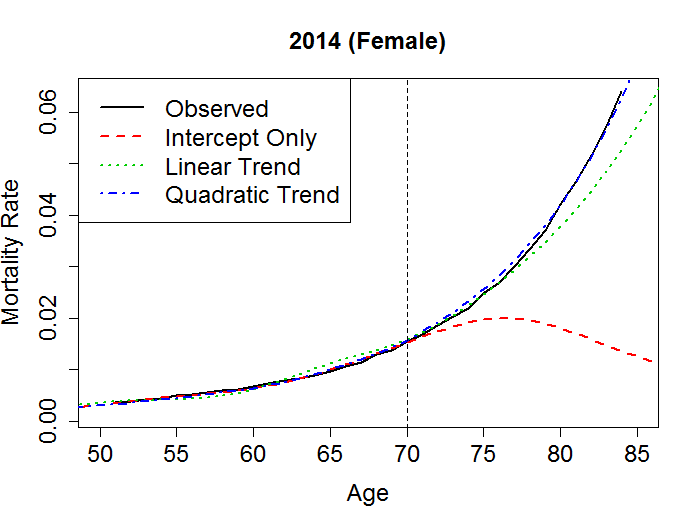}
\caption{ \label{fig:trendcomparison-fem} Comparison of mean function choices in extrapolating mortality rates at old ages for US Females.  Models are fit to years {1999--2010} and ages 50--70 (Subset III), with estimates made for Age 50--85 in {2011 and 2014}. The vertical line indicates the boundary of the training dataset in $x_{ag}$. The mean functions are given in Table \ref{table:trendfunctions-fem}.}
\end{figure}

 \begin{figure}[ht]
  \centering
  \begin{tabular}{cc} \hspace*{-20pt}
  \includegraphics[scale=0.18]{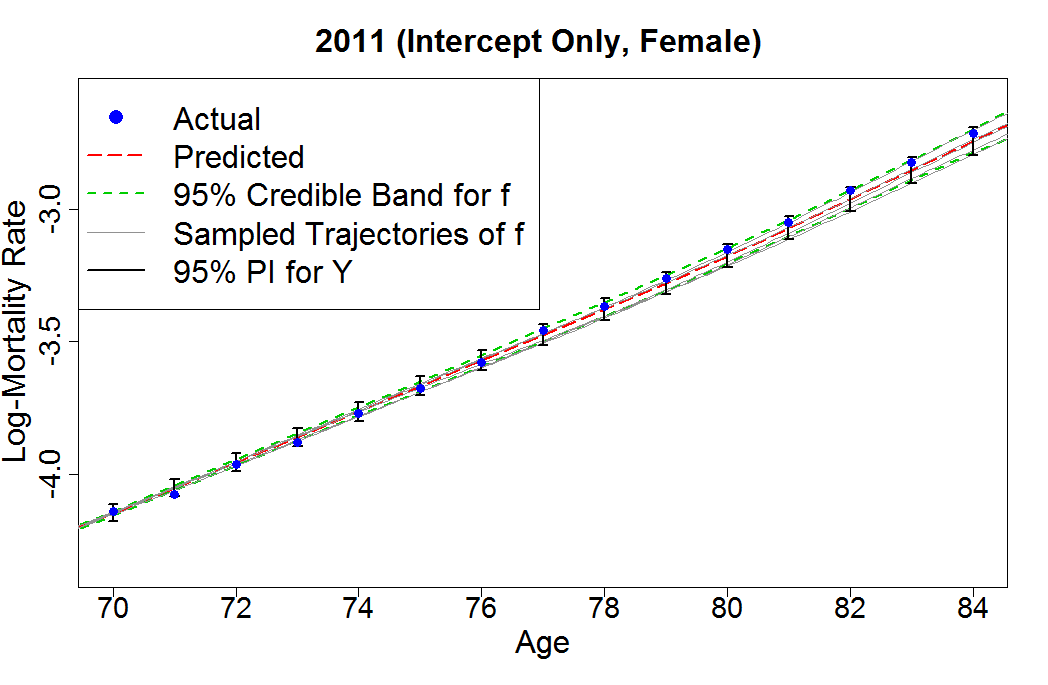} &
  \includegraphics[scale=0.18]{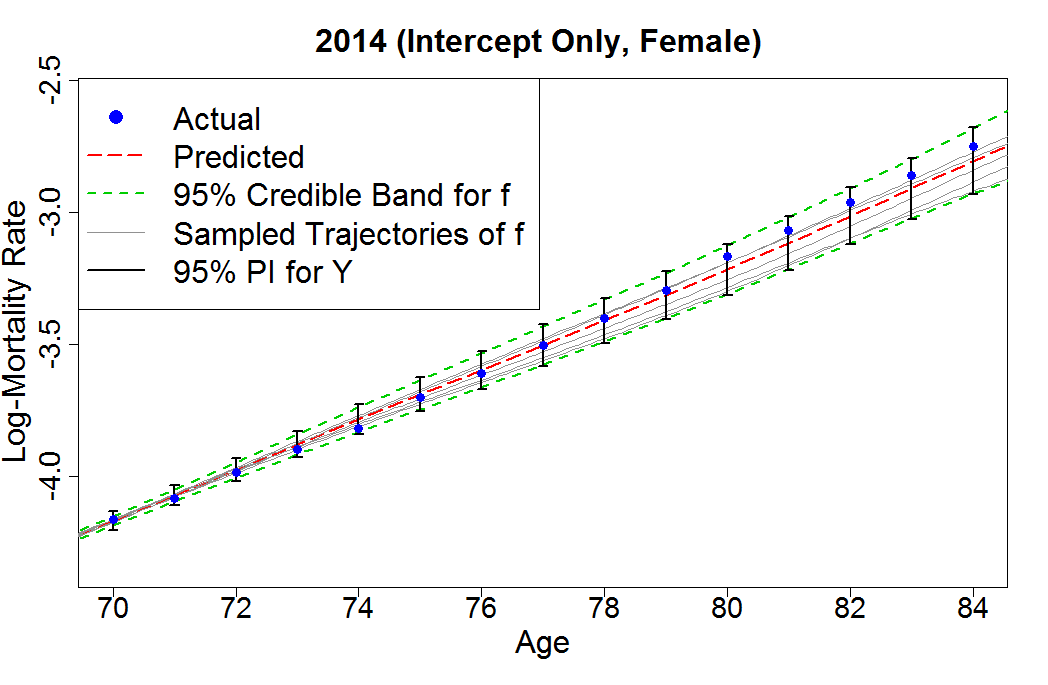}
  \end{tabular}
  \caption{ \label{fig:predict-2010-fem} Mortality rate prediction for years 2011 and 2014 and ages 71--84.  Model is fit on Subset II US Female data with intercept-only  mean function and squared-exponential kernel.}
\end{figure}

\begin{table}[ht]
\centering
\begin{tabular}{lllllllll} \hline
\multicolumn{8}{c}{Quadratic $m(x) = \beta_0 + \beta_1^{ag}x_{ag} + \beta_1^{yr}x_{yr} + \beta_2^{ag} x_{ag}^2$}\\ \hline
Ages Fit & $\beta_0$ & $\beta_{1}^{ag}$ & $\beta_2^{ag}$ &  $\beta_1^{yr}$& $\eta^2$ & $\sigma^2$ & $\theta_{ag}$ & $\theta_{yr}$ \\ \hline
Extended $[1,84]$ & -25.224  &  -0.008  &  8.721e-04  &  8.678e-03  &  2.170e-01  &  1.187e-03  &  4.095  &  13.040  \\
Younger $[50,69]$ & 1.128  &  0.080  &  3.912e-05  &  -5.471e-03  &  4.311e-03  &  2.907e-04  &  5.695  &  5.487  \\
Older $[70,84]$ &  17.272  &  -0.038  &  9.071e-04  &  -1.151e-02  &  2.543e-03  &  1.334e-04  &  44.118  &  6.856  \\   \hline
All $[50,84]$ &  7.473  &  0.035  &  4.186e-04  &  -7.980e-03  &  2.814e-03  &  2.236e-04  &  5.574  &  5.249  \\   \hline
\end{tabular}
\caption{GP models fitted by age groups with US Female data. All models used squared-exponential kernel and years 1999--2014. \label{table:agesegmented-fem}}
\end{table}

 \begin{figure}[ht]
  \centering
  \includegraphics[width=0.35\textwidth,height=2in,trim=0.15in 0.15in 0.15in 0.15in]{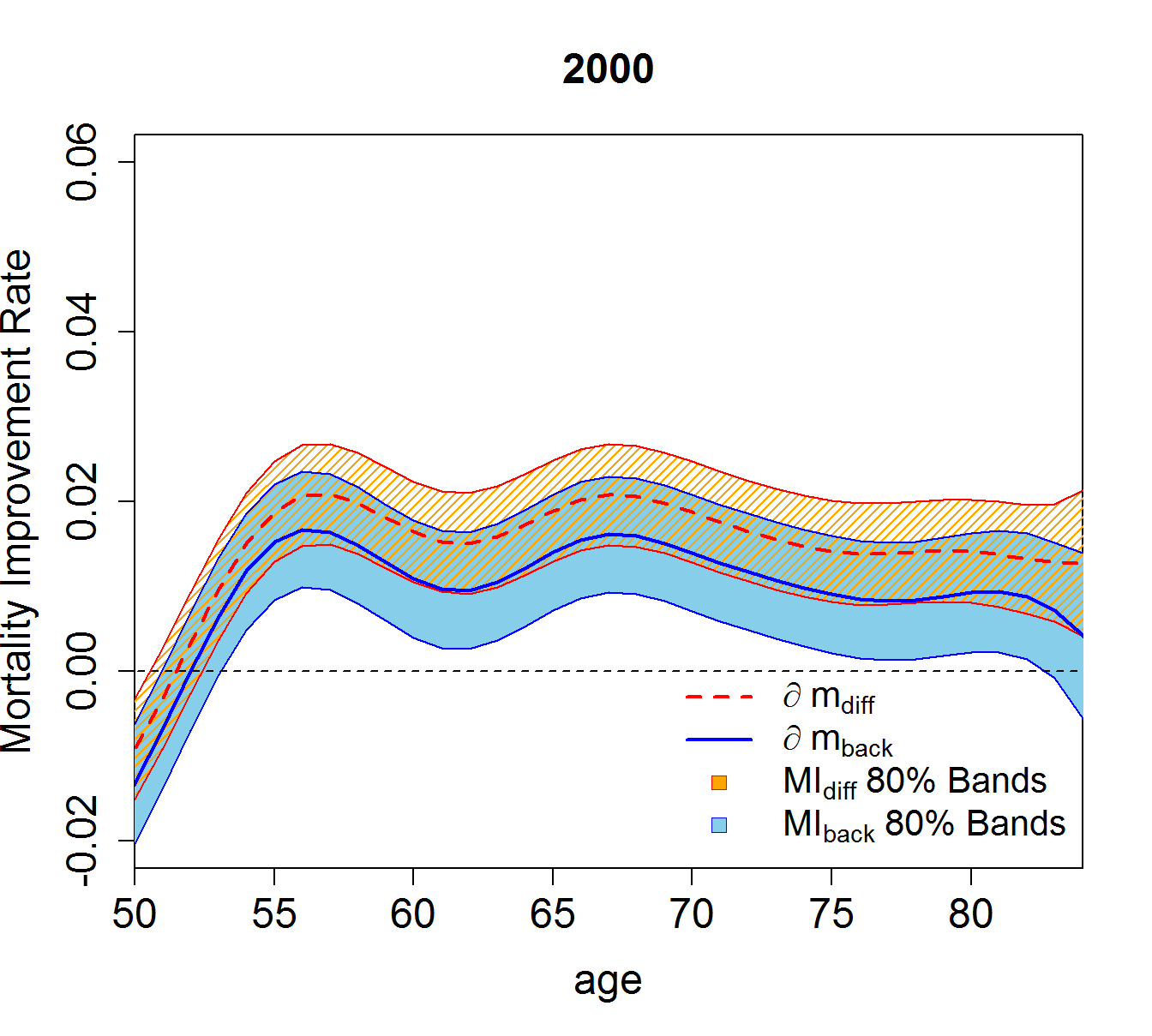}
  \includegraphics[width=0.35\textwidth,height=2in,trim=0.15in 0.15in 0.15in 0.15in]{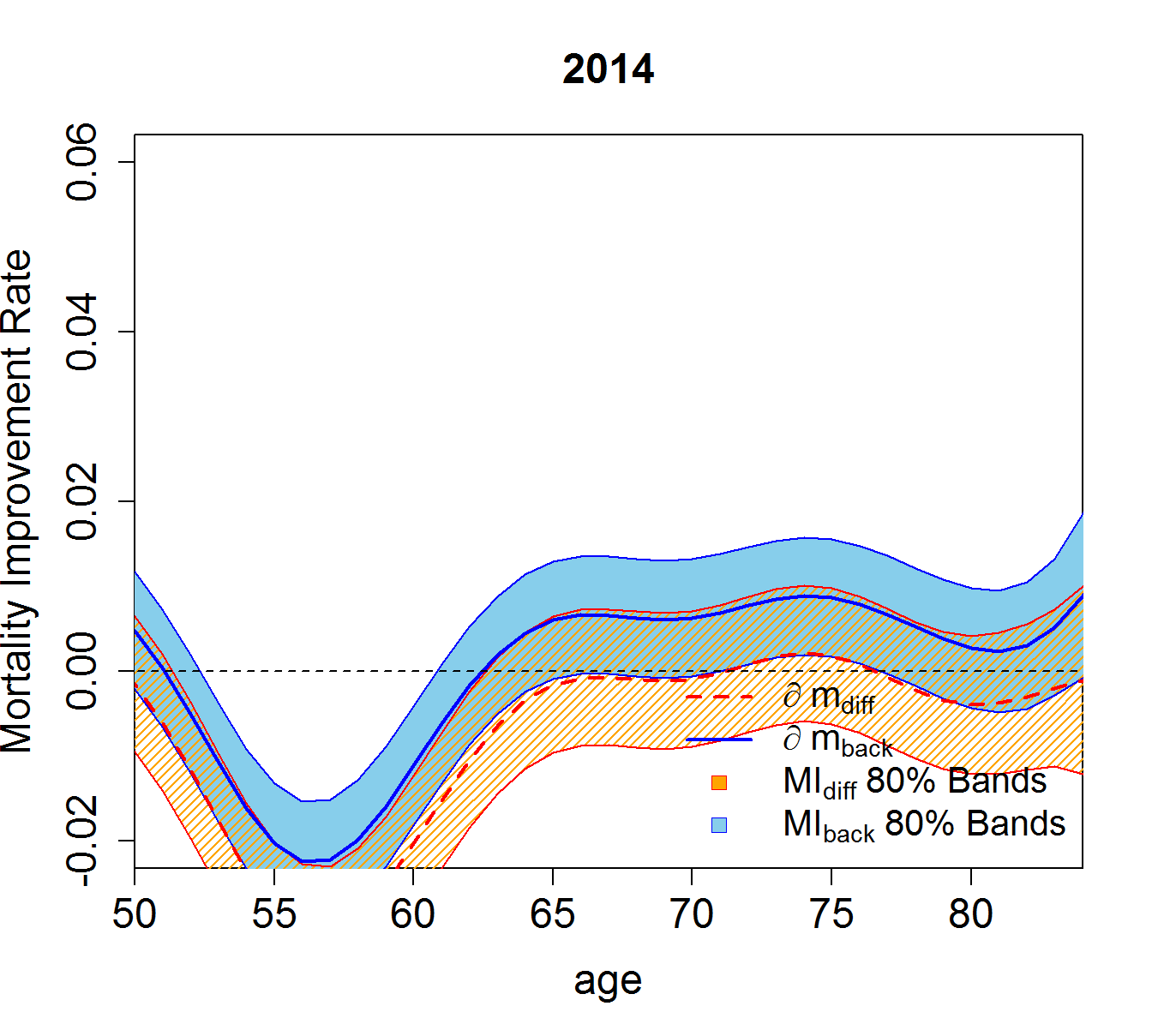}
  \caption{Estimated US female mortality improvement using the differential GP model (instantaneous improvement) and the YoY improvement from the original GP model. We show the means and 80\% uncertainty bands for $MI_{diff}^{GP}$ and $MI_{back}^{GP}$ for ages 50--84 and years 2000 \& 2014.  Models used are fit to All Data . \label{fig:improvbandplotF}}
\end{figure}

\clearpage
\pagestyle{empty}

\section{Tables and Figures for Japan and UK Male/Female Data}

\begin{table}[ht]
\centering
\begin{tabular}{l|rrrrrr}
& \multicolumn{1}{c}{Japan M} & \multicolumn{1}{c}{Japan F} & \multicolumn{1}{c}{UK M} & \multicolumn{1}{c}{UK F}\\ \hline
$\theta_{ag}$ & 10.0969 & 24.1946 & 31.3212 & 16.9089 \\
$\theta_{yr}$ & 11.4233 & 13.6124 & 5.2799 & 25.8392 \\
$\eta^2$ & 2.571 & 4.6069 & 3.0681 & 3.0043\\
$\sigma^2$ & 1.257e-03 & 2.239e-03 & 1.569e-03 & 2.135e-03\\
$\beta_0$ & -4.9898 & -5.5825 & -4.2687 & -4.2945\\
\end{tabular}
\caption{ Hyperparameter estimates based on maximum likelihood (\texttt{DiceKriging}).  The GP is fitted to all data and uses squared-exponential covariance kernel \eqref{eq:sqExp} with prior mean $m(x) = \beta_0$. \label{table:app-hyperparameters}}
\end{table}


 \begin{figure}[!hb]
  \centering
  \begin{tabular}{cc} \hline
  Japan Males & Japan Females\\
  \includegraphics[width=0.23\textwidth,height=0.24\textheight,trim=0in 0.1in 0in 0.1in]{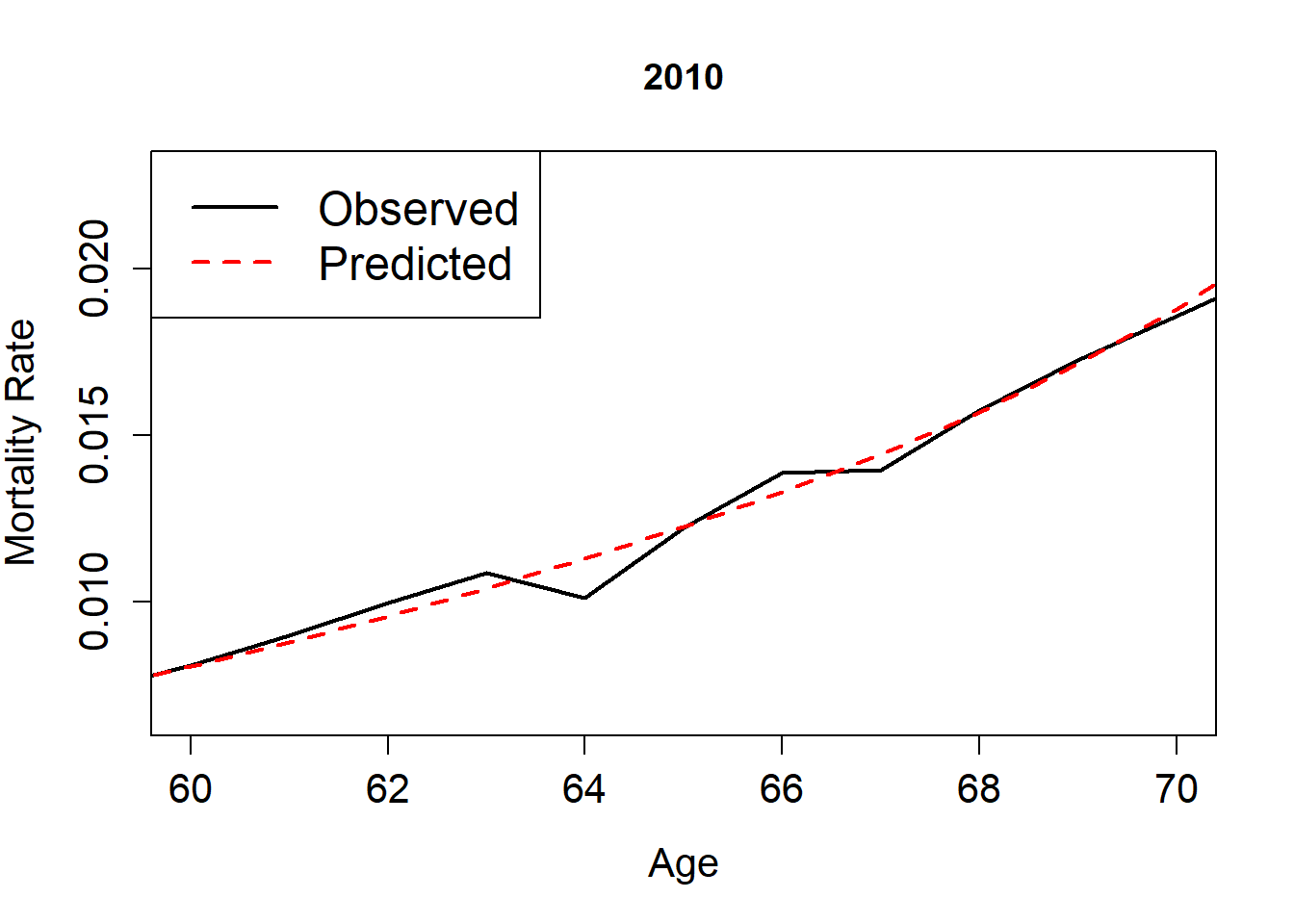}  \includegraphics[width=0.23\textwidth,height=0.24\textheight,trim=0in 0.1in 0in 0.1in]{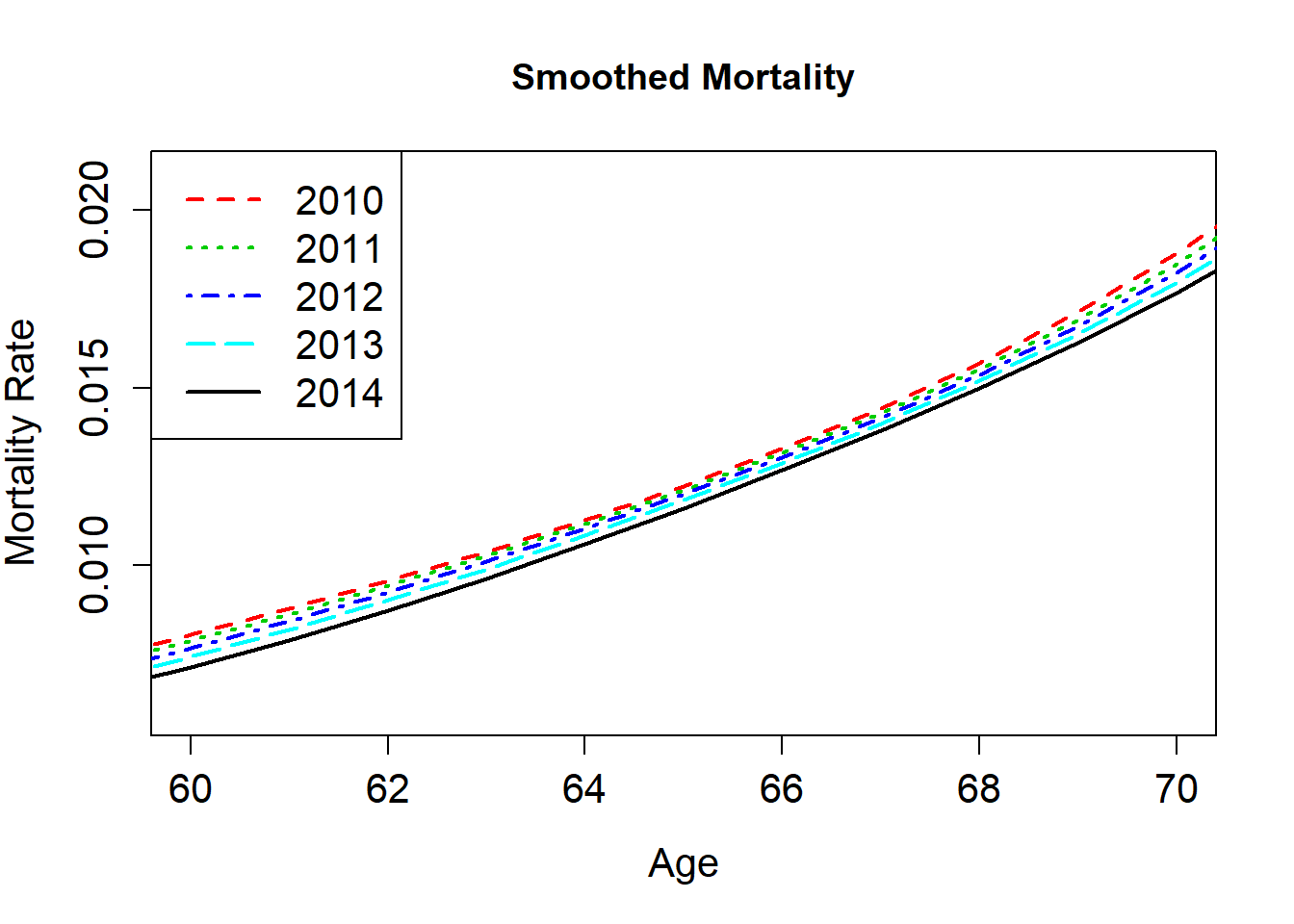} &
  \includegraphics[width=0.23\textwidth,height=0.24\textheight,trim=0in 0.1in 0in 0.1in]{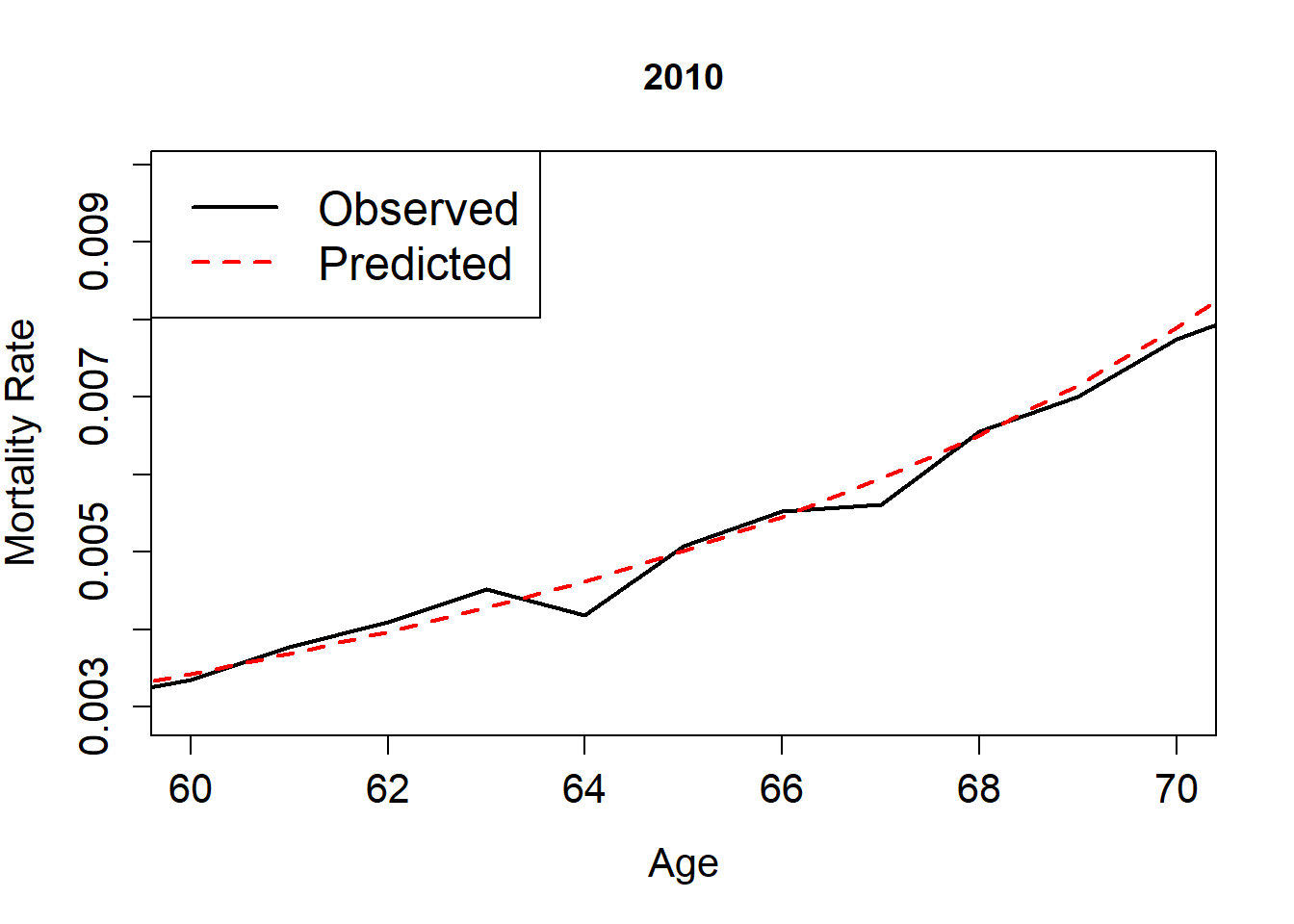}   \includegraphics[width=0.23\textwidth,height=0.24\textheight,trim=0in 0.1in 0in 0.1in]{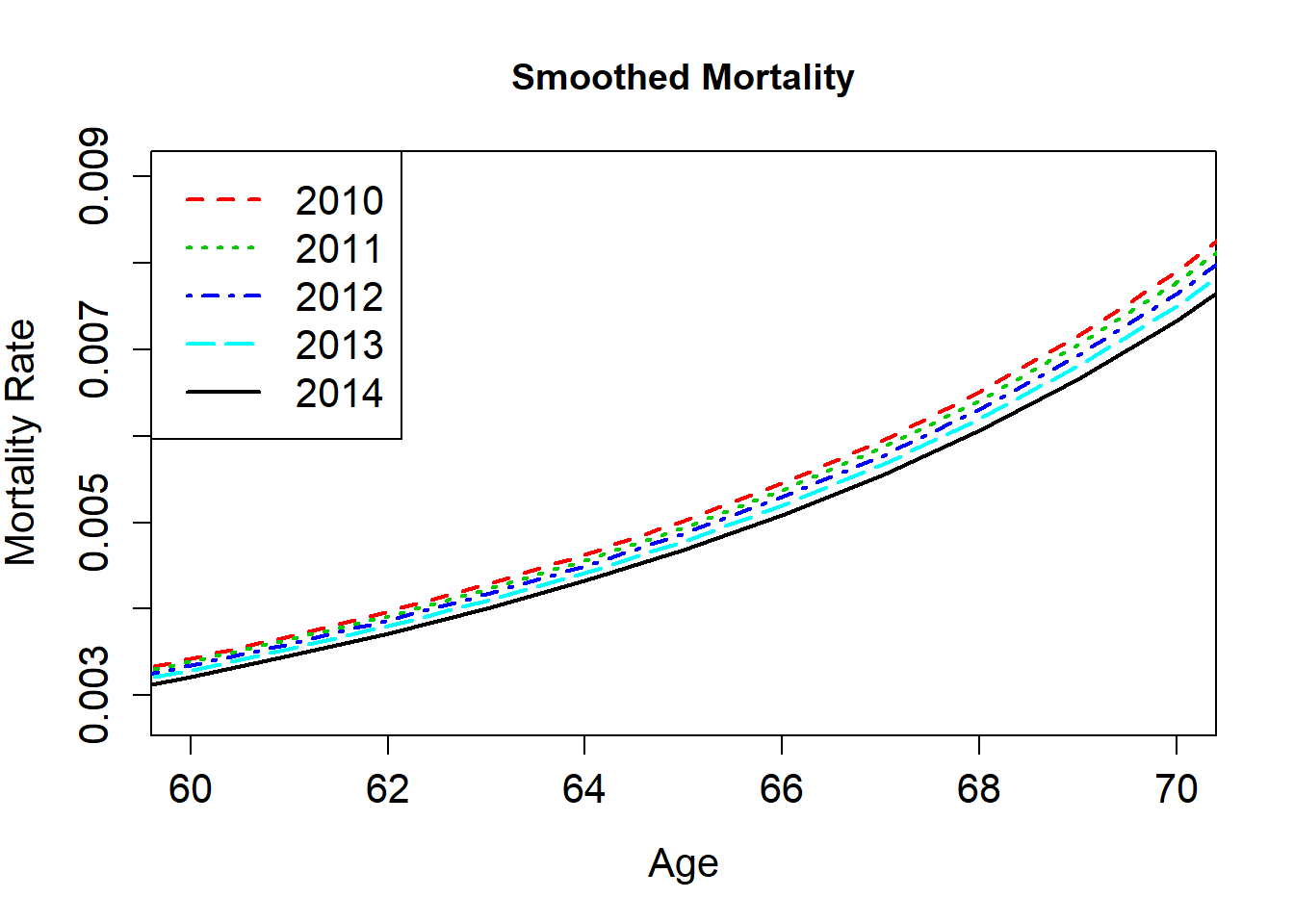} \\ \hline
  UK Males & UK Females\\
  \includegraphics[width=0.23\textwidth,height=0.24\textheight,trim=0in 0.1in 0in 0.1in]{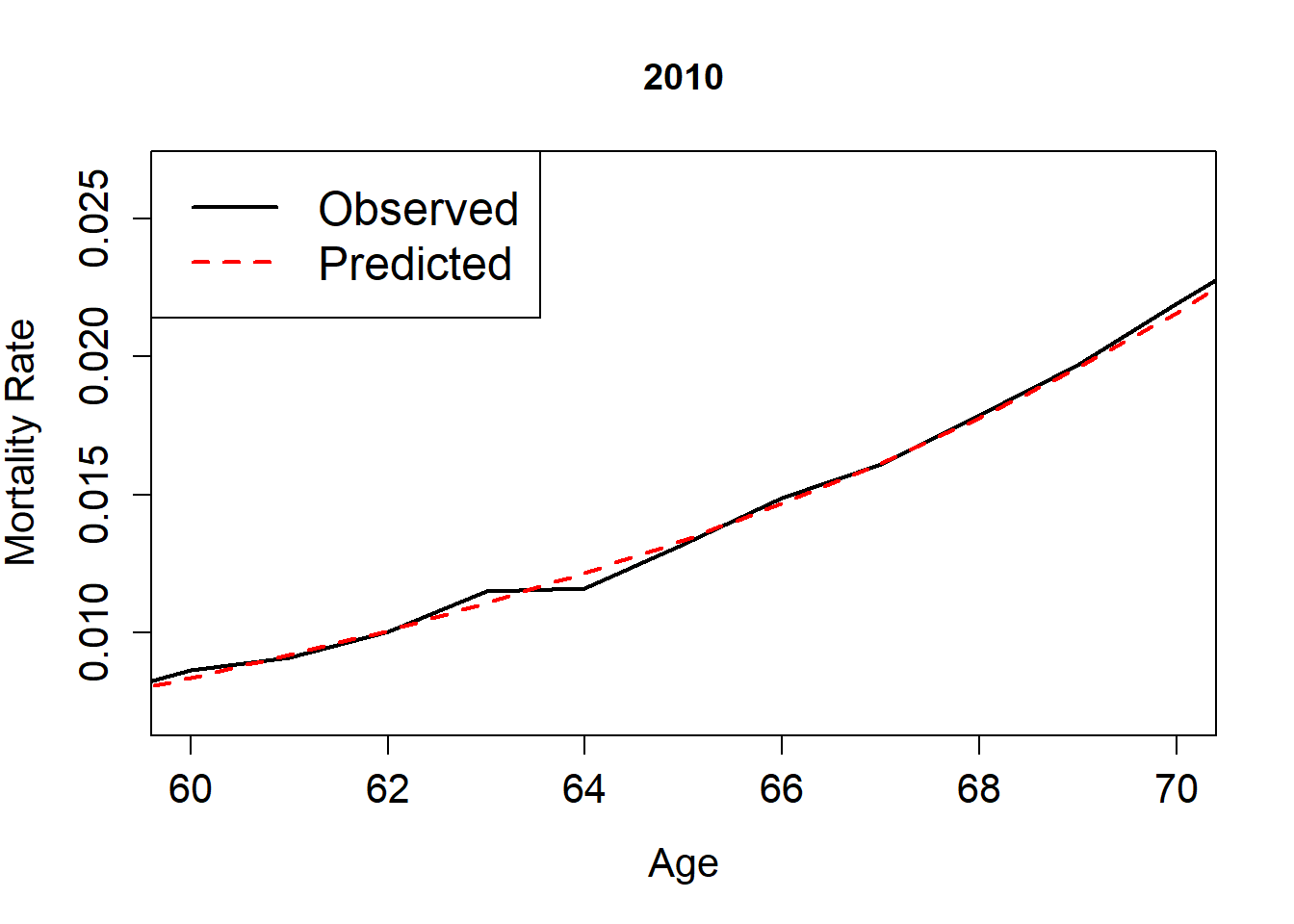}  \includegraphics[width=0.23\textwidth,height=0.24\textheight,trim=0in 0.1in 0in 0.1in]{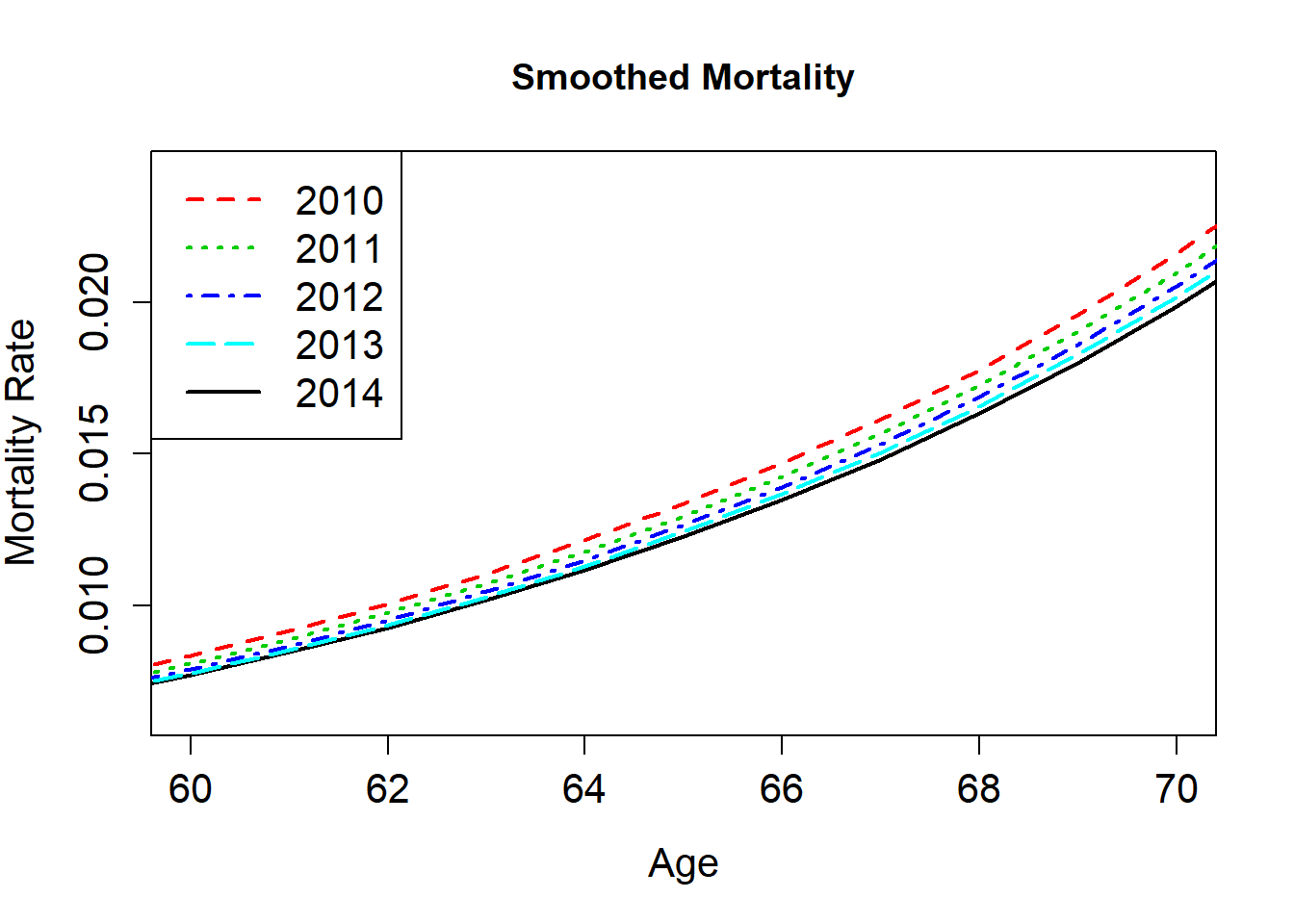} &
  \includegraphics[width=0.23\textwidth,height=0.24\textheight,trim=0in 0.1in 0in 0.1in]{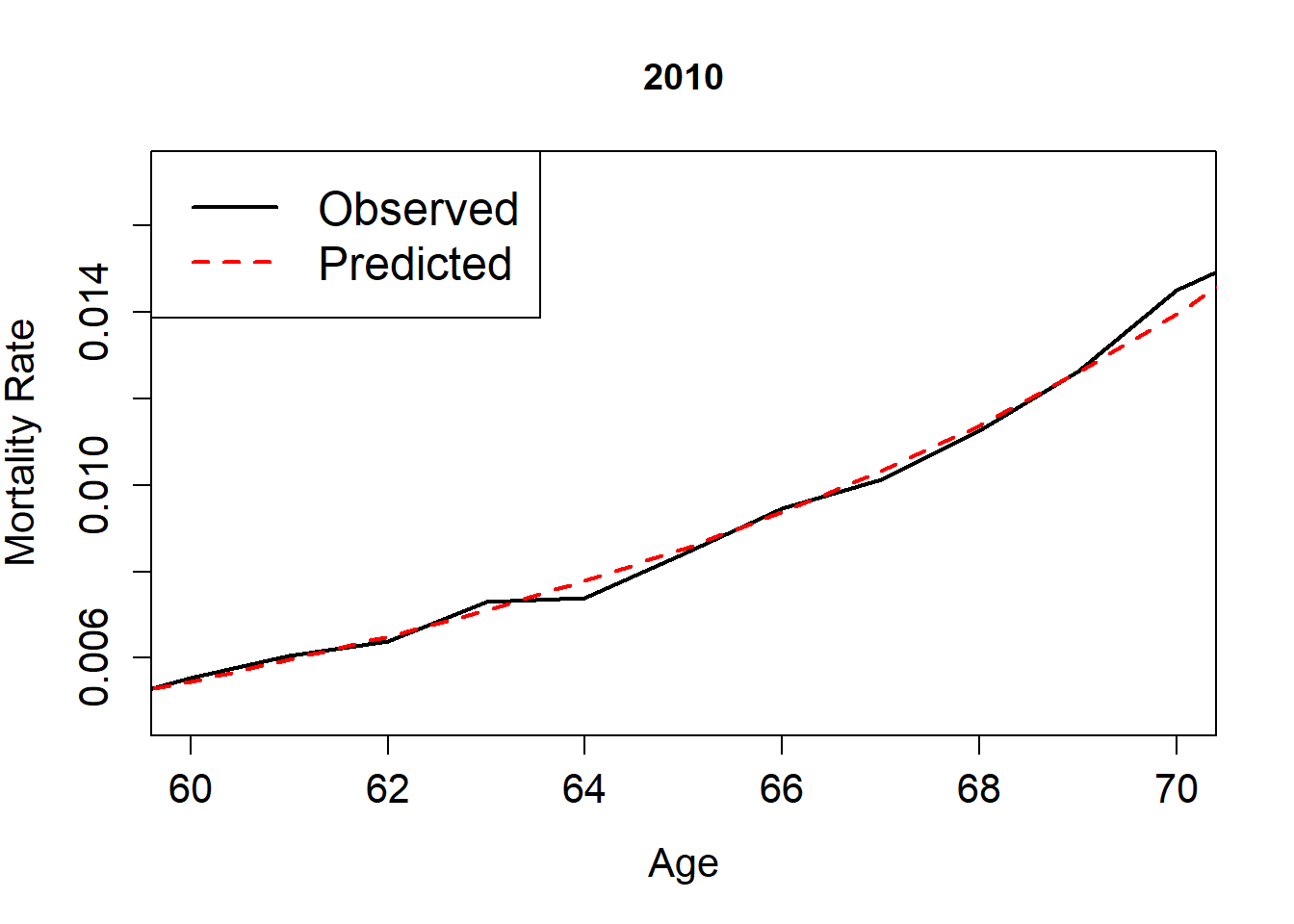}   \includegraphics[width=0.23\textwidth,height=0.24\textheight,trim=0in 0.1in 0in 0.1in]{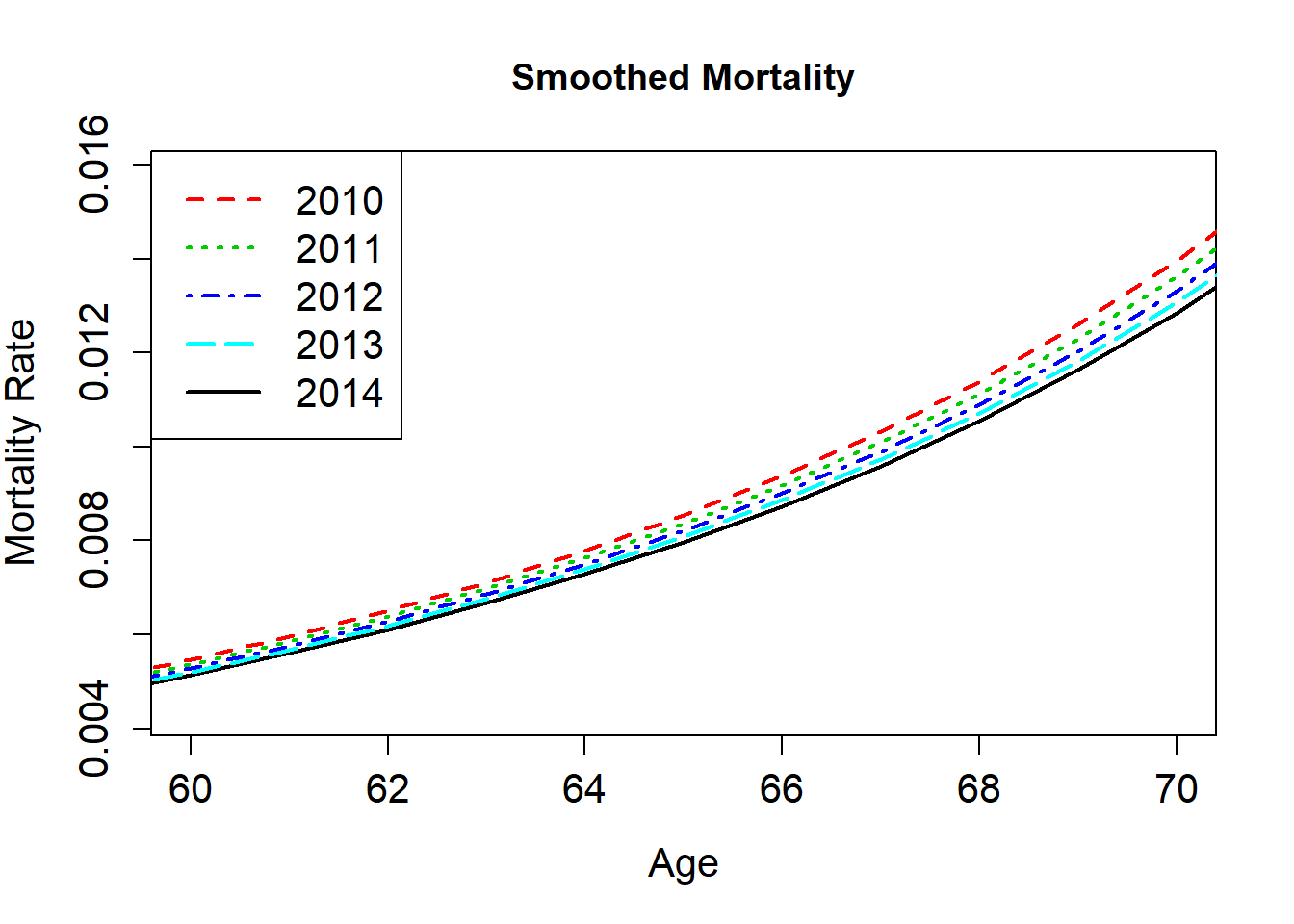}
  \end{tabular}

  \caption{\small Mortality rates for individuals aged 60--70 during the year 2010. Raw (solid) vs.~smoothed (dashed) mortality curves. Models are fit to 1999--2014 HMD data for Ages 50--84 (All data). Mean function $m(x)$ is intercept-only, $m(x)=\beta_0$.}\label{fig:app-GP-smooth-multiyear}
\end{figure}

 \begin{figure}[ht]
  \centering
\begin{tabular}{c} \hline
  Japan Males \\
   \includegraphics[width=0.30\textwidth,height=0.20\textheight]{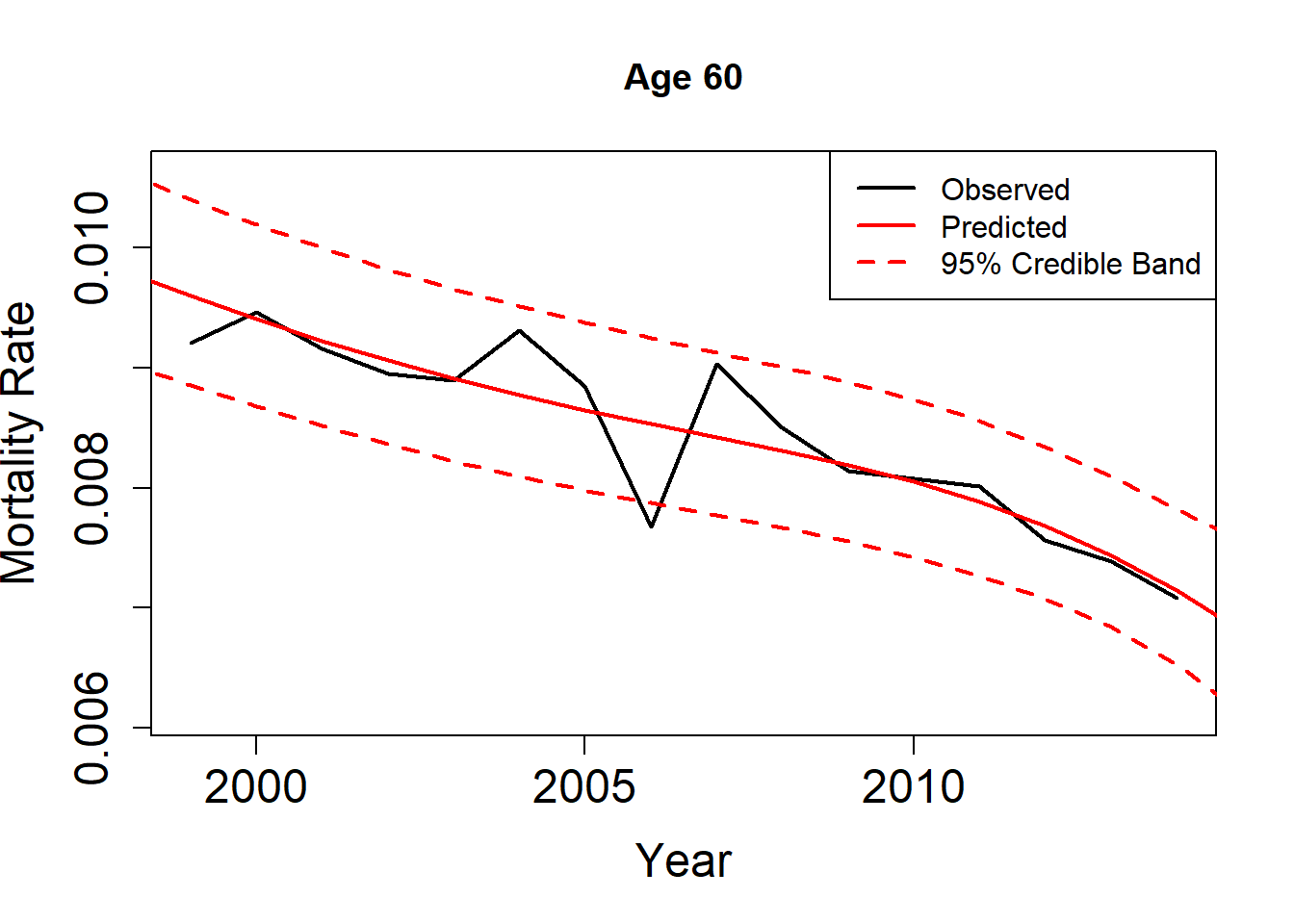}  \includegraphics[width=0.30\textwidth,height=0.20\textheight]{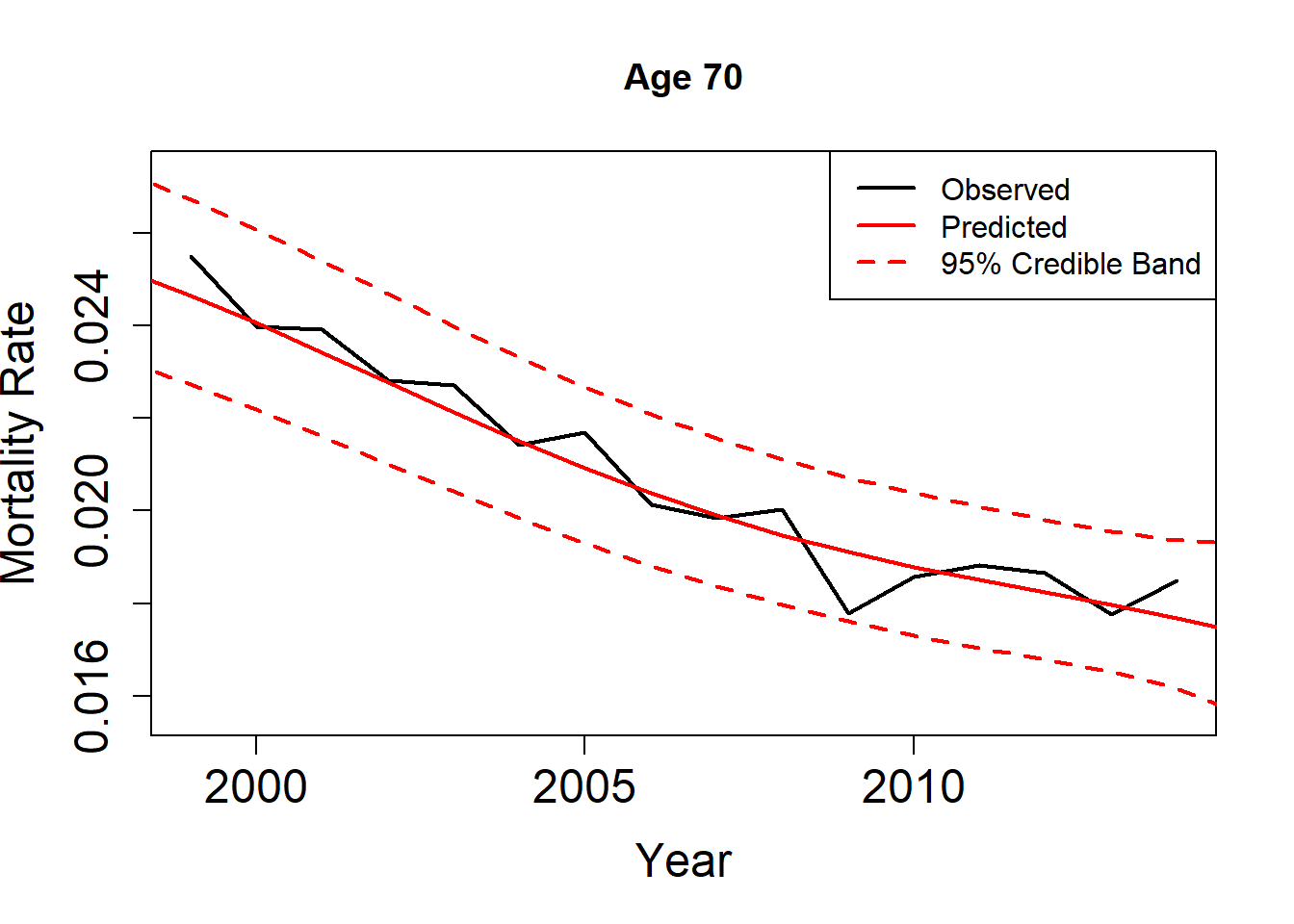}
  \includegraphics[width=0.30\textwidth,height=0.20\textheight]{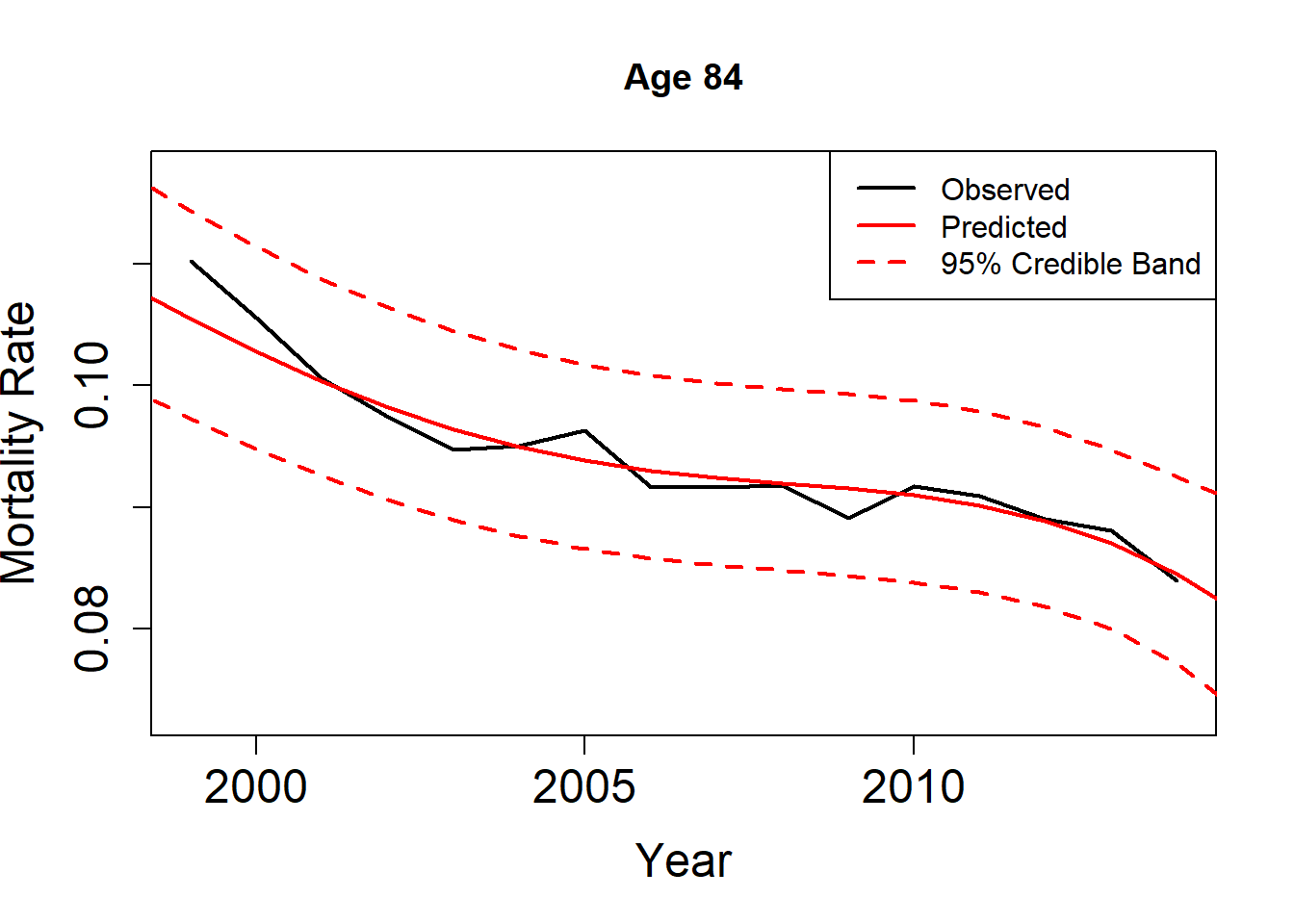} \\ \hline
  Japan Females \\
   \includegraphics[width=0.30\textwidth,height=0.20\textheight]{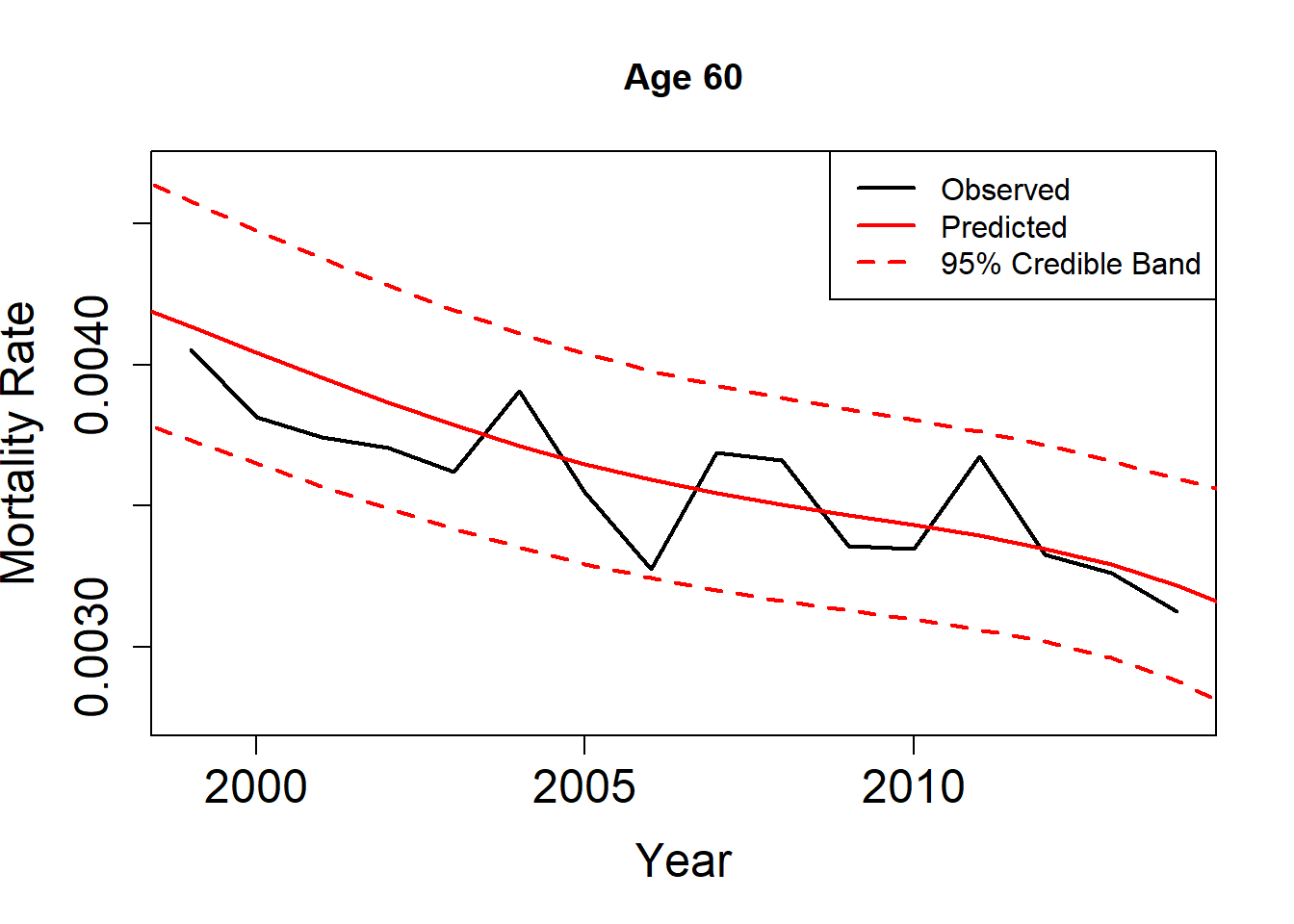}  \includegraphics[width=0.30\textwidth,height=0.20\textheight]{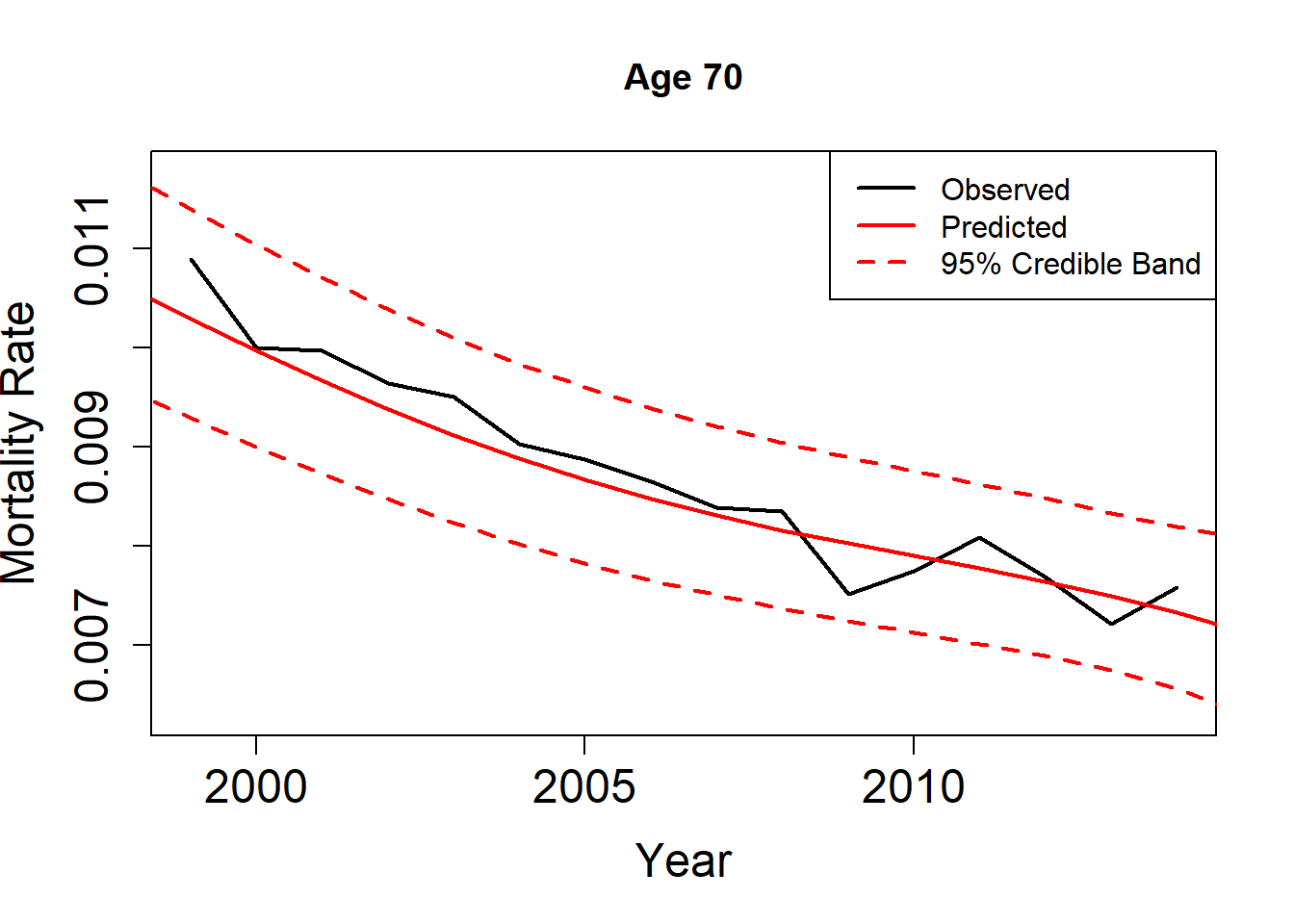}
  \includegraphics[width=0.30\textwidth,height=0.20\textheight]{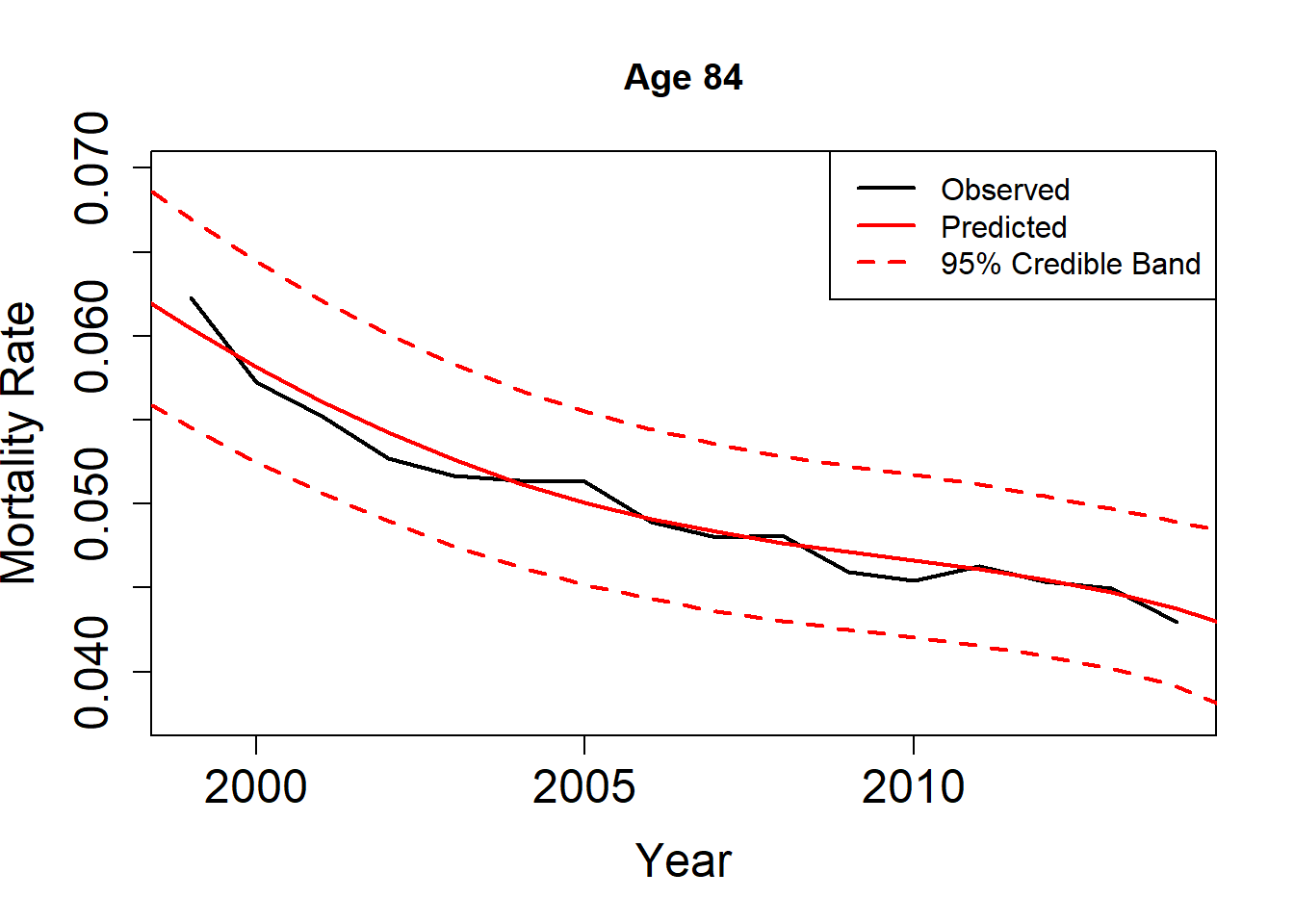} \\ \hline
  UK Males \\
   \includegraphics[width=0.30\textwidth,height=0.20\textheight]{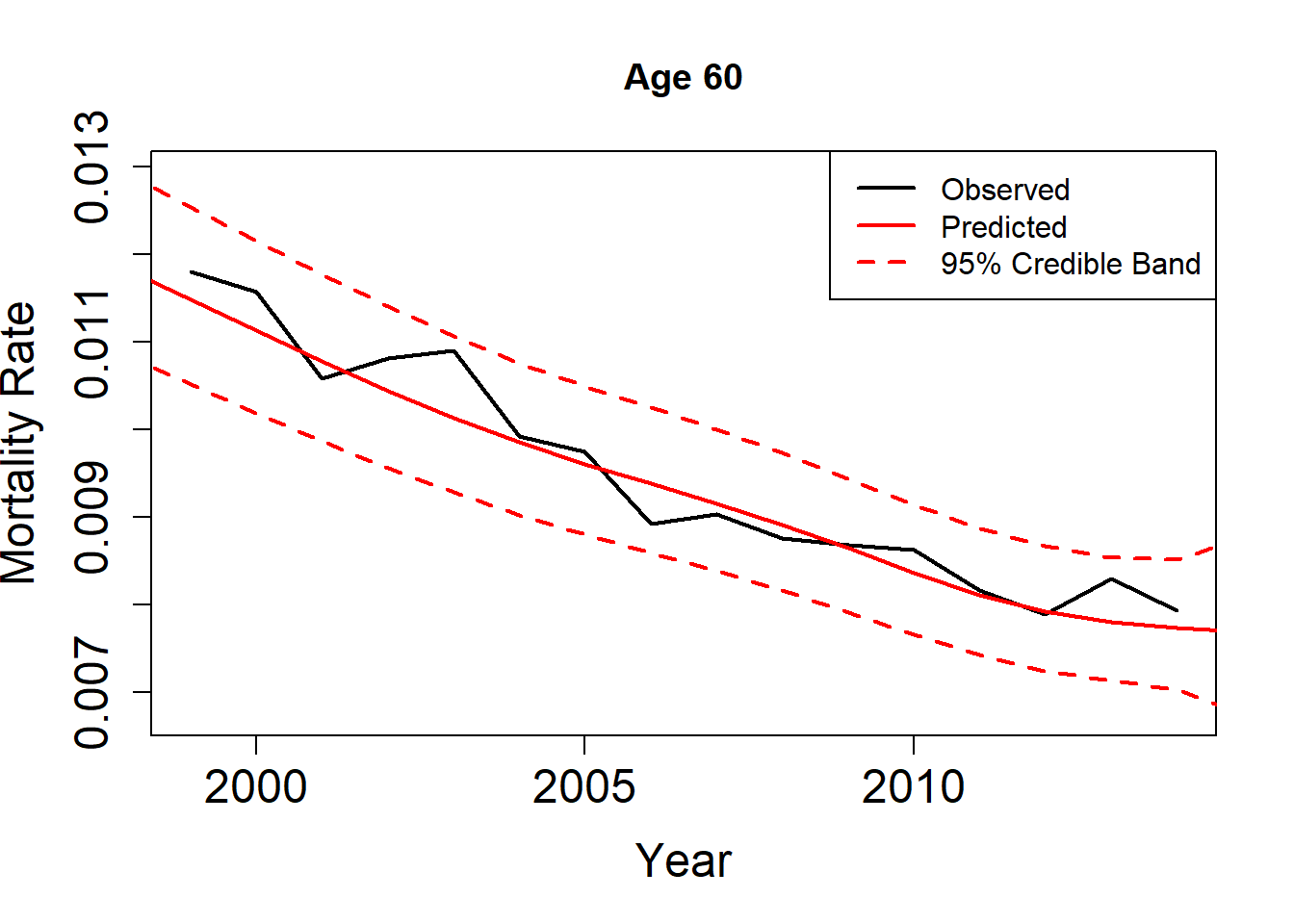}  \includegraphics[width=0.30\textwidth,height=0.20\textheight]{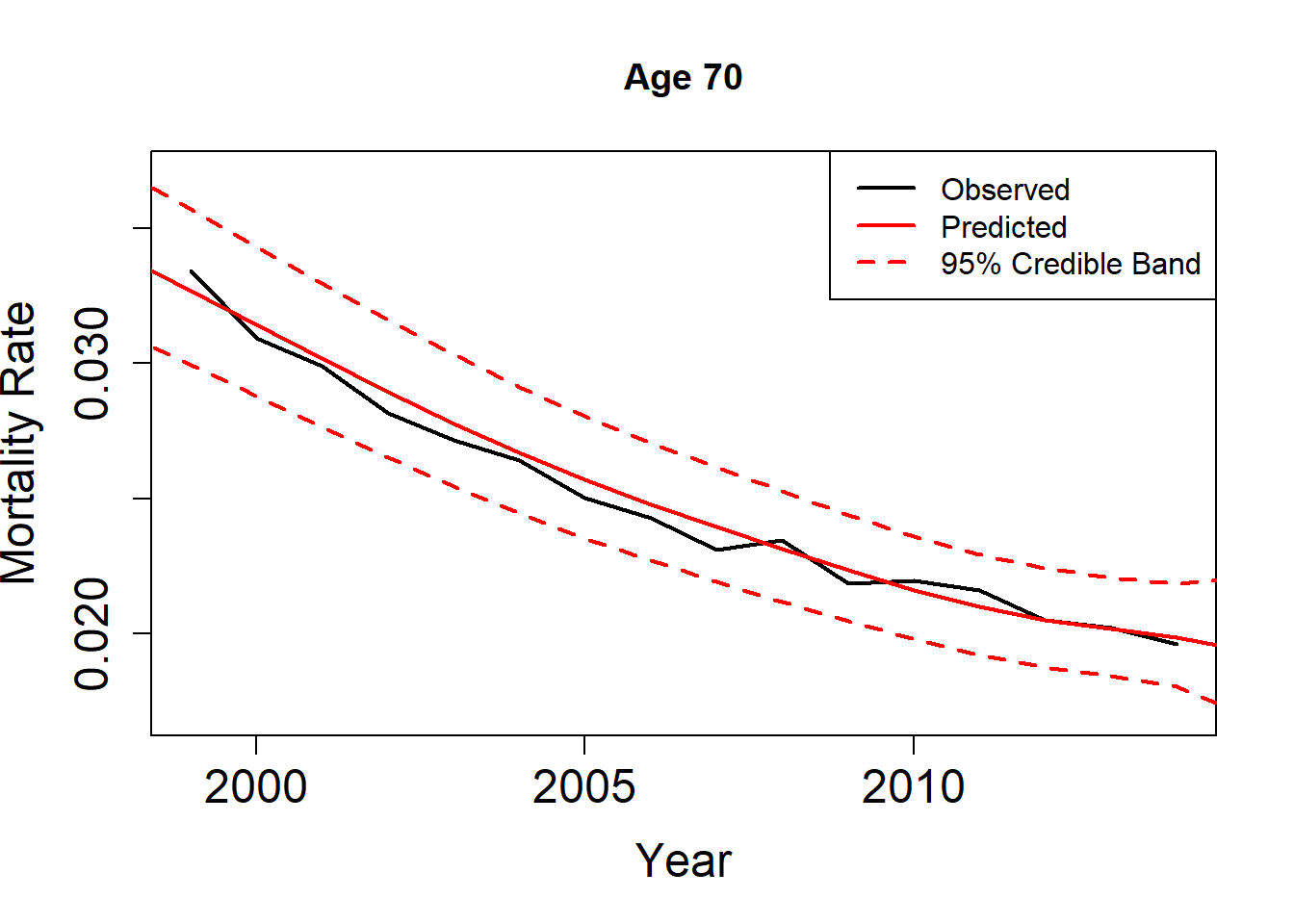}
  \includegraphics[width=0.30\textwidth,height=0.20\textheight]{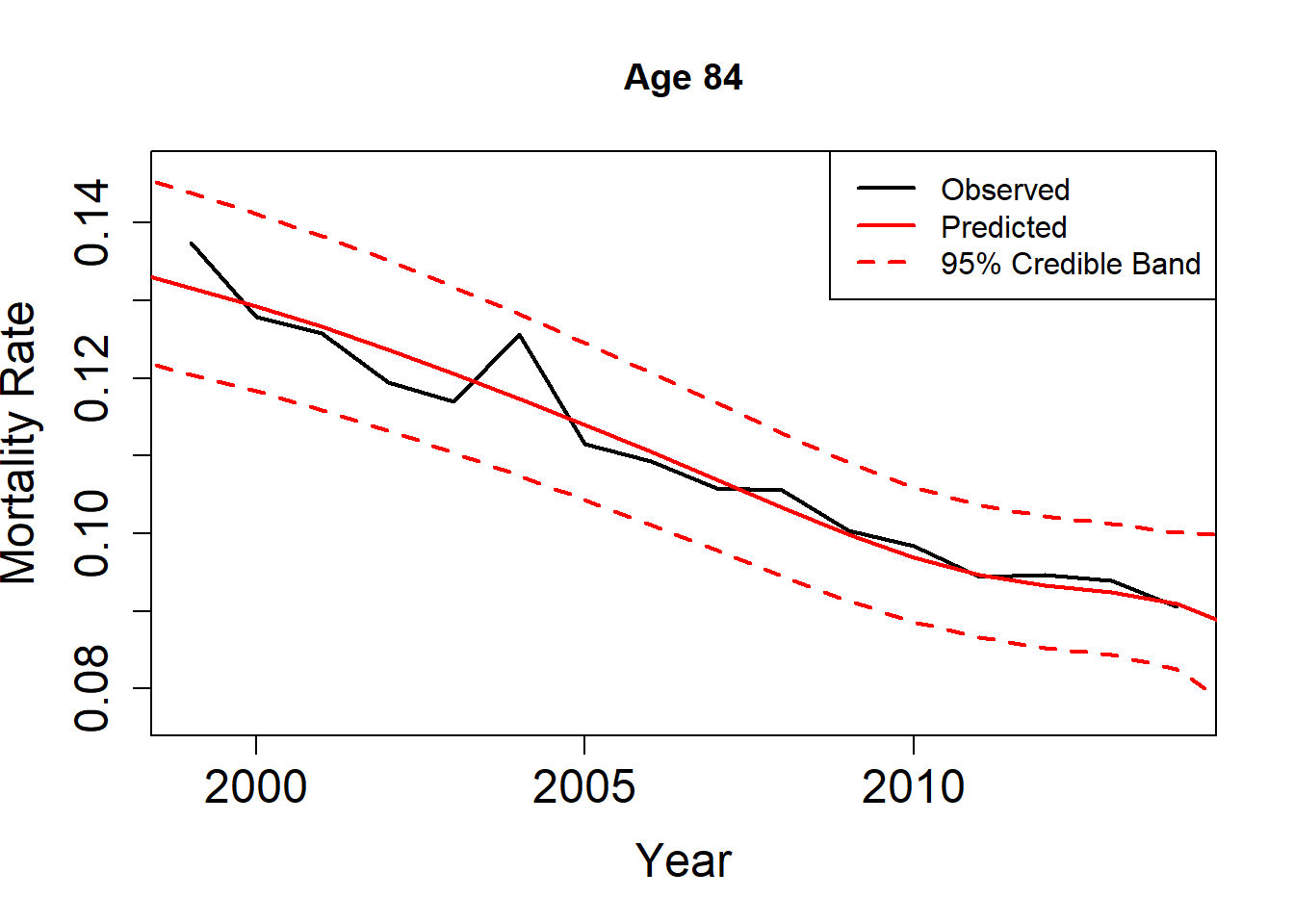} \\ \hline
  UK Females \\
   \includegraphics[width=0.30\textwidth,height=0.20\textheight]{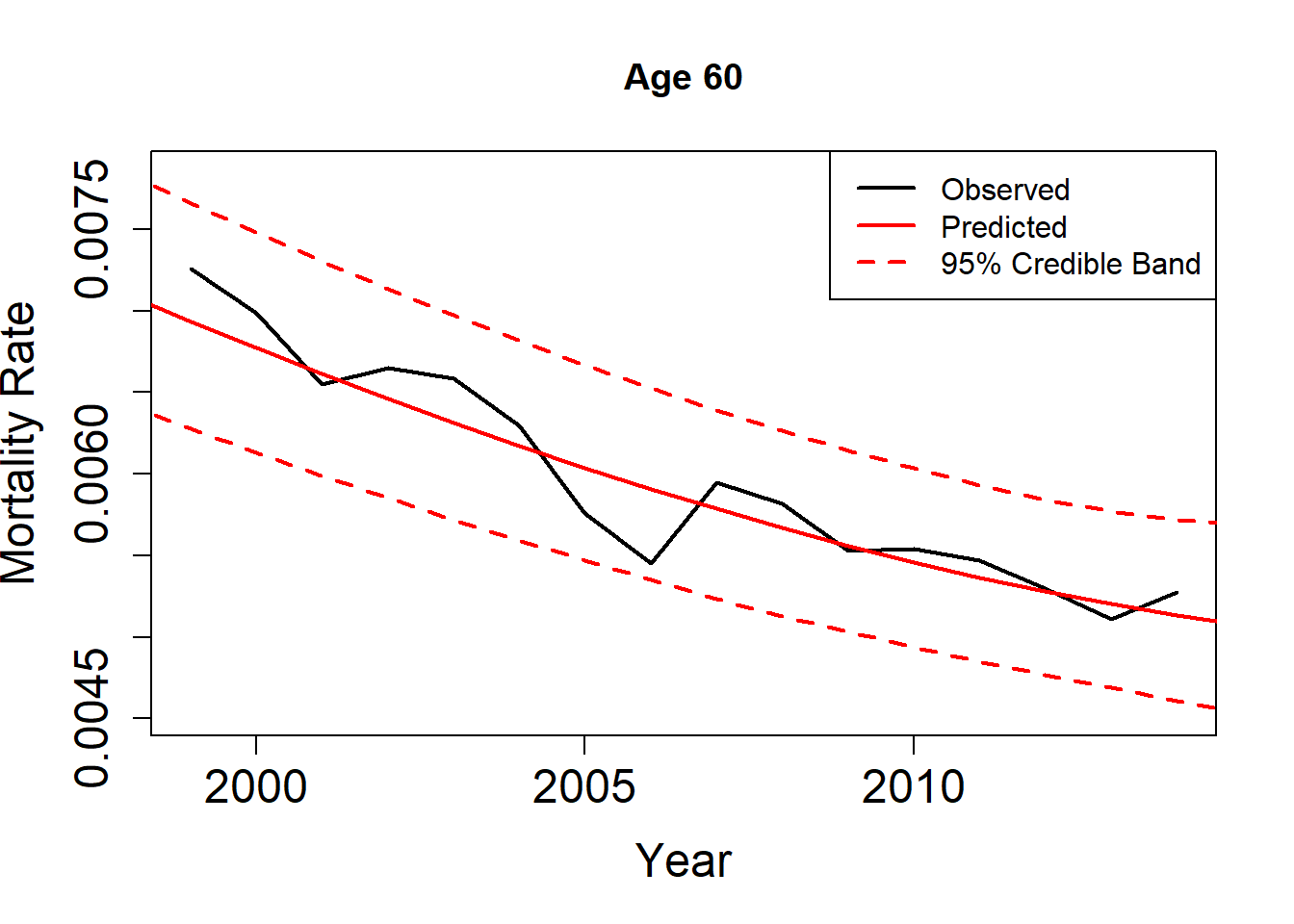}  \includegraphics[width=0.30\textwidth,height=0.20\textheight]{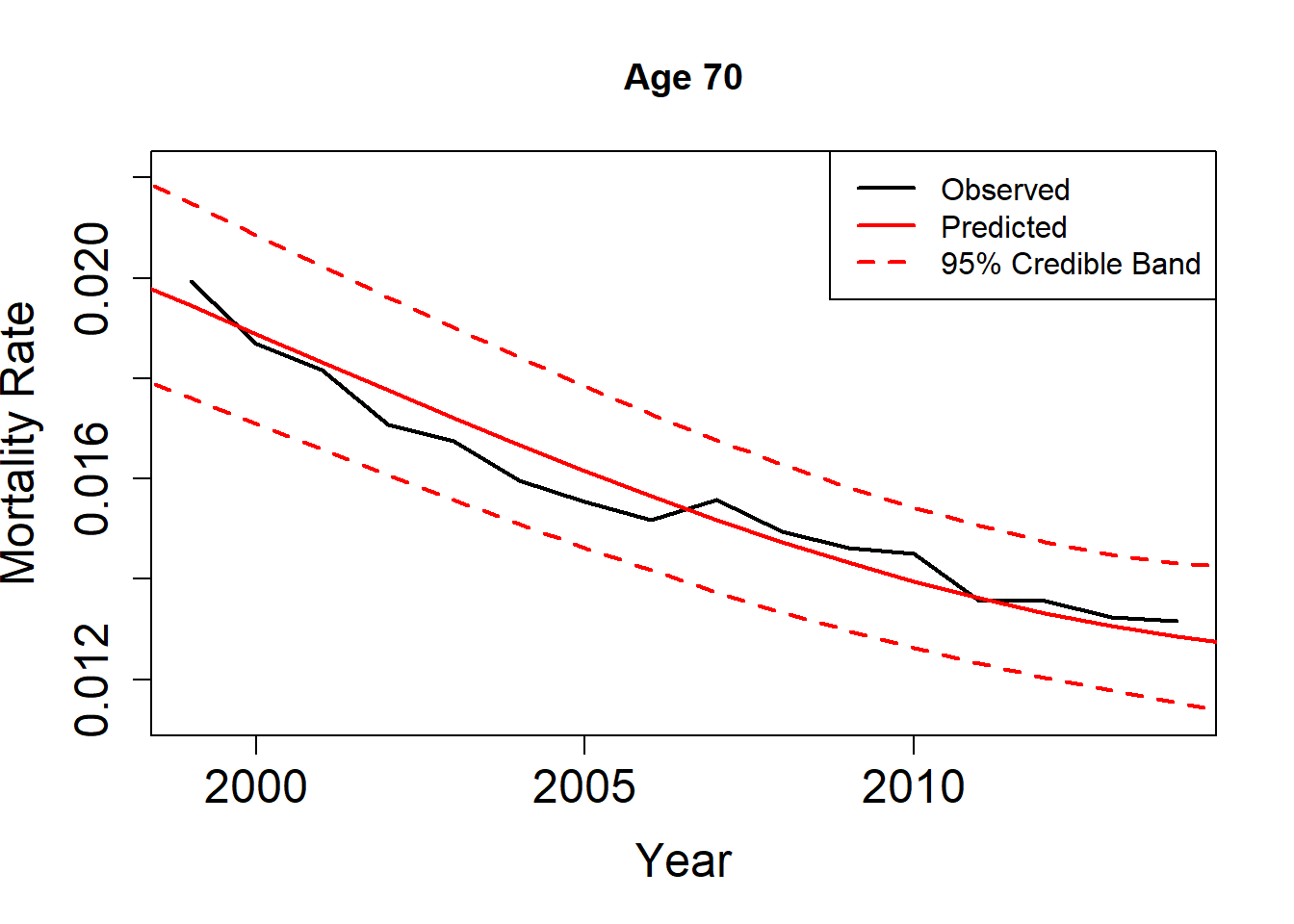}
  \includegraphics[width=0.30\textwidth,height=0.20\textheight]{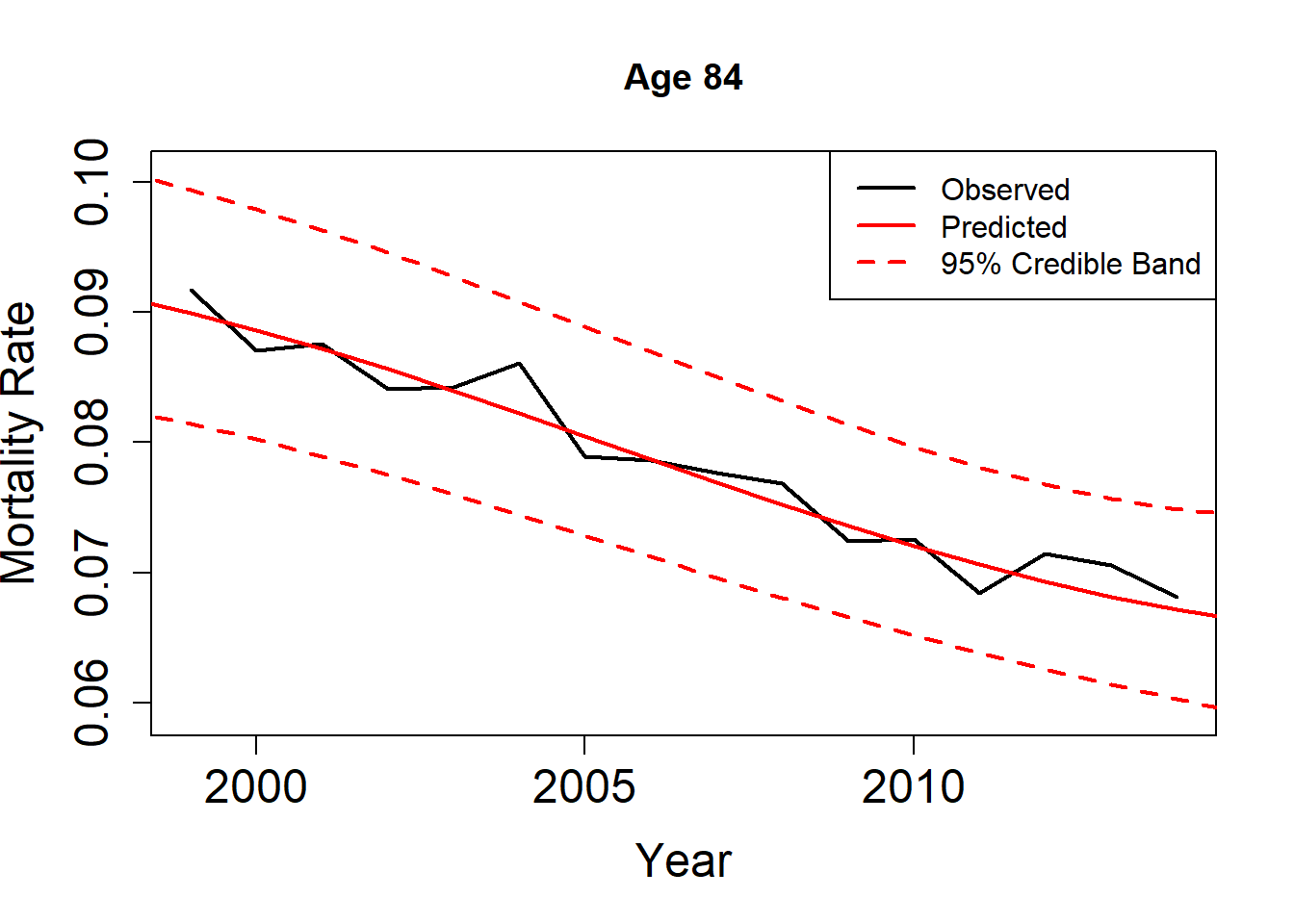} 
  \end{tabular}

  \caption{Mortality rates for individuals aged 60, 70 and 84 over time. The plots show raw mortality rates (solid black) for years 1999--2014, as well as predicted mean of the smoothed mortality surface (solid red) and its 95\% credible band, for 1999--2016. Mean function  is intercept-only, $m(x)=\beta_0$.\label{fig:app-MortalityVSTime} }
\end{figure}

\begin{figure}[ht]
  \centering
\begin{tabular}{c} \hline
  Japan Males \\
   \includegraphics[width=0.470\textwidth,height=0.20\textheight]{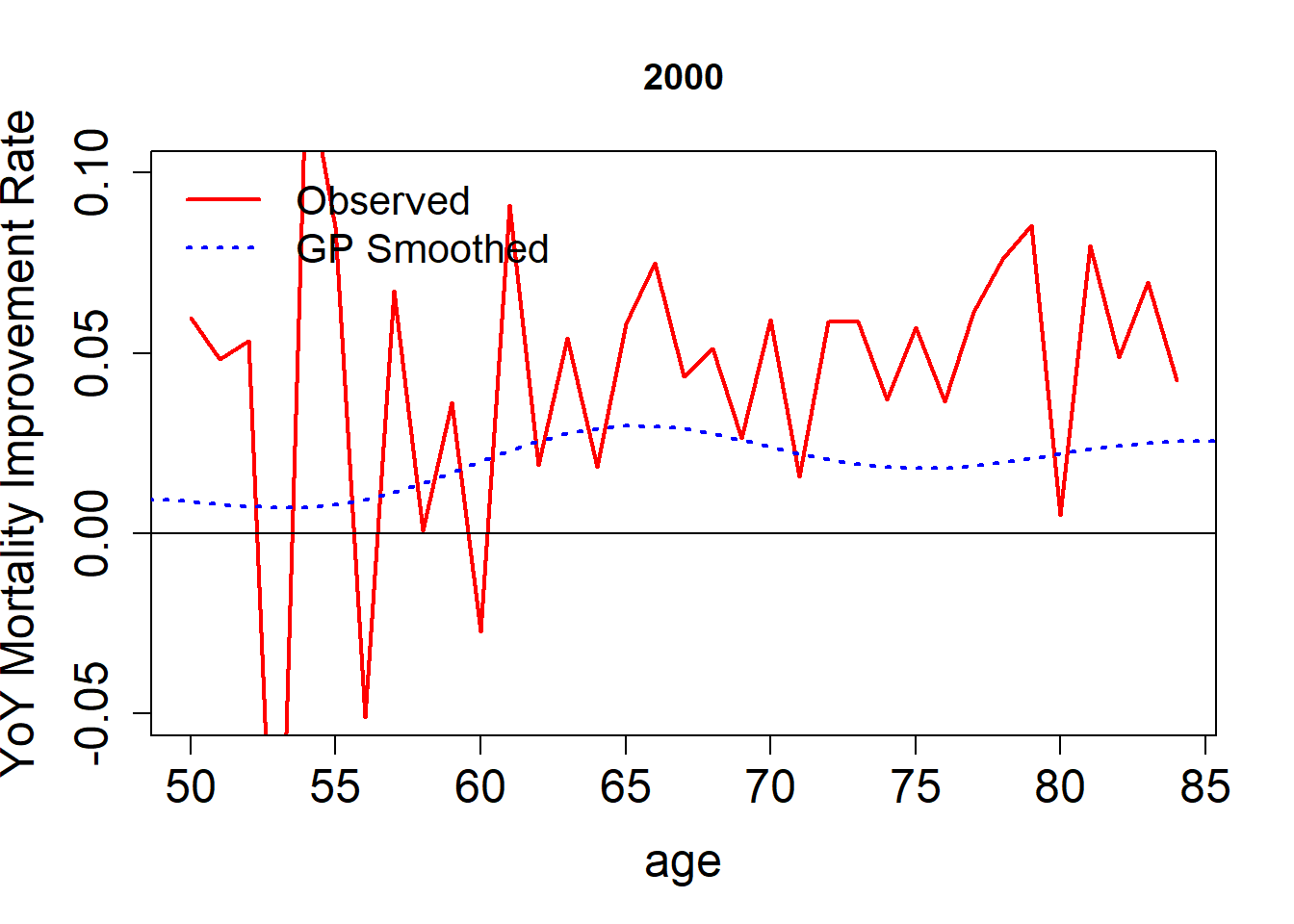}  \includegraphics[width=0.47\textwidth,height=0.20\textheight]{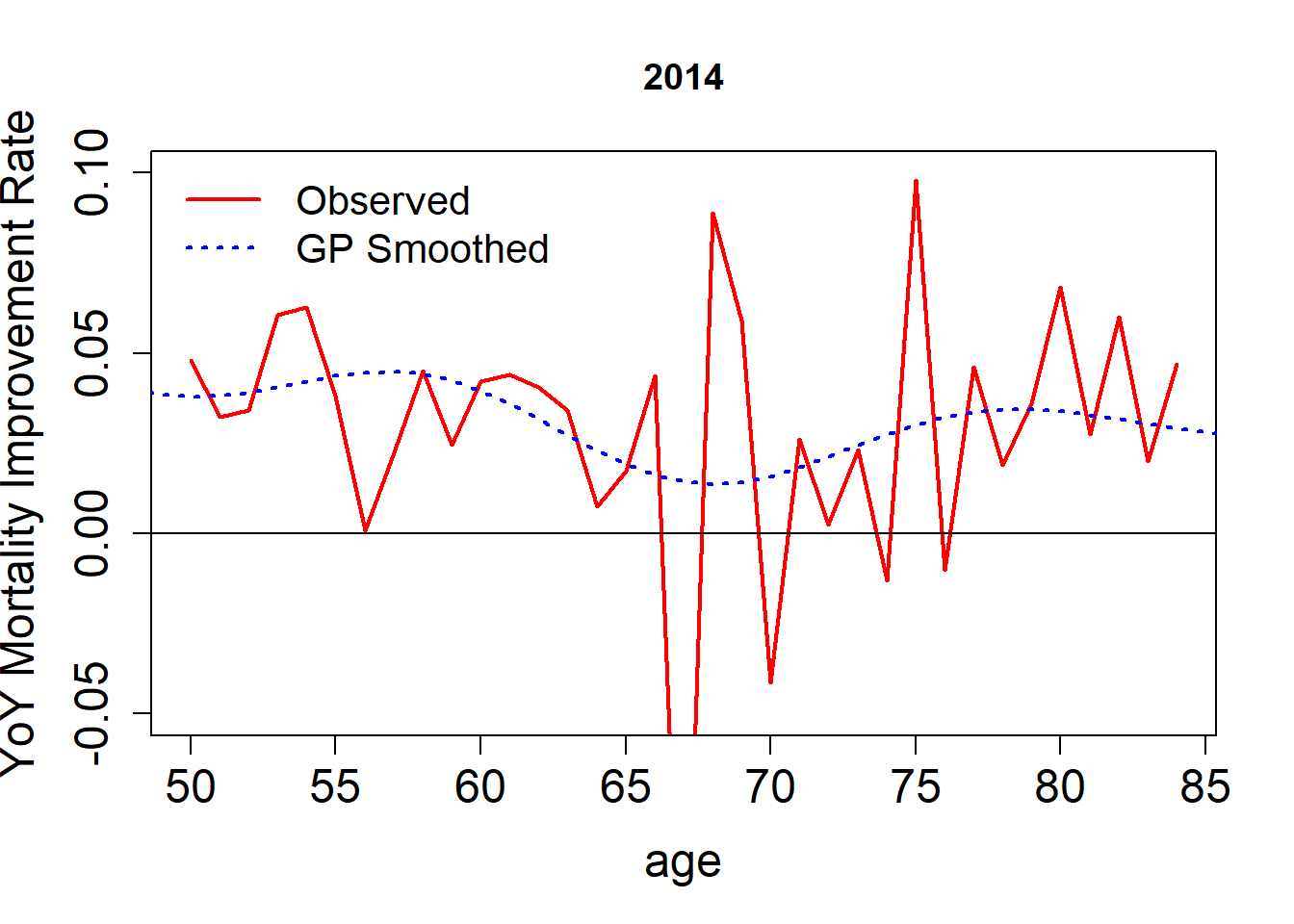}  \\ \hline
  Japan Females \\
   \includegraphics[width=0.47\textwidth,height=0.20\textheight]{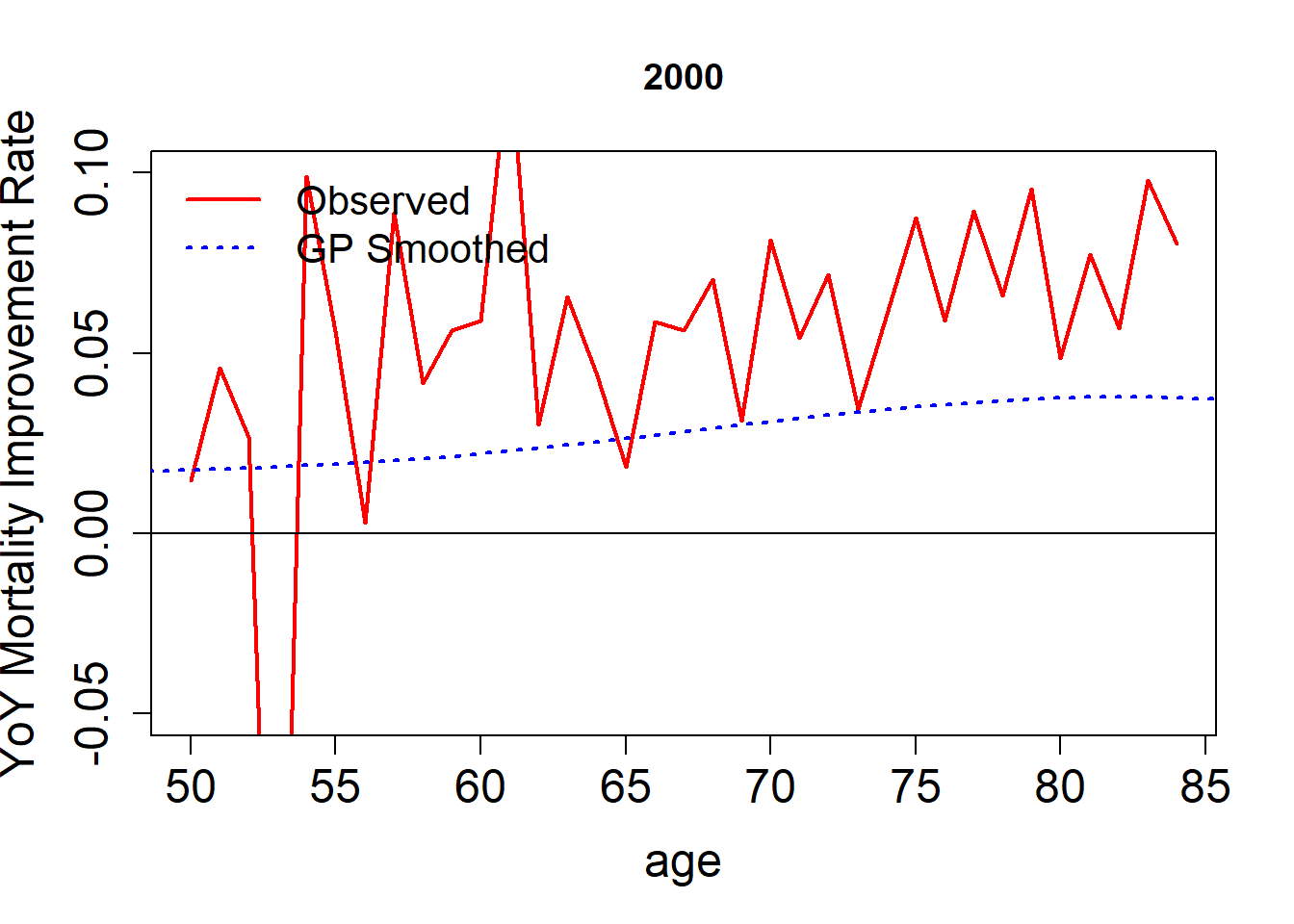}  \includegraphics[width=0.47\textwidth,height=0.20\textheight]{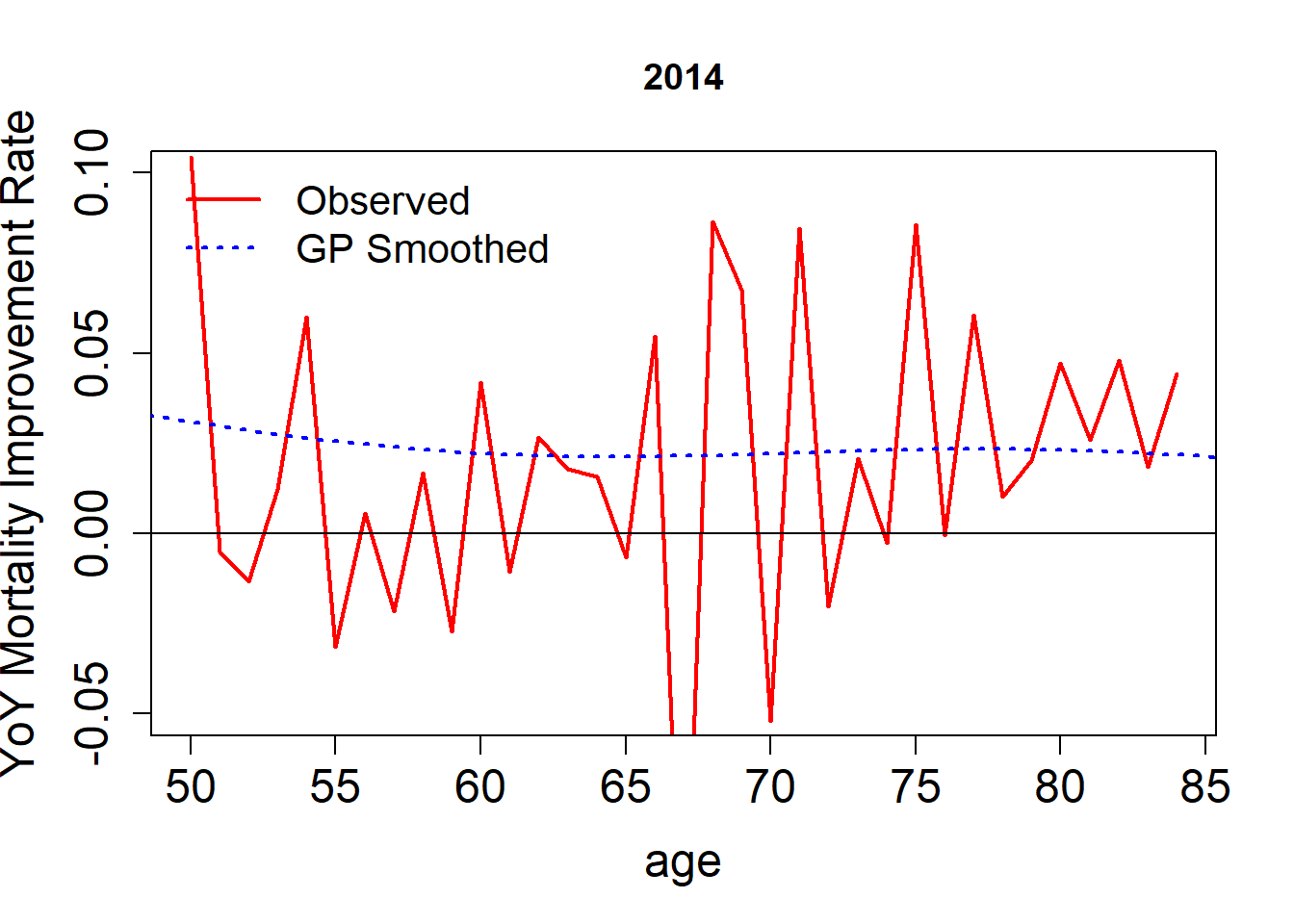}  \\ \hline
  UK Males \\
   \includegraphics[width=0.47\textwidth,height=0.20\textheight]{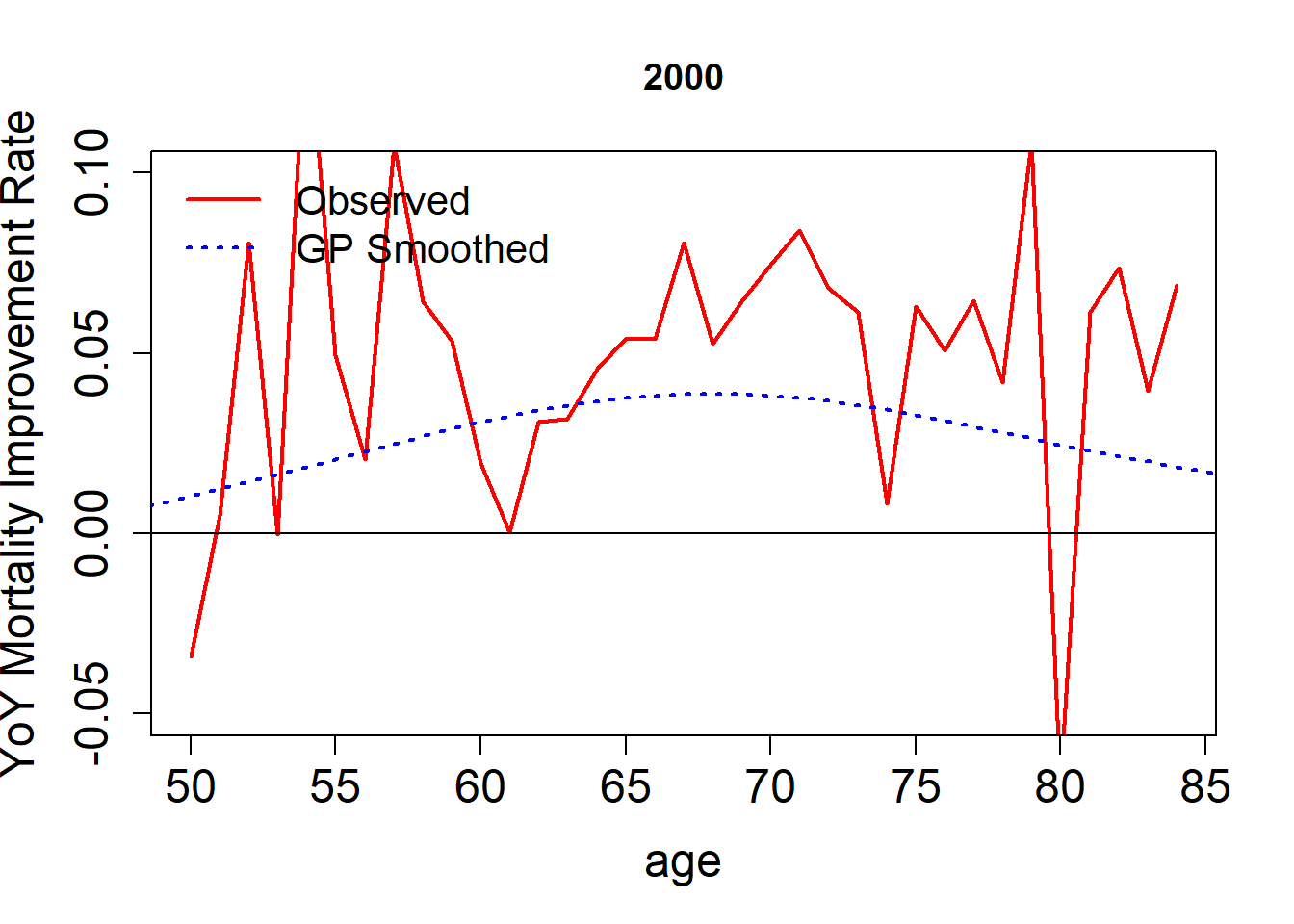}  \includegraphics[width=0.47\textwidth,height=0.20\textheight]{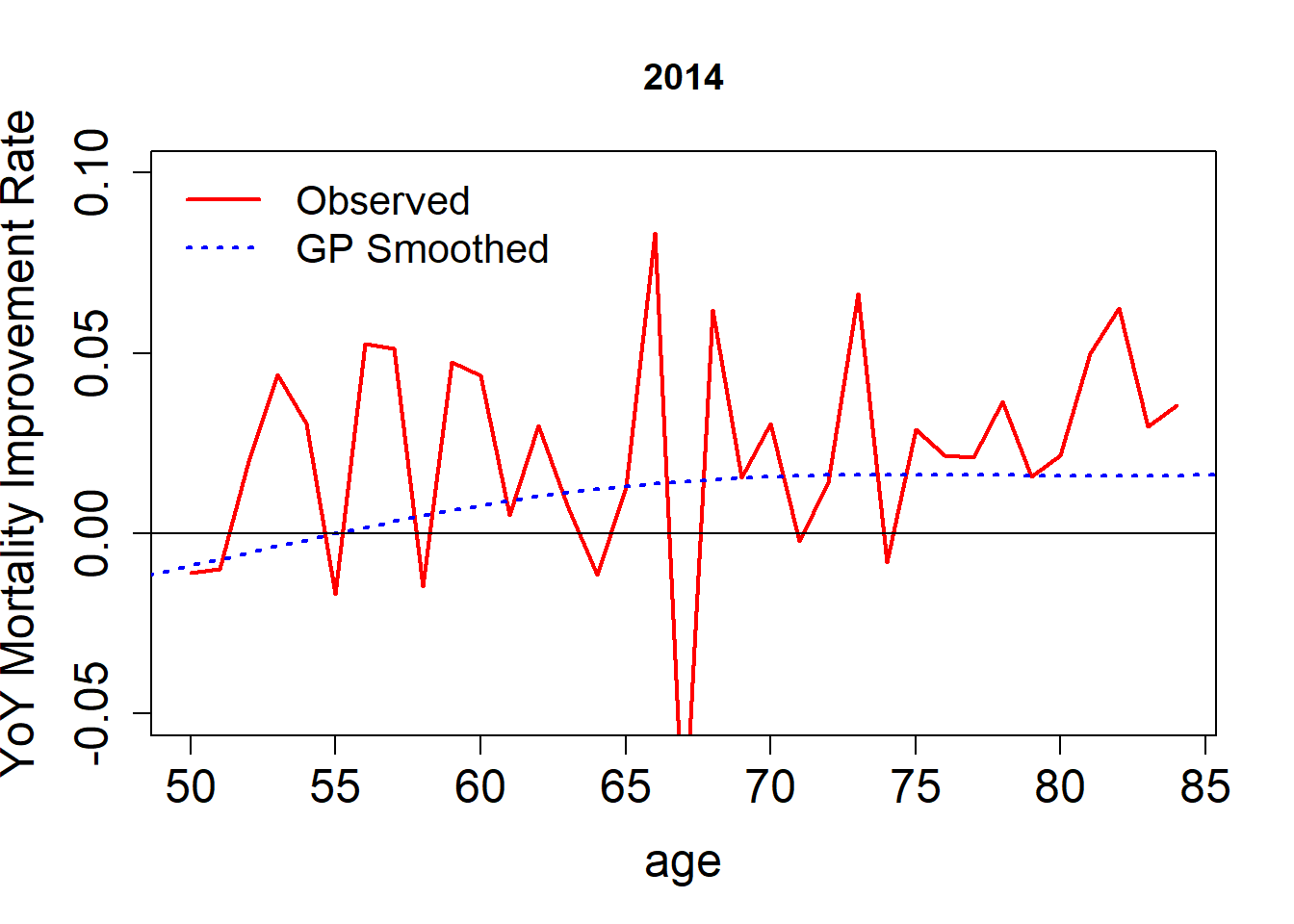}  \\ \hline
  UK Females \\
   \includegraphics[width=0.47\textwidth,height=0.20\textheight]{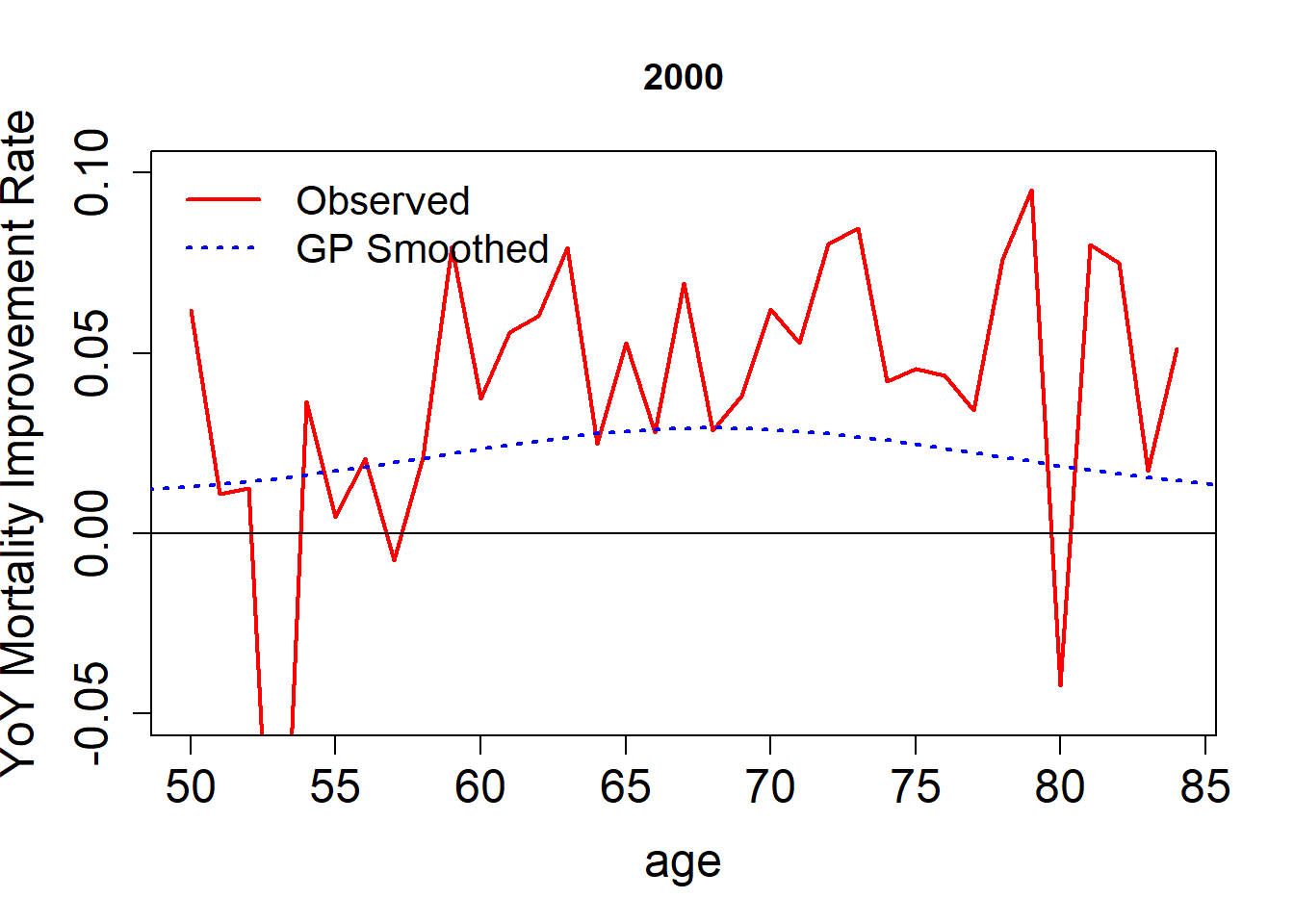}  \includegraphics[width=0.47\textwidth,height=0.20\textheight]{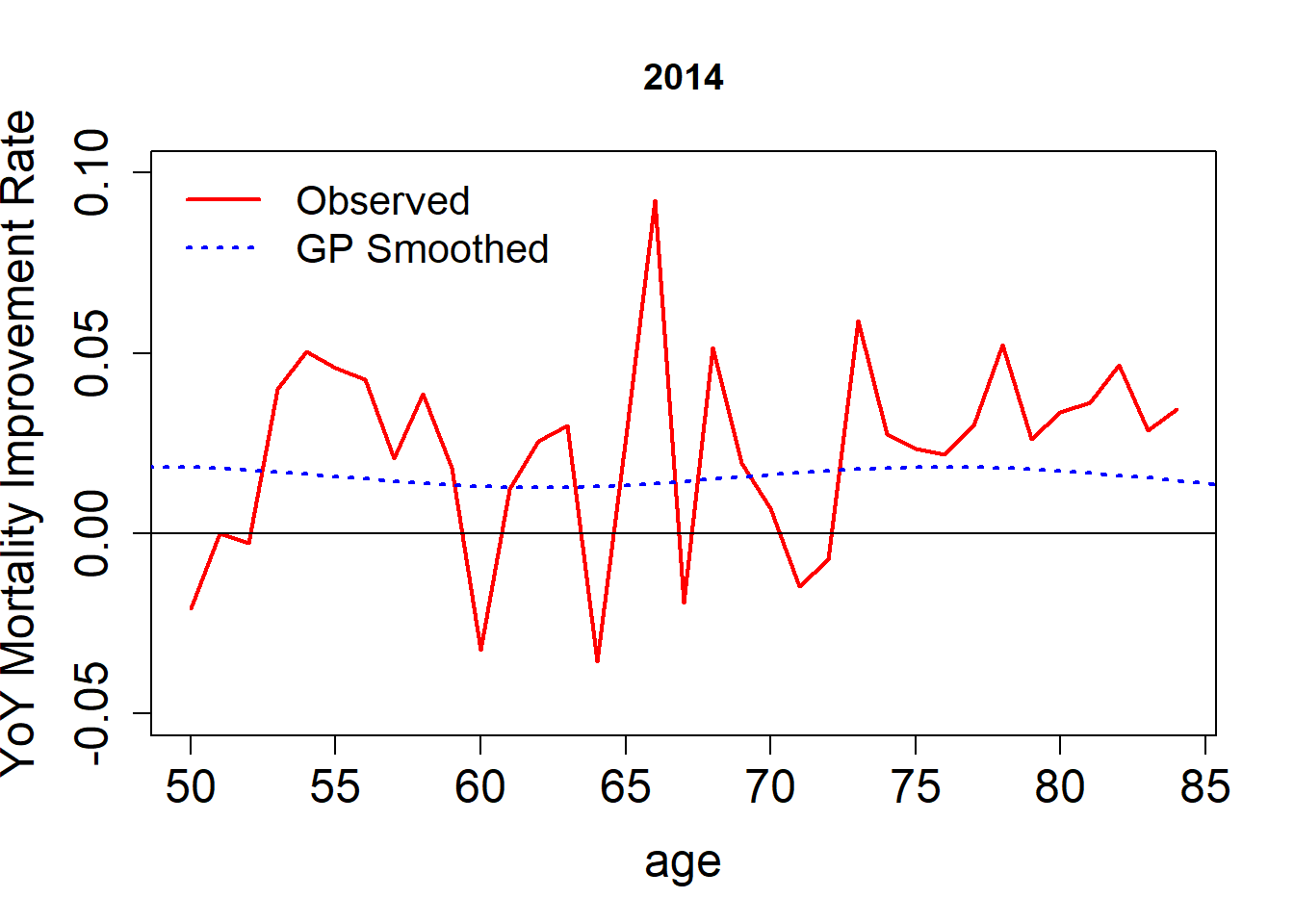} 
  \end{tabular}

  \caption{Mortality improvement factors for All Data. Solid lines indicate the empirical mortality experience $MI_{back}^{obs}(\cdot; yr)$ for years $yr \in \{2000, 2014\}$, the dotted lines are $m_{back}^{GP}(\cdot; yr)$ from \eqref{eq:MIback}. \label{fig:app-MP-multiyear} }
\end{figure}

\begin{figure}[ht]
  \centering
  \begin{tabular}{cc} \hline
  Japan Males & Japan Females\\
  \includegraphics[width=0.47\textwidth]{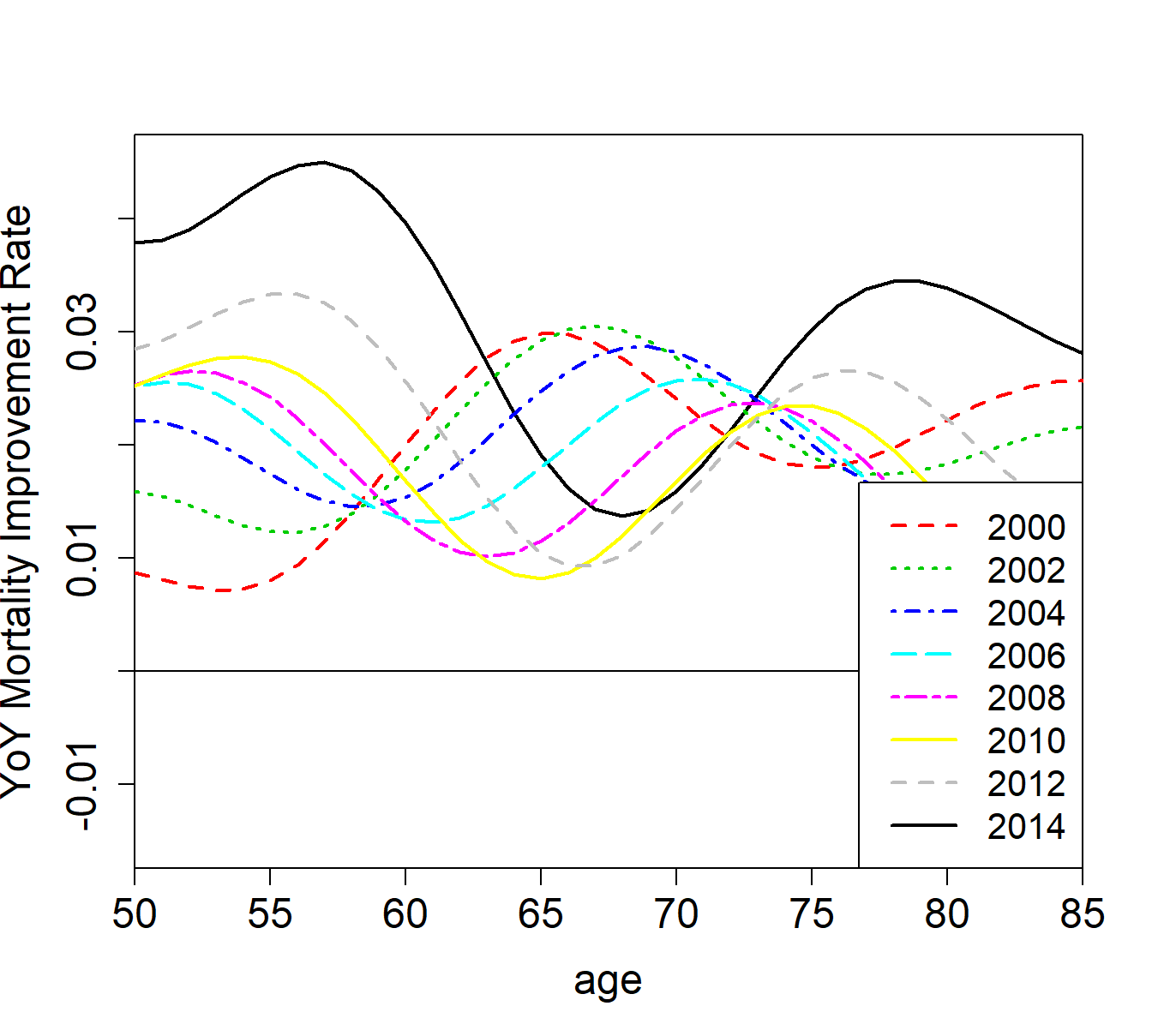}  &
  \includegraphics[width=0.47\textwidth]{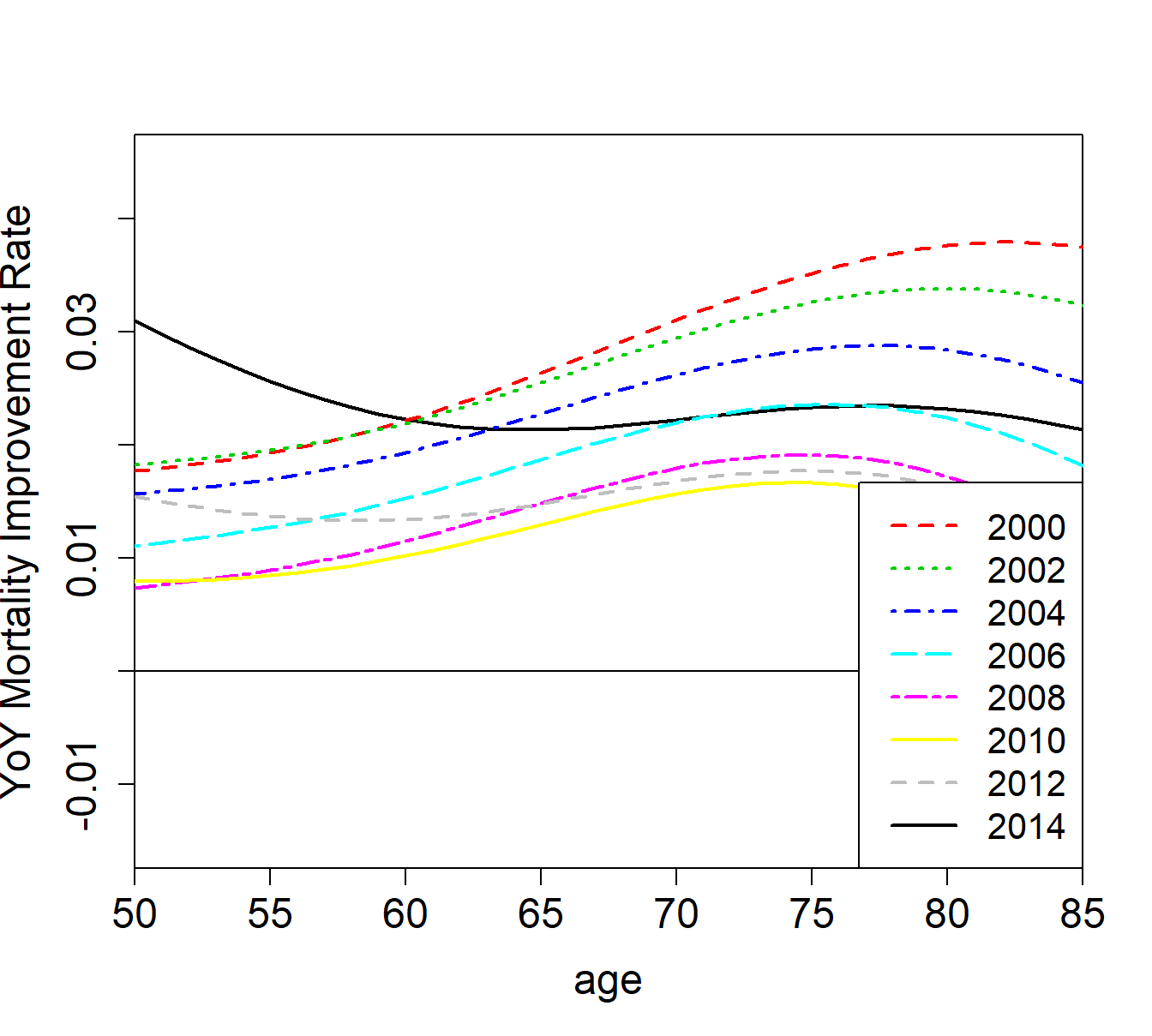}   \\ \hline
  UK Males & UK Females\\
  \includegraphics[width=0.47\textwidth]{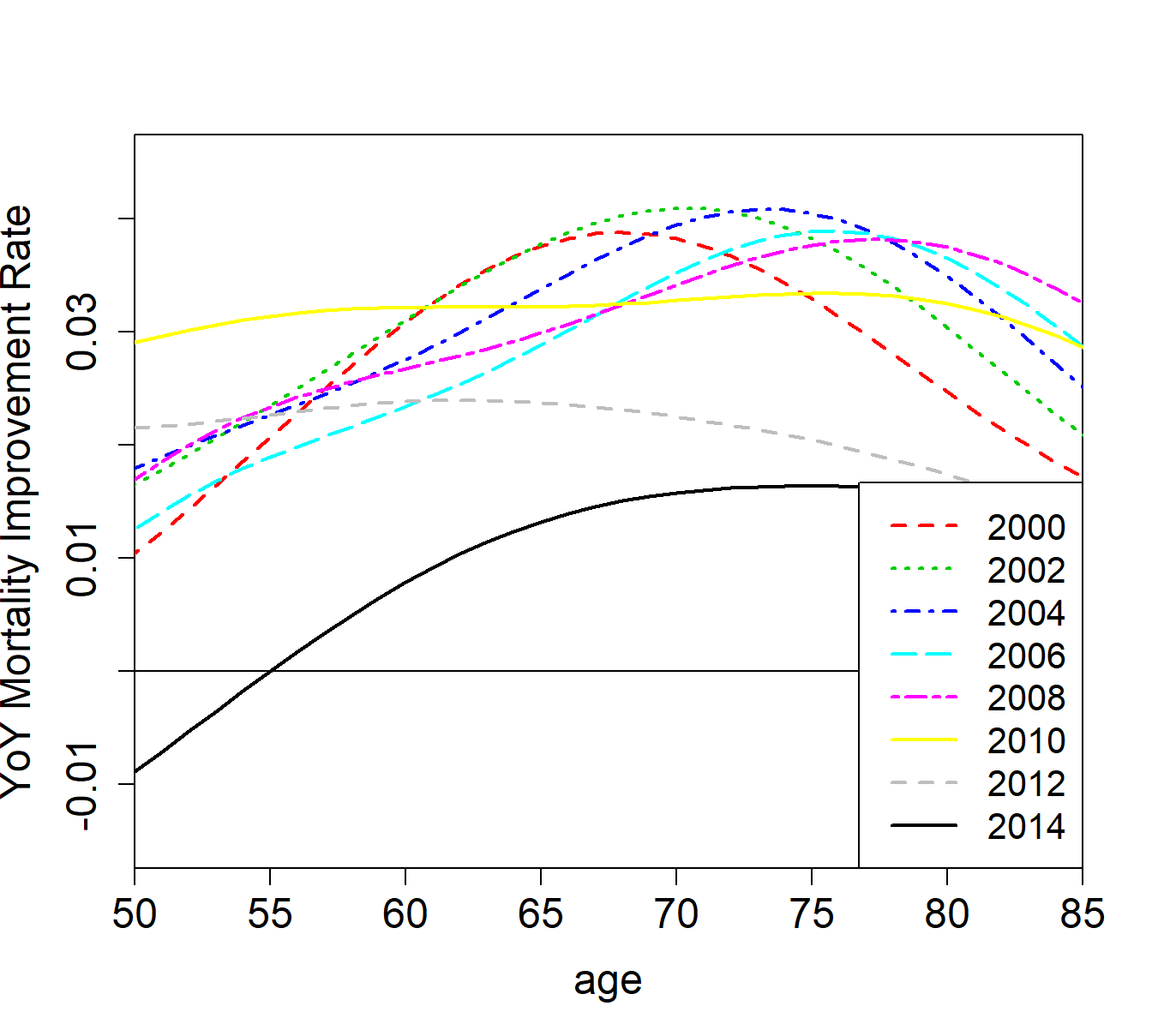}  &
  \includegraphics[width=0.47\textwidth]{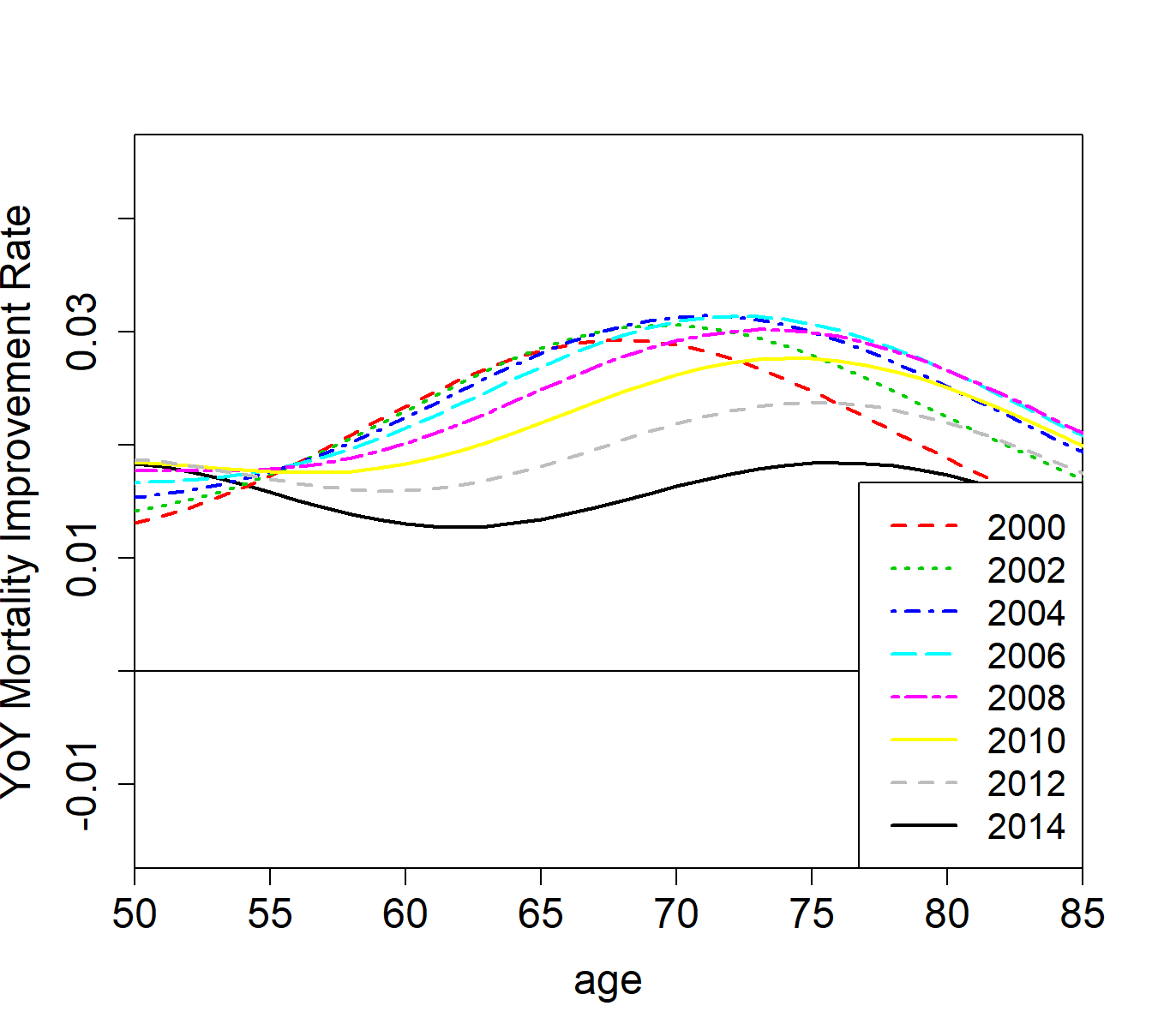} \\ \hline
  \end{tabular}

  \caption{Comparison of smoothed yearly mortality improvement factors $m_{back}^{GP}(x_{ag}; yr)$ from \eqref{eq:MIback} for Males using All data and $yr=2000, \ldots, 2014$. The curves for 2000 and 2014 are the same as in Figure \ref{fig:app-MP-multiyear}.\label{fig:app-mortTrend} }
\end{figure}

 \begin{figure}[ht]
  \centering
  \begin{tabular}{c} \hline
  Japan Males \\
  \includegraphics[width=0.47\textwidth,height=1.6in]{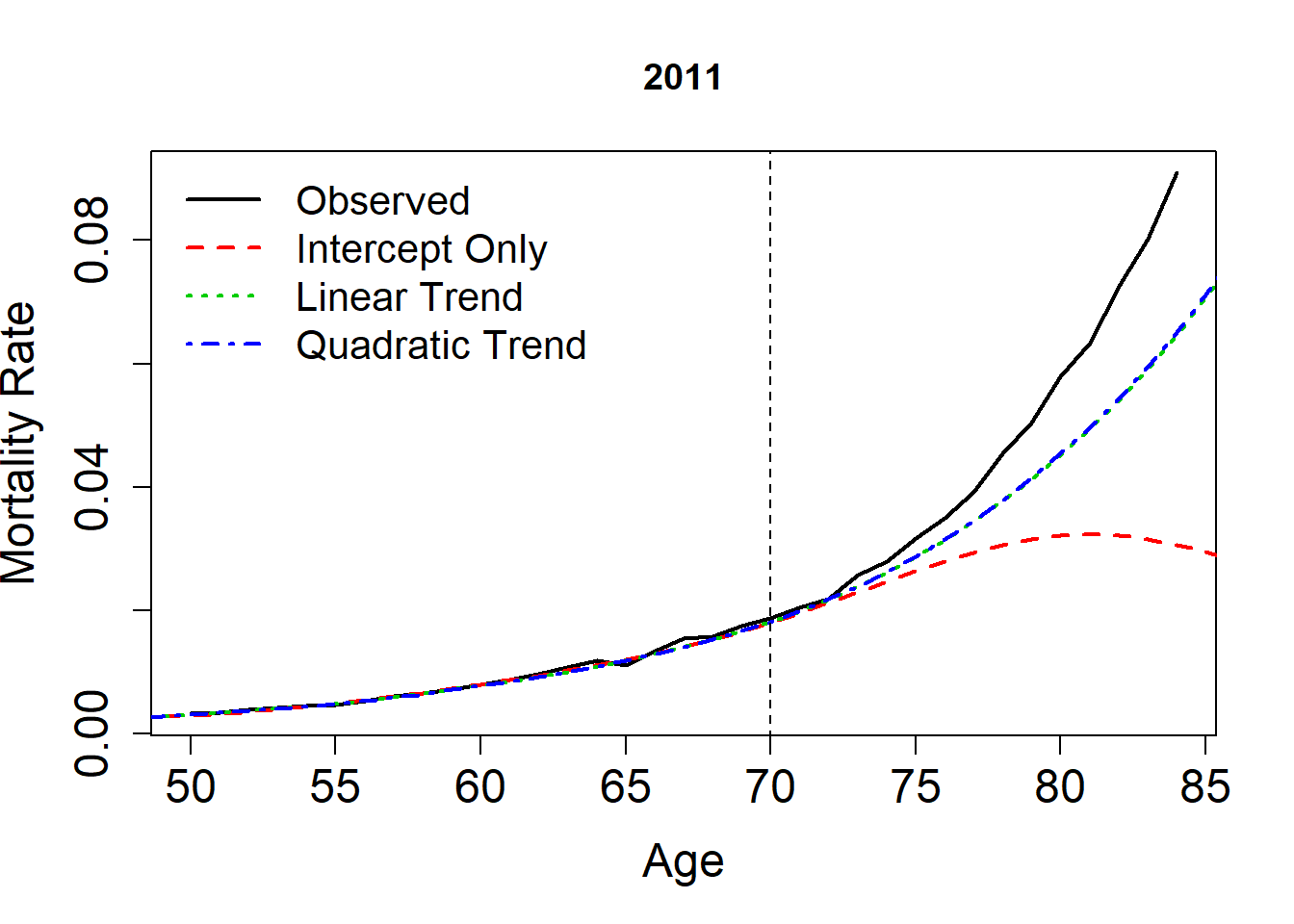}  \includegraphics[width=0.47\textwidth,height=1.6in]{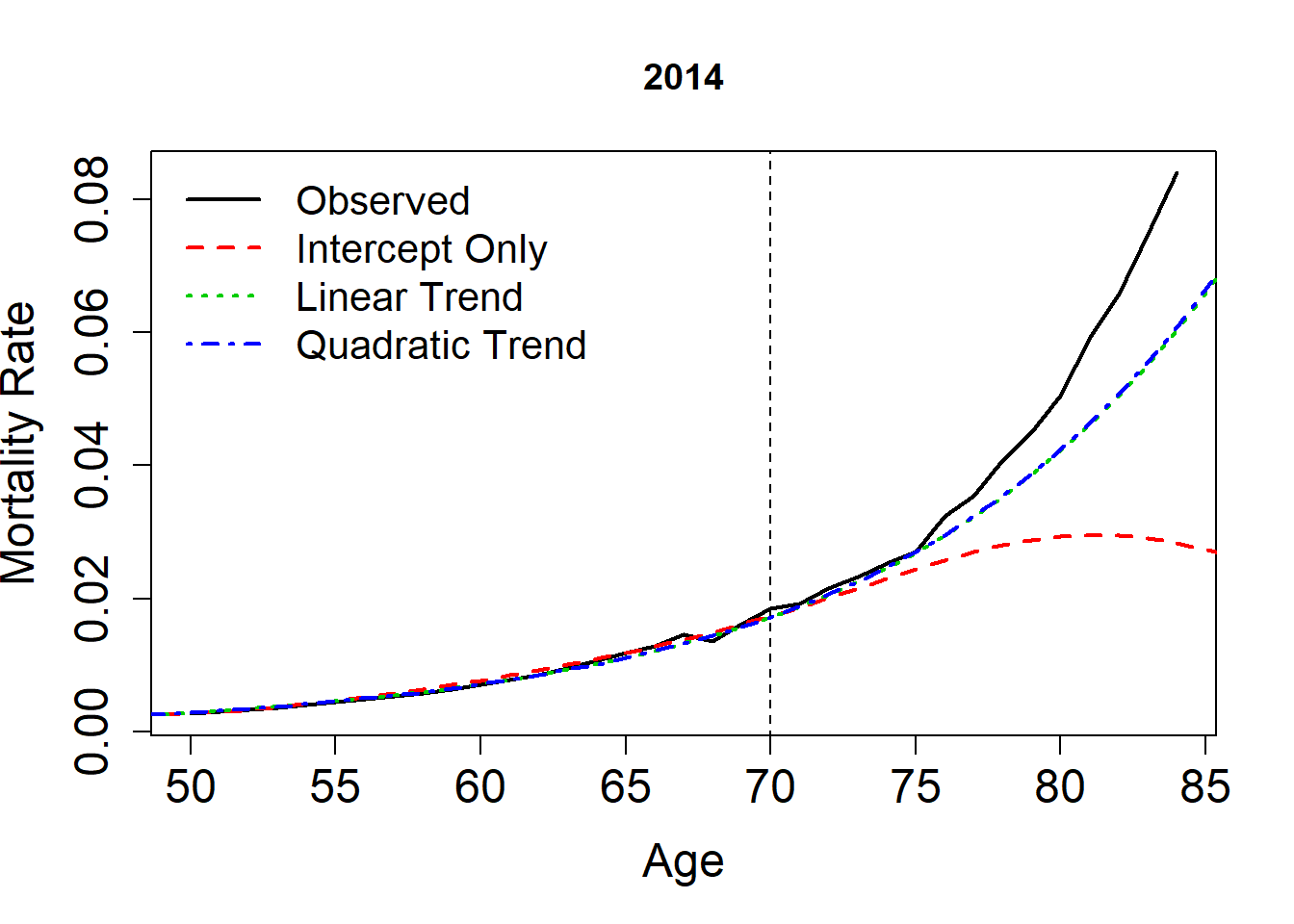} \\ \hline  Japan Females \\
  \includegraphics[width=0.47\textwidth,height=1.6in]{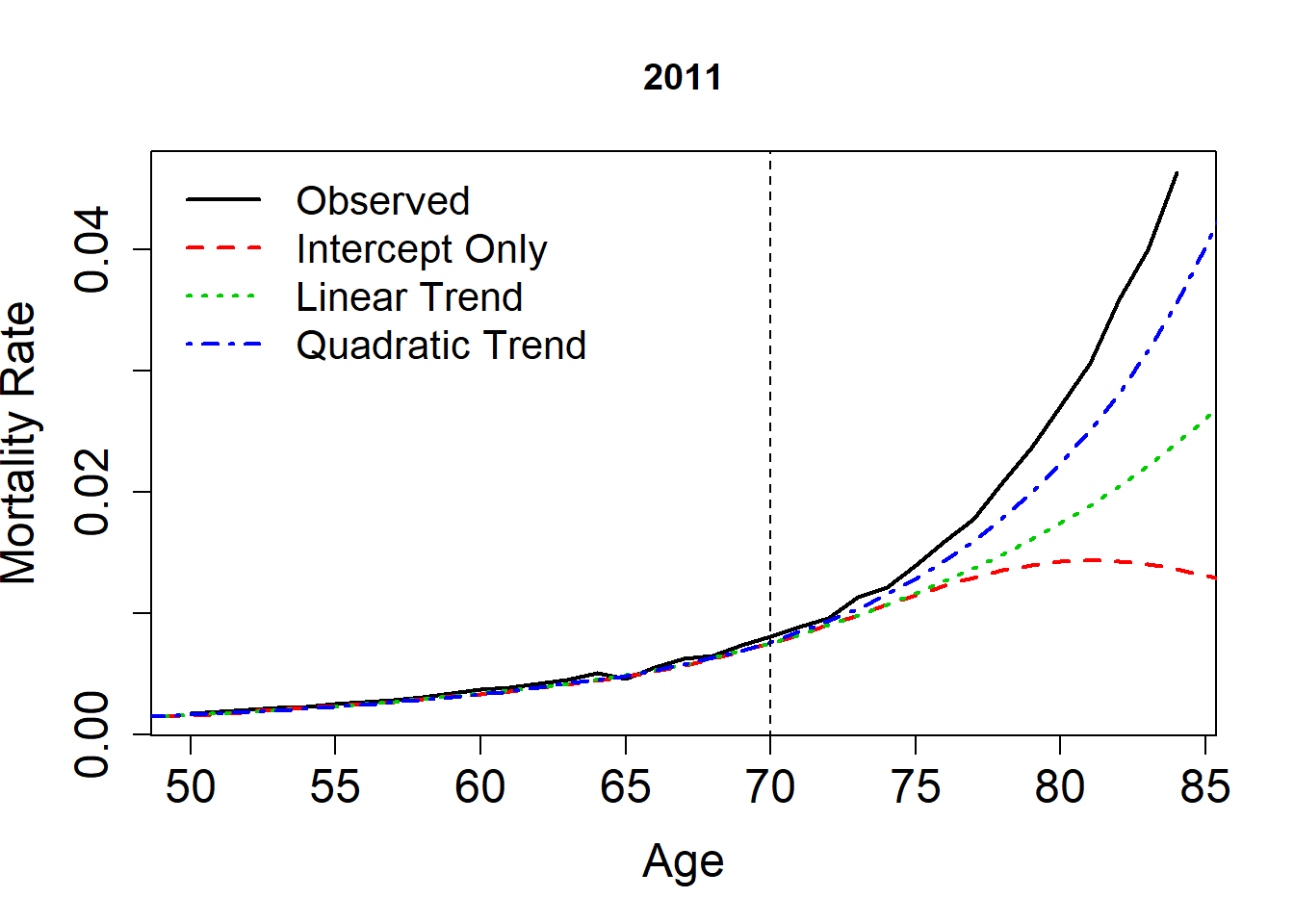}   \includegraphics[width=0.47\textwidth,height=1.6in]{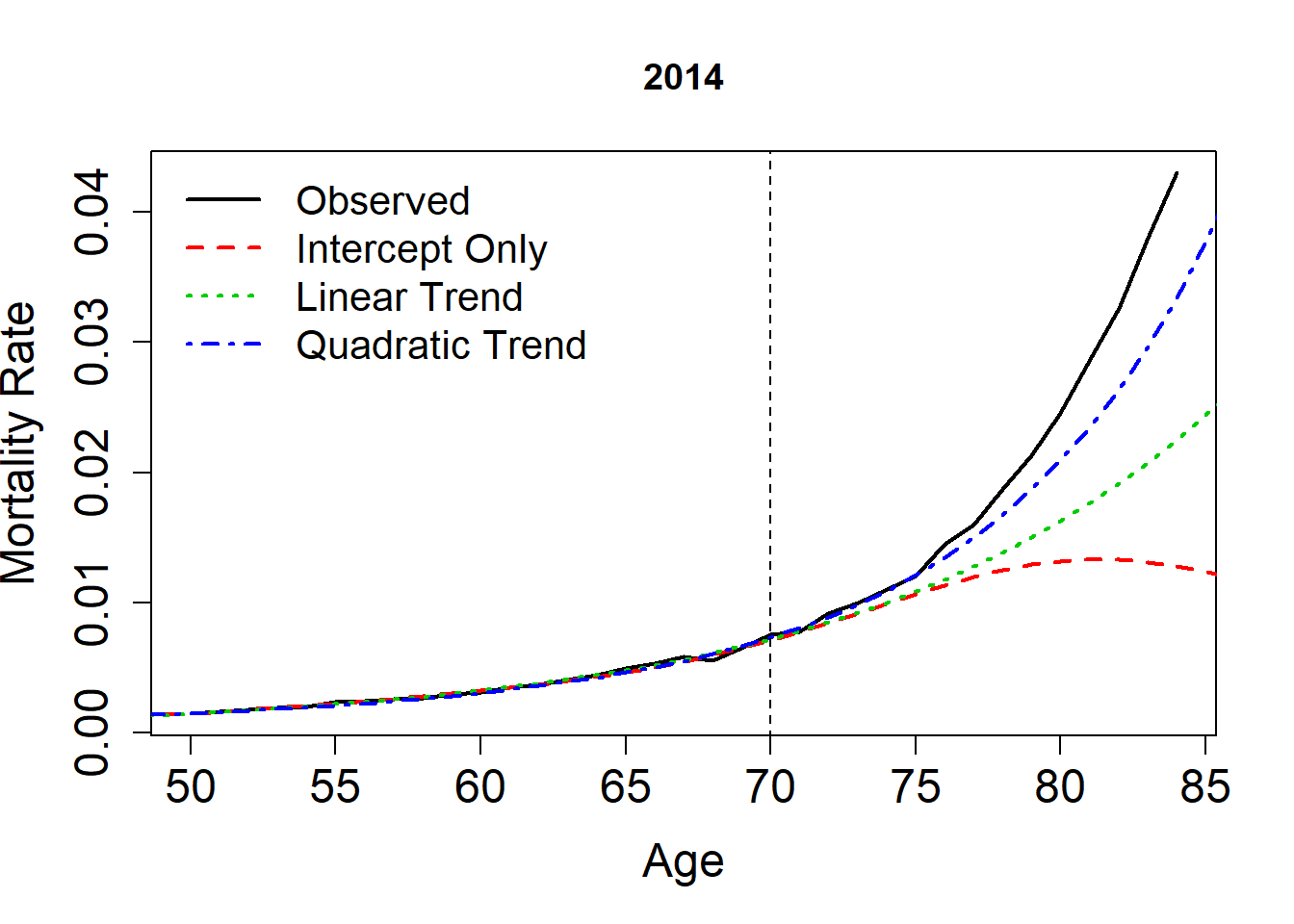} \\ \hline
  UK Males \\
  \includegraphics[width=0.47\textwidth,height=1.6in]{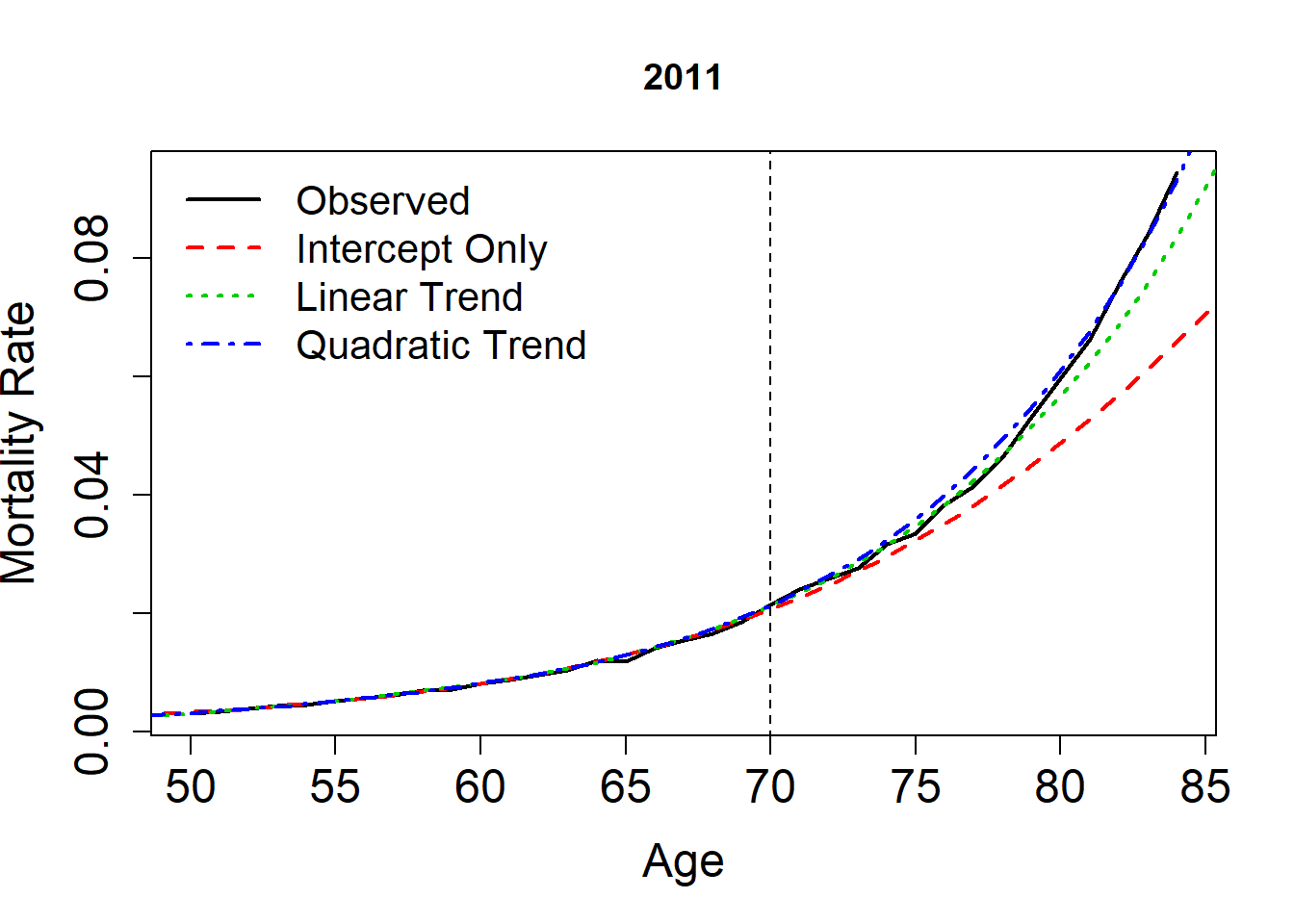}  \includegraphics[width=0.47\textwidth,height=1.6in]{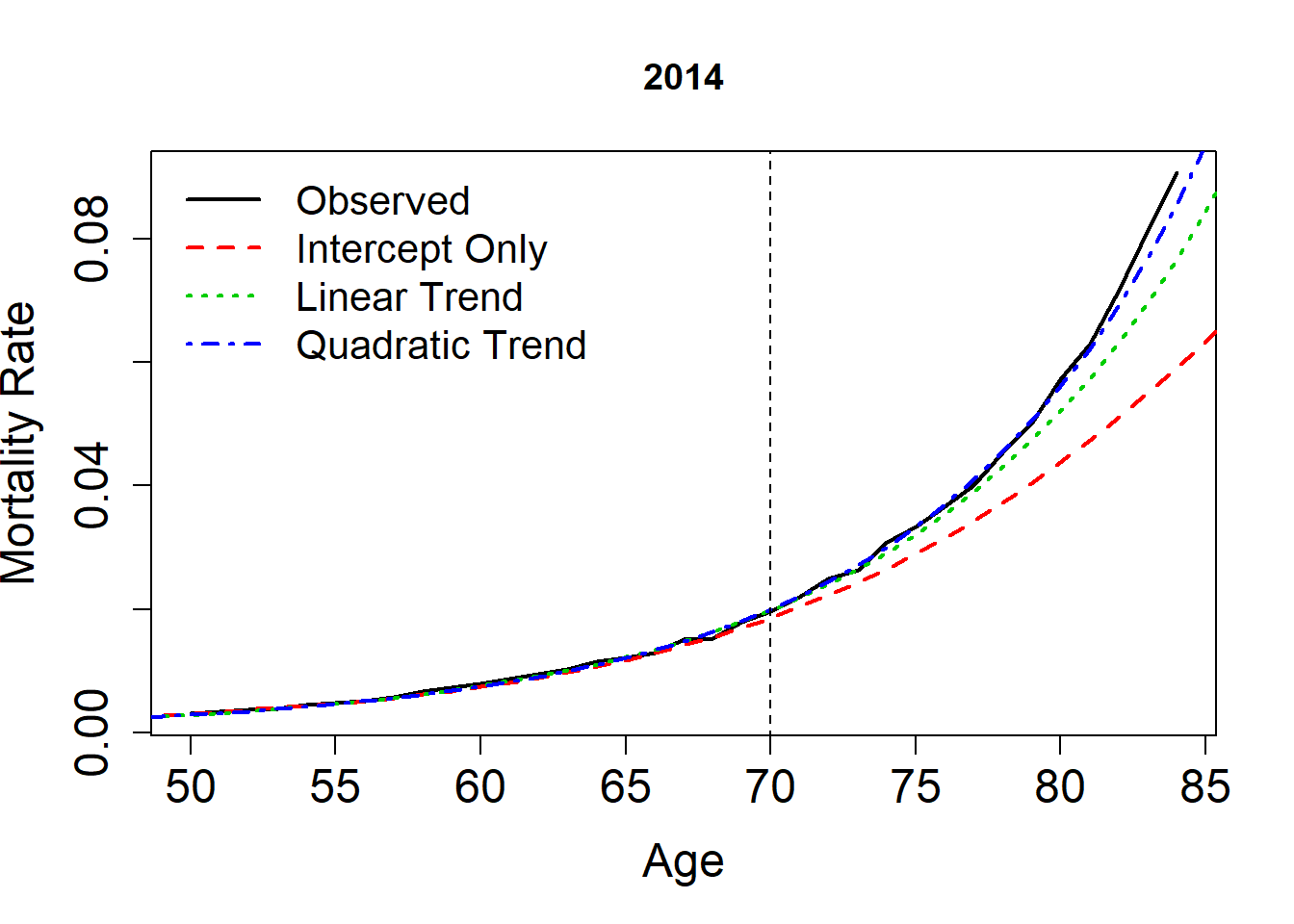} \\
  \hline  UK Females \\ 
  \includegraphics[width=0.47\textwidth,height=1.6in]{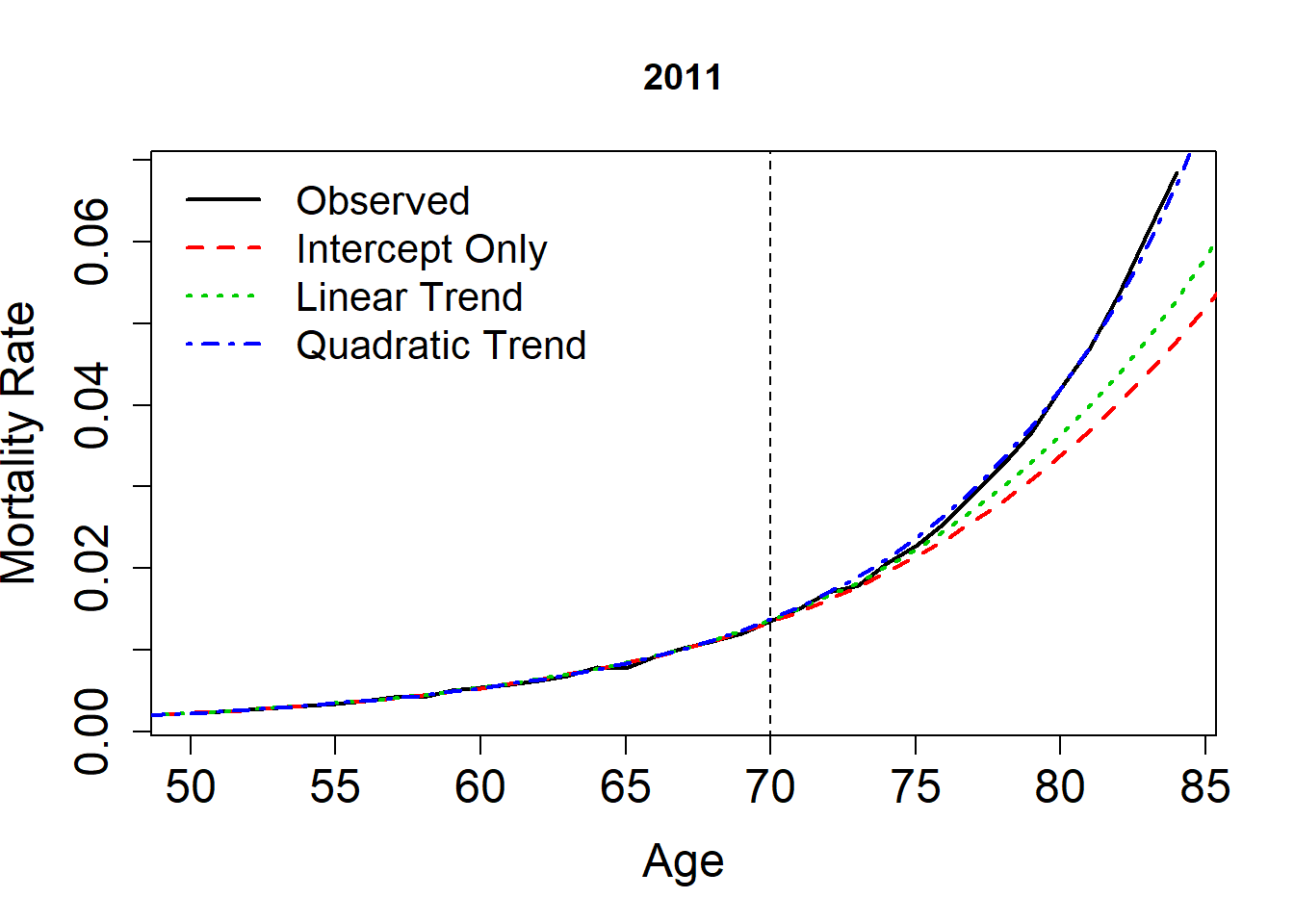}   \includegraphics[width=0.47\textwidth,height=1.6in]{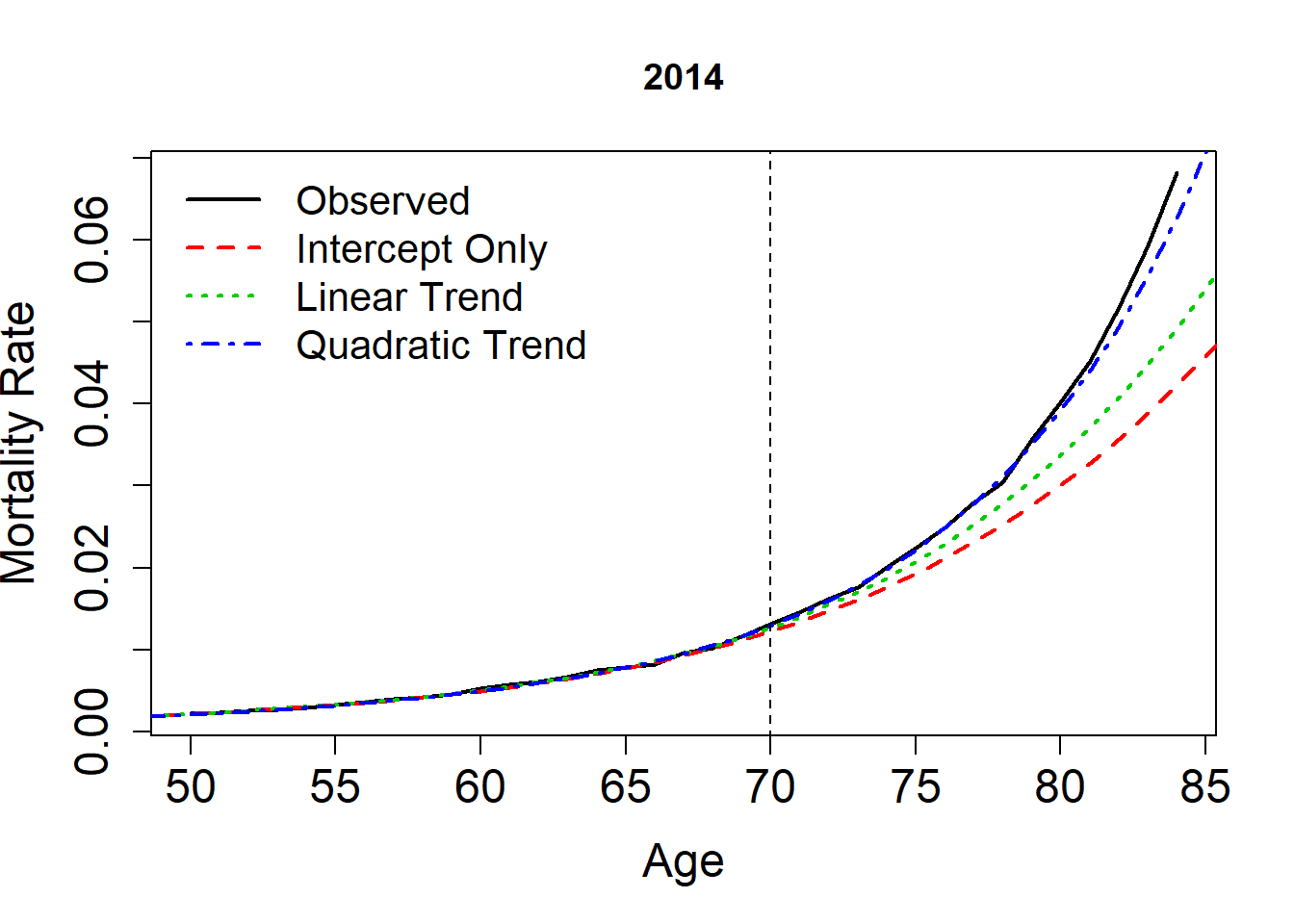} \\ \hline
  \end{tabular}

  \caption{ \label{fig:app-trendcomparison} Comparison of mean function choices in extrapolating mortality rates at old ages.  Models are fit to years {1999--2010} and ages 50--70 (Subset III), with estimates made for Age 50--85 in {2011 and 2014}. The vertical line indicates the boundary of the training dataset in $x_{ag}$. The mean functions are given in Table \ref{table:trendfunctions}.}
\end{figure}

 \begin{figure}[ht]
  \centering
  \begin{tabular}{cc} \hline
  Japan Males & Japan Females\\
  \includegraphics[width=0.23\textwidth,height=0.33\textheight]{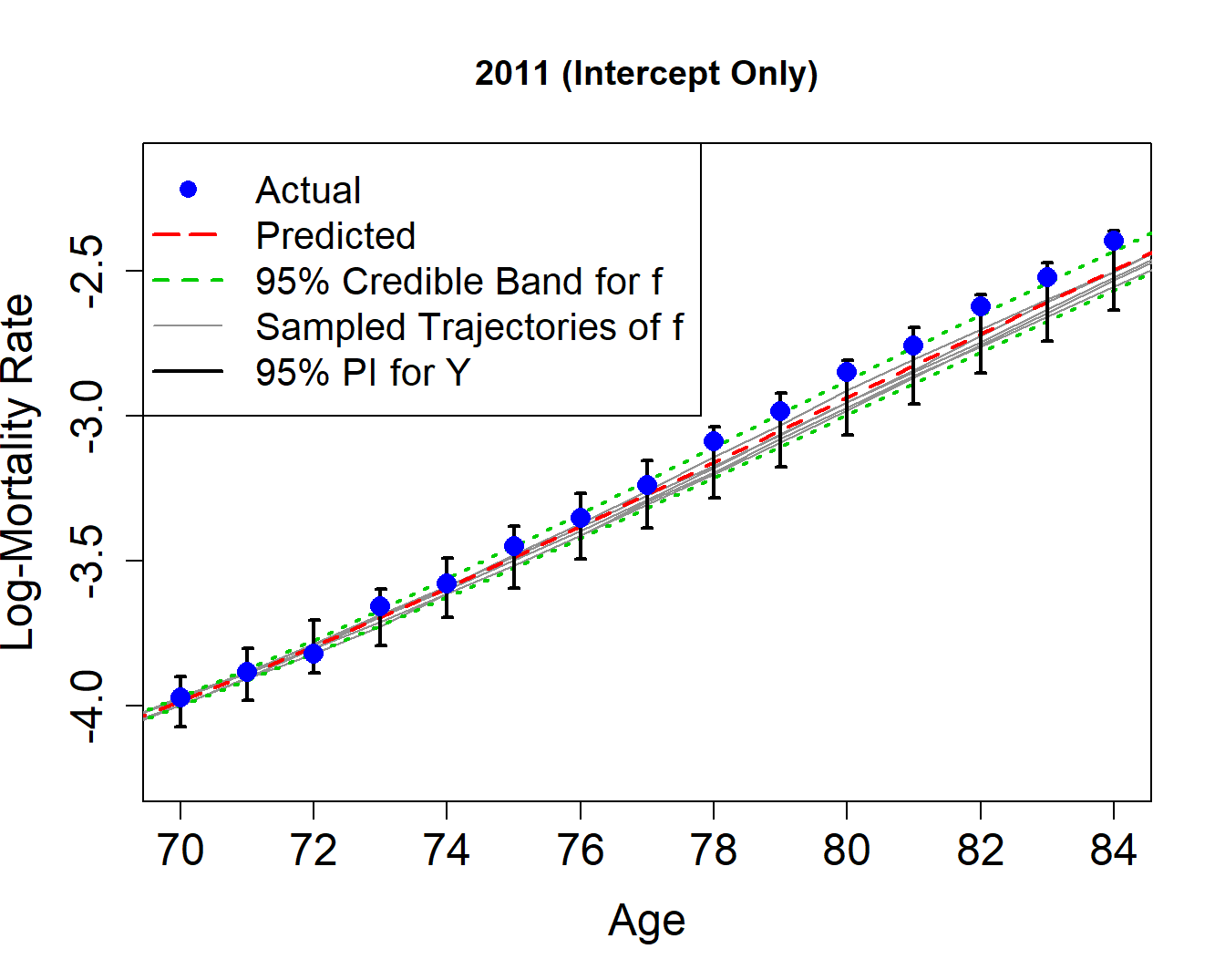}  \includegraphics[width=0.23\textwidth,height=0.33\textheight]{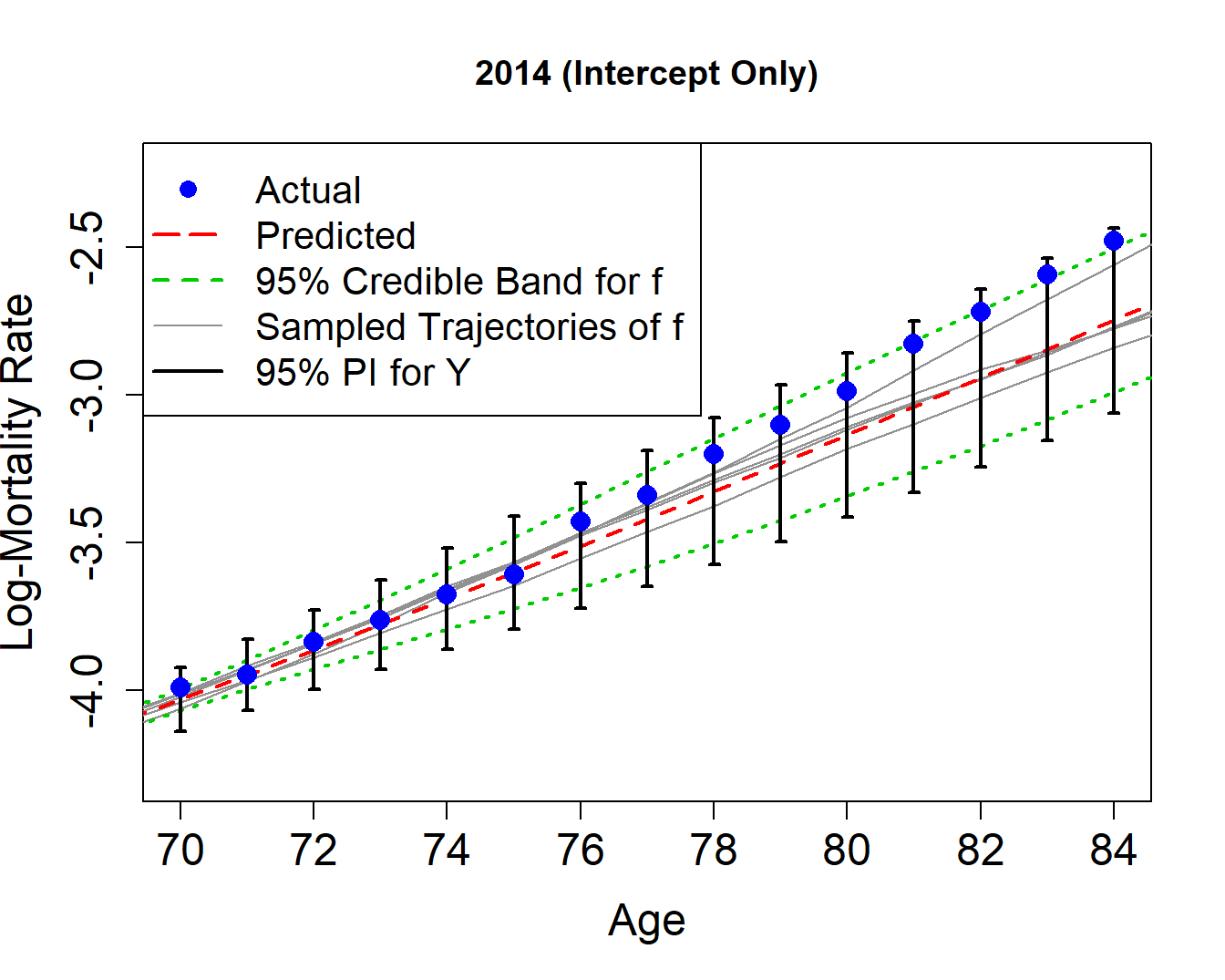} &
  \includegraphics[width=0.23\textwidth,height=0.33\textheight]{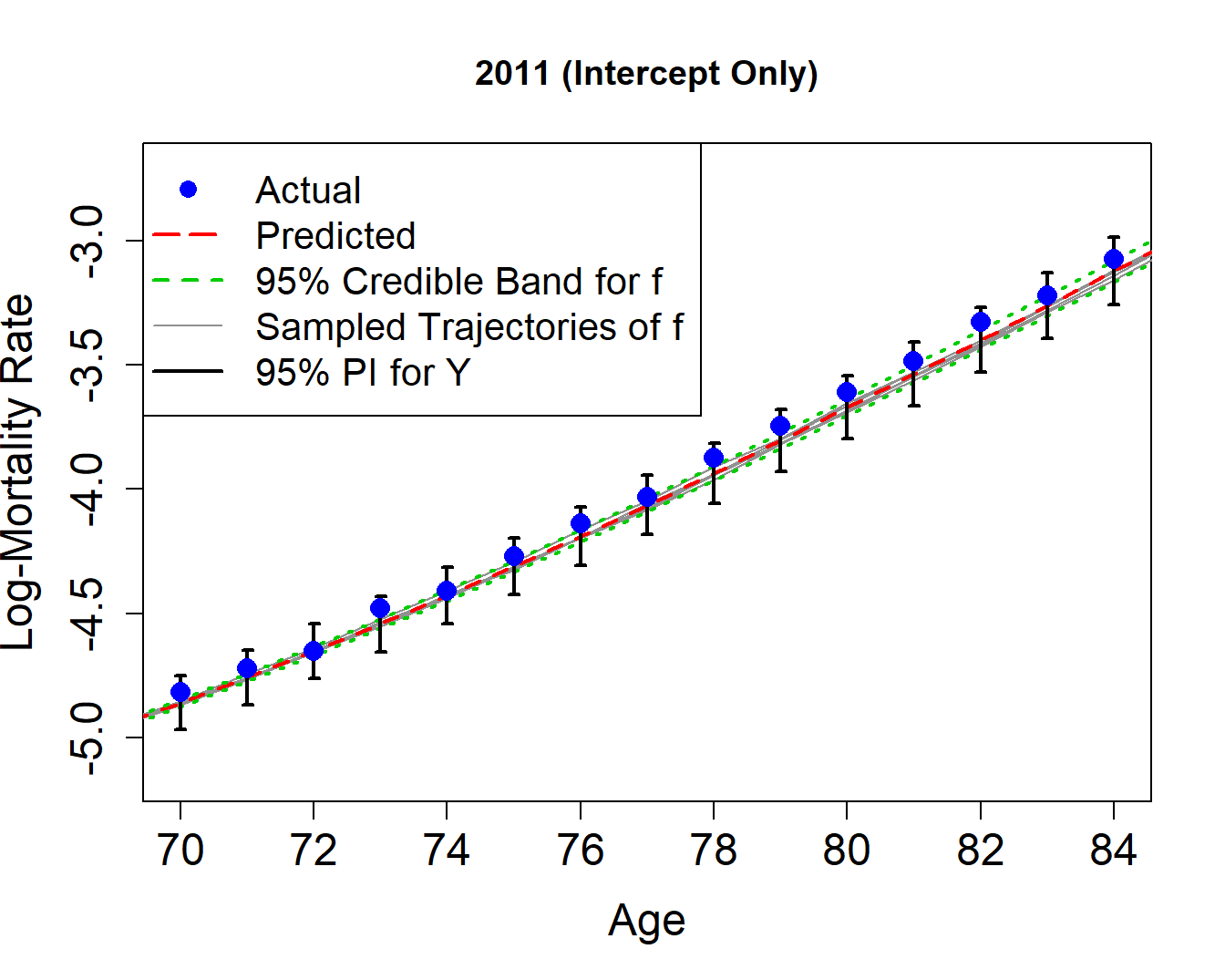}   \includegraphics[width=0.23\textwidth,height=0.33\textheight]{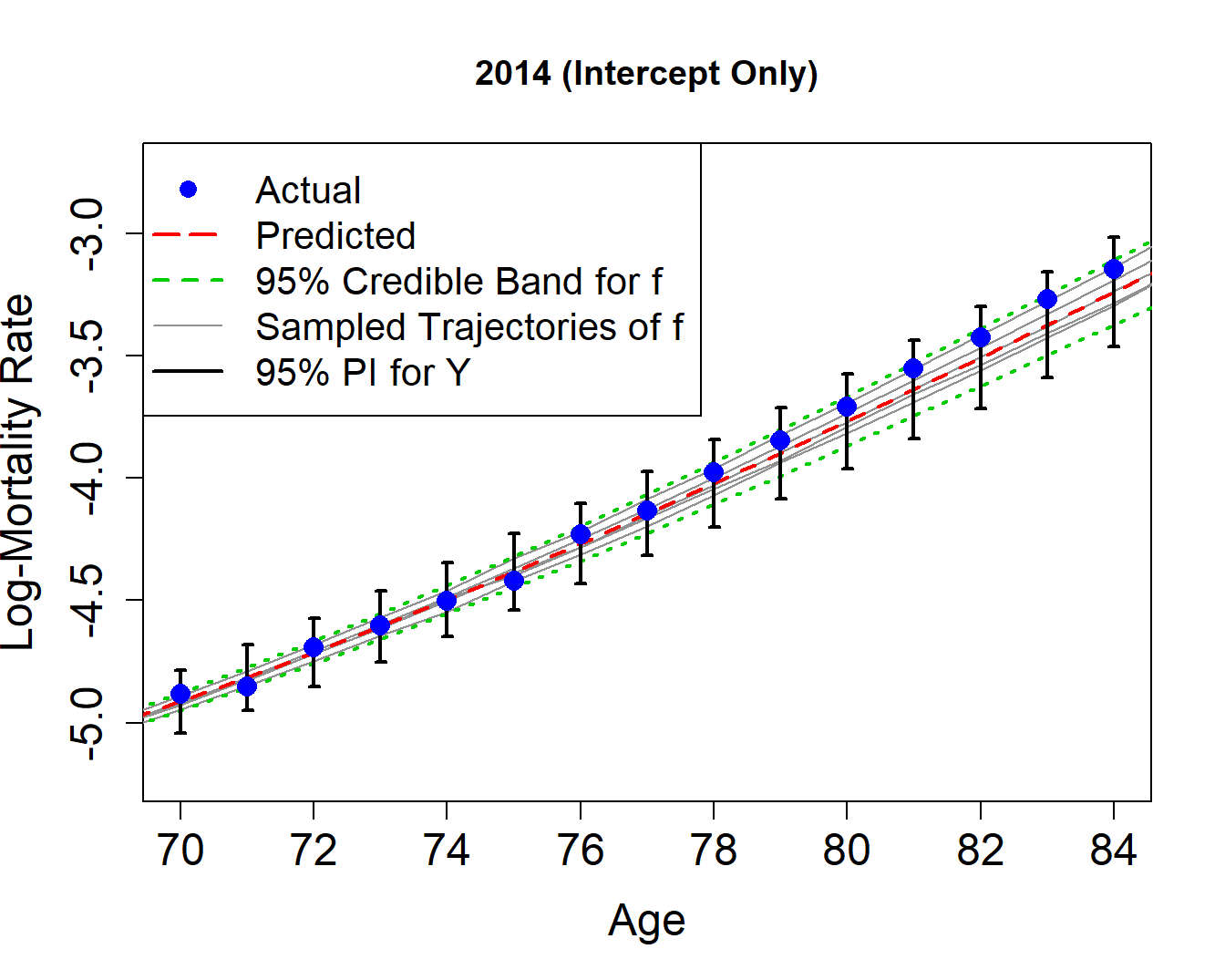} \\ \hline
  UK Males & UK Females\\
  \includegraphics[width=0.23\textwidth,height=0.33\textheight]{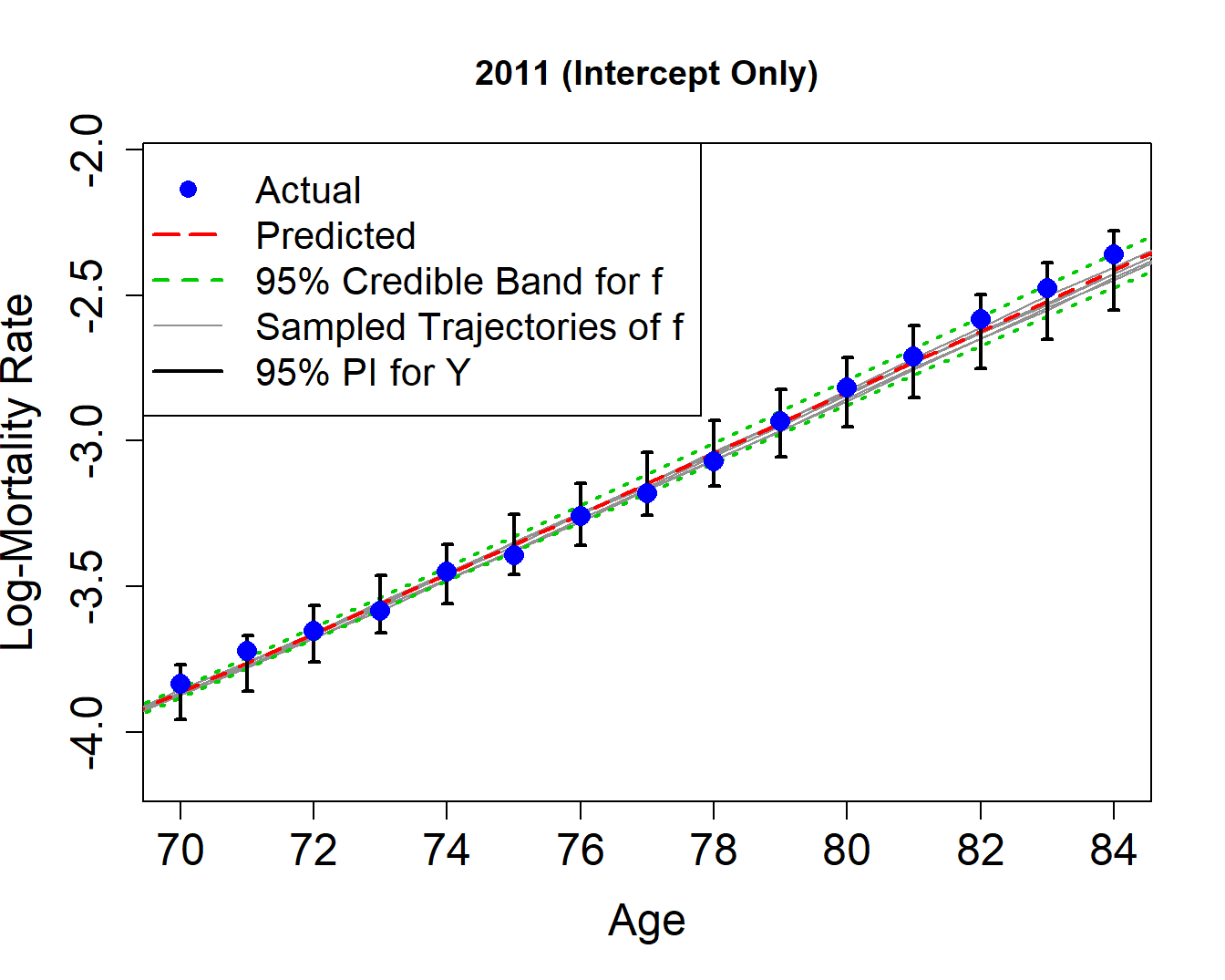}  \includegraphics[width=0.23\textwidth,height=0.33\textheight]{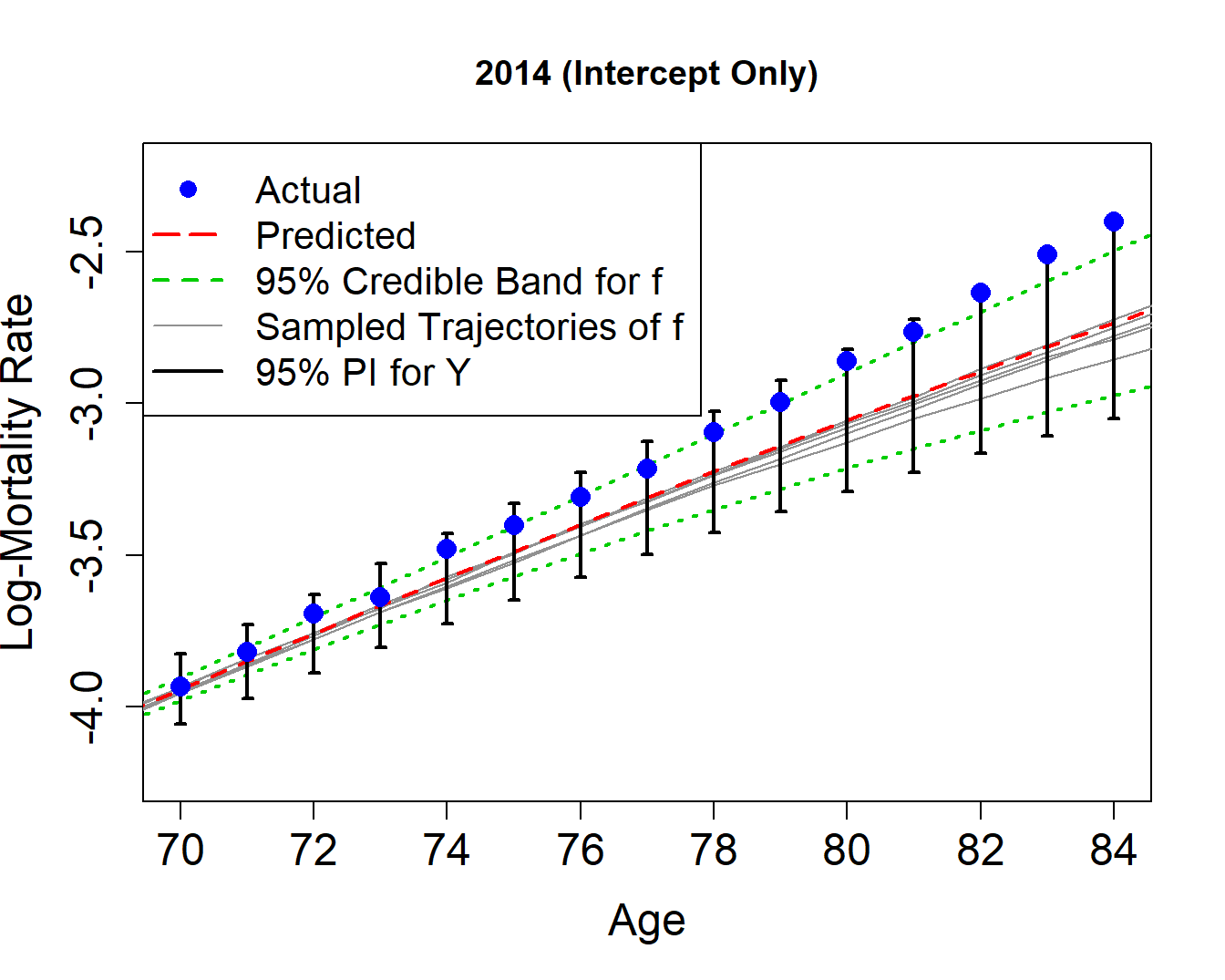} &
  \includegraphics[width=0.23\textwidth,height=0.33\textheight]{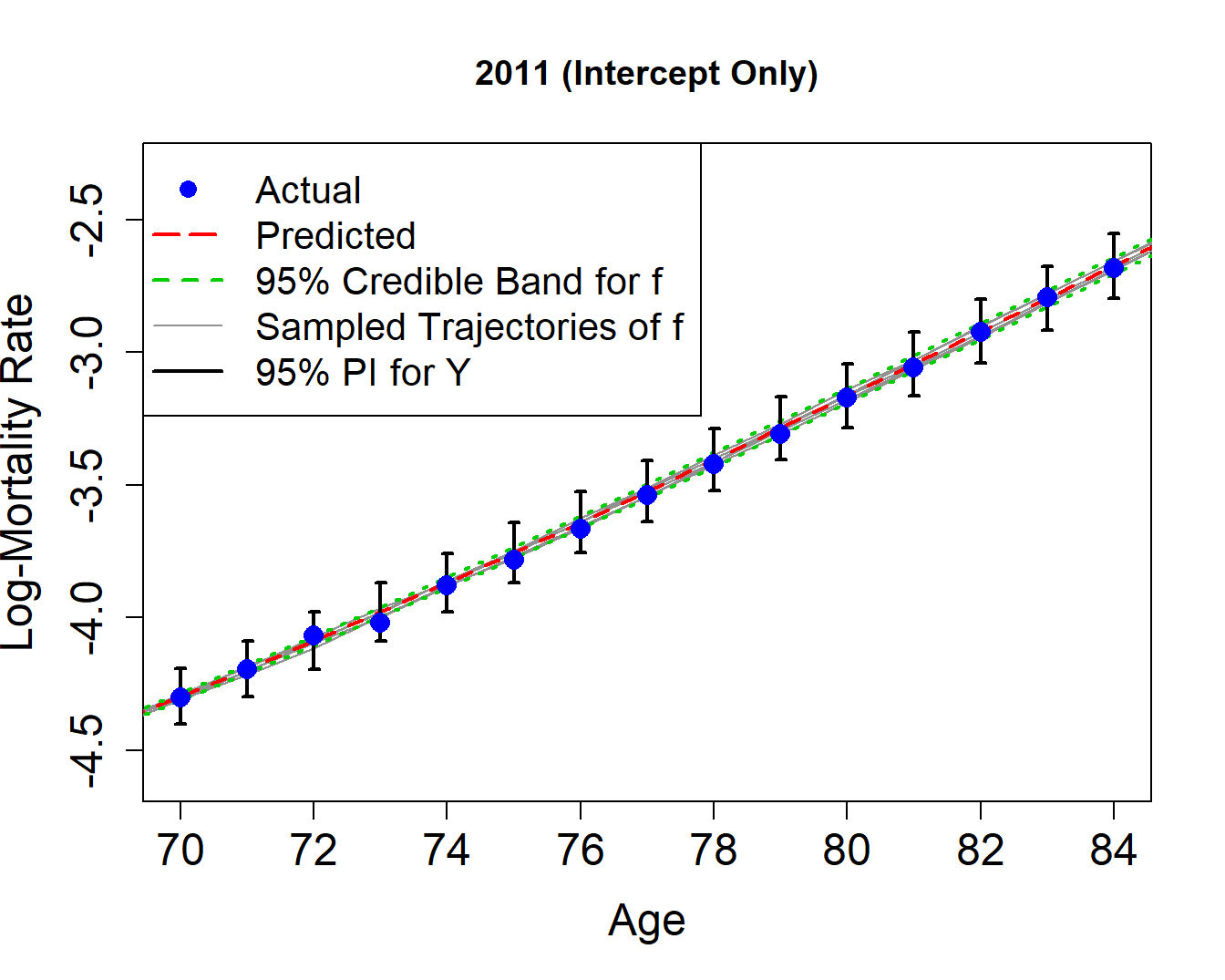}   \includegraphics[width=0.23\textwidth,height=0.33\textheight]{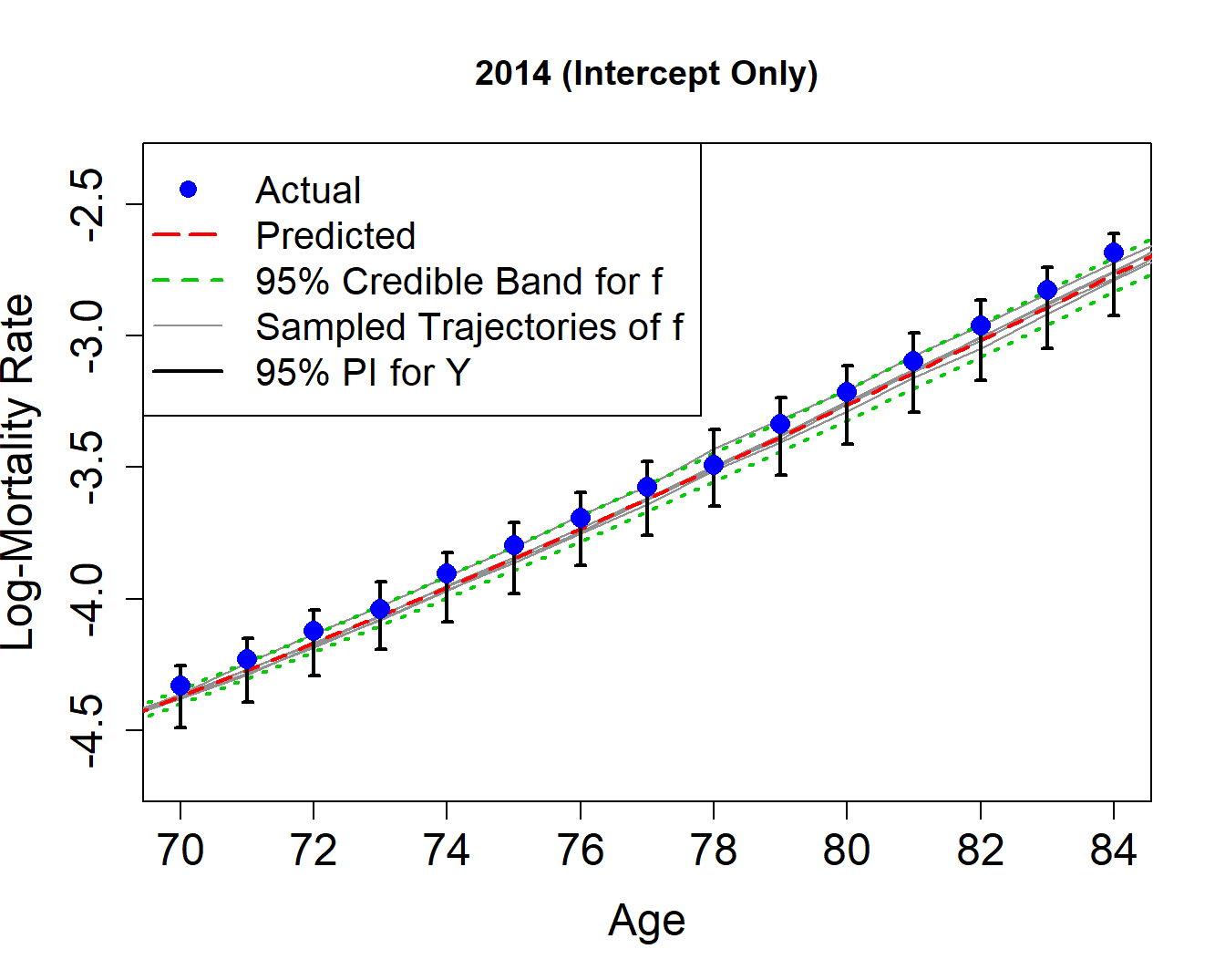} \\ \hline
  \end{tabular}

  \caption{ \label{fig:app-predict-2010} Mortality rate prediction for years 2011 and 2014 and ages 71--84.  Model is fit with Subset II data with intercept-only mean functions and squared-exponential kernel.  ``Simulated paths of $f$'' refers to simulated trajectories of the latent $\mathbf{f}_*$.  Credible bands are for the mortality surface $\mathbf{f}_*$; vertical intervals are for predicted observable mortality experience $\mathbf{y}_*$. }
\end{figure}

\begin{figure}[ht]
  \centering
\begin{tabular}{c} \hline
  Japan Males \\
   \includegraphics[width=0.470\textwidth,height=0.20\textheight]{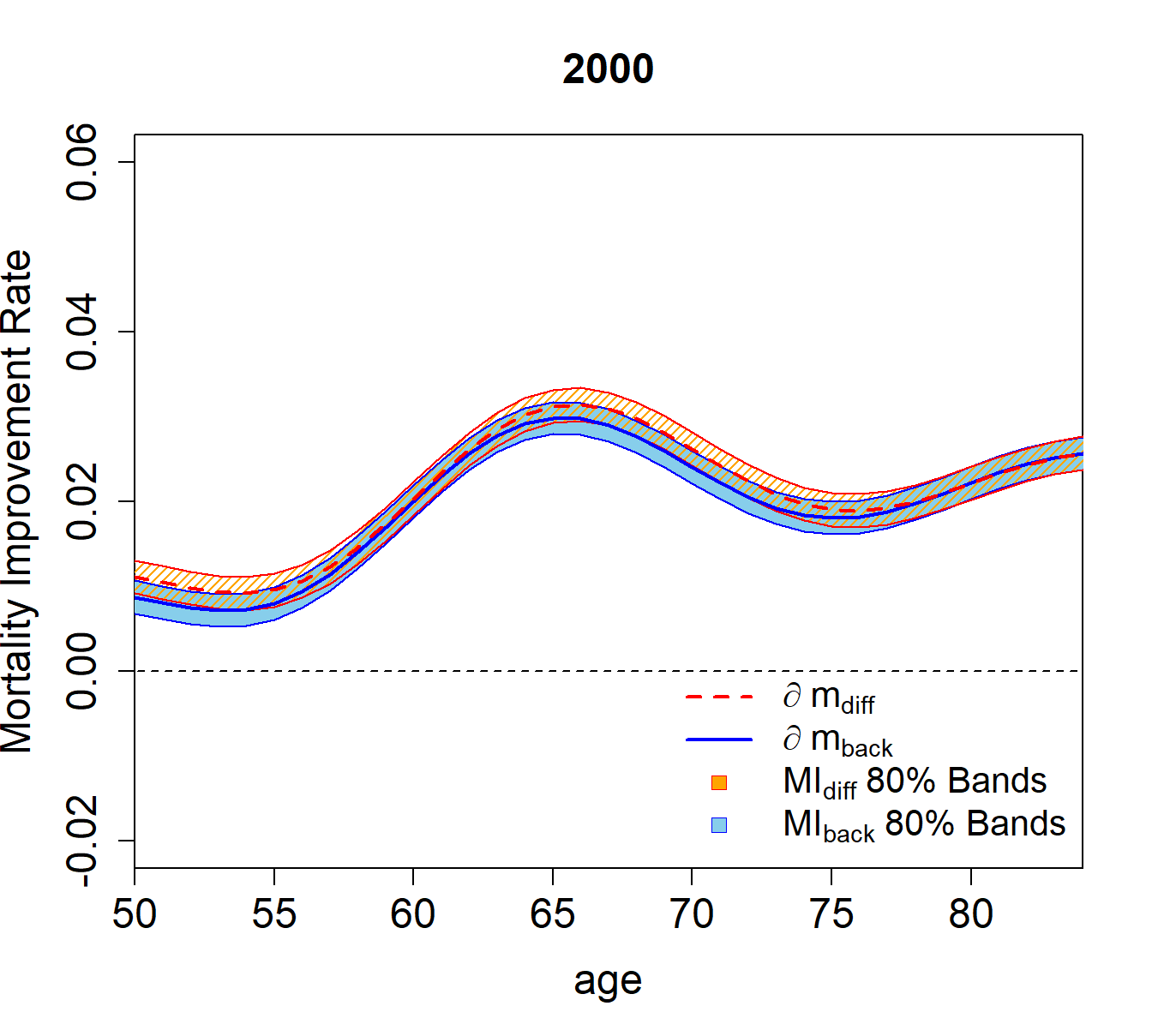}  \includegraphics[width=0.47\textwidth,height=0.20\textheight]{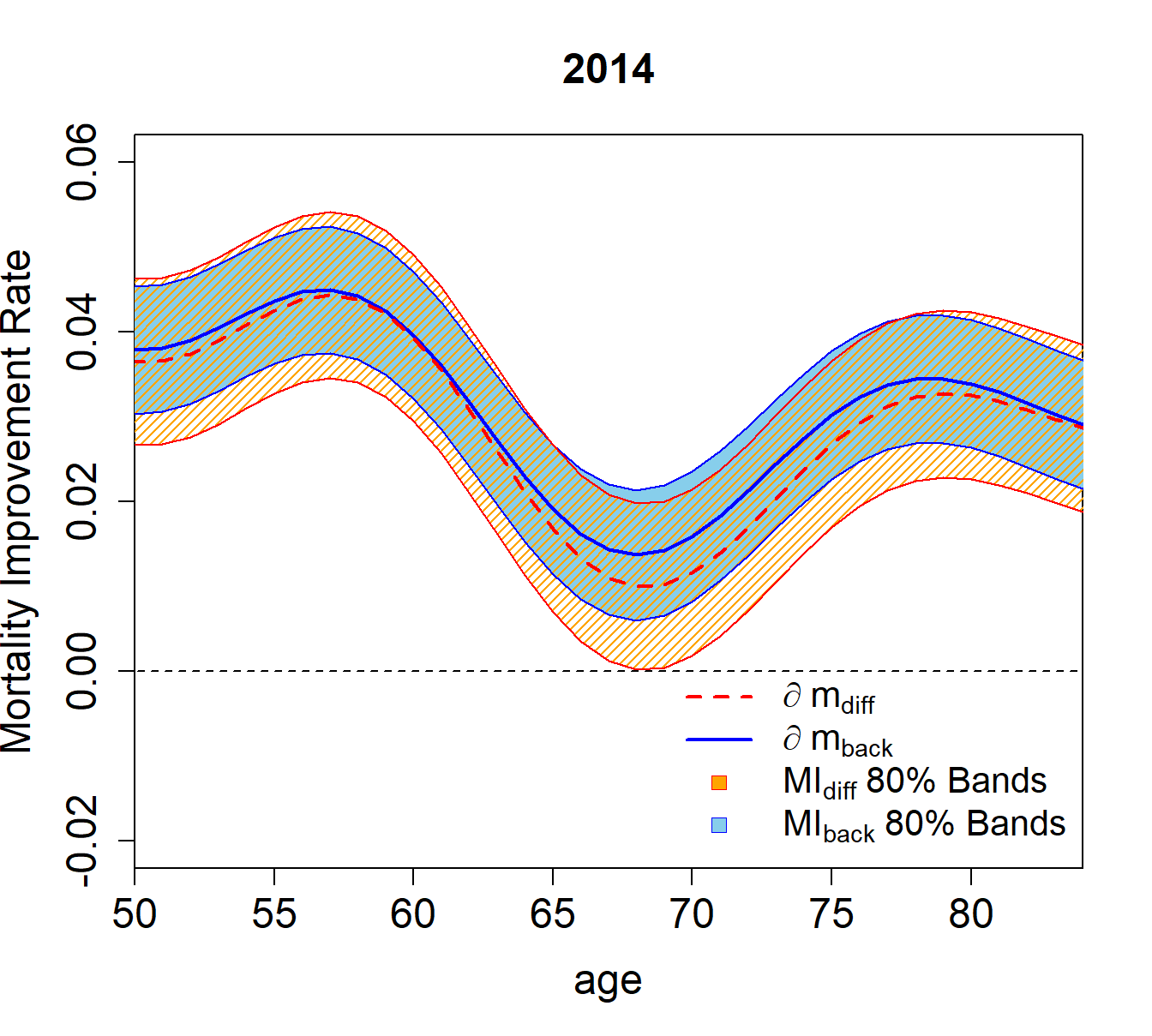}  \\ \hline
  Japan Females \\
   \includegraphics[width=0.47\textwidth,height=0.20\textheight]{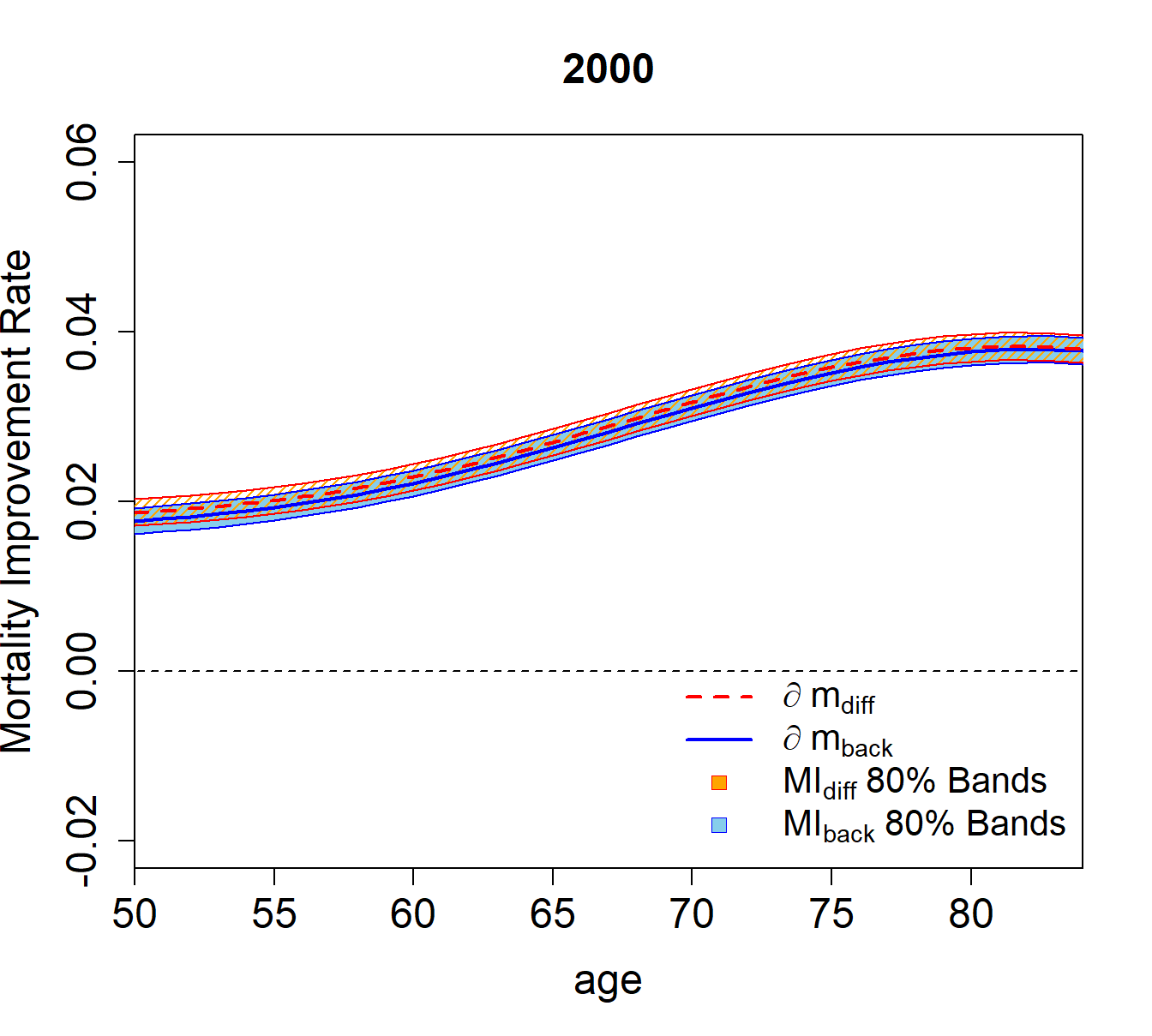}  \includegraphics[width=0.47\textwidth,height=0.20\textheight]{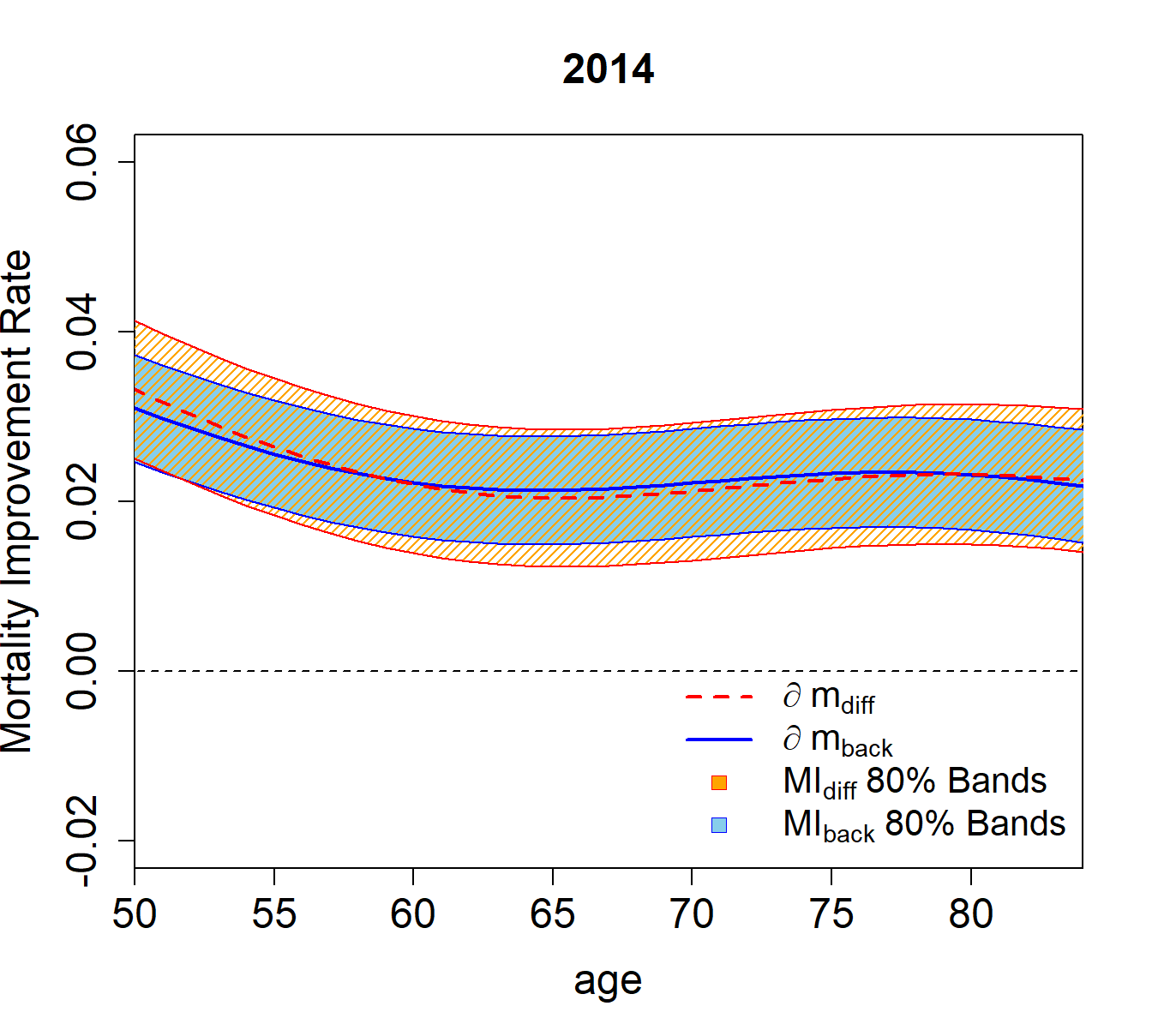}  \\ \hline
  UK Males \\
   \includegraphics[width=0.47\textwidth,height=0.20\textheight]{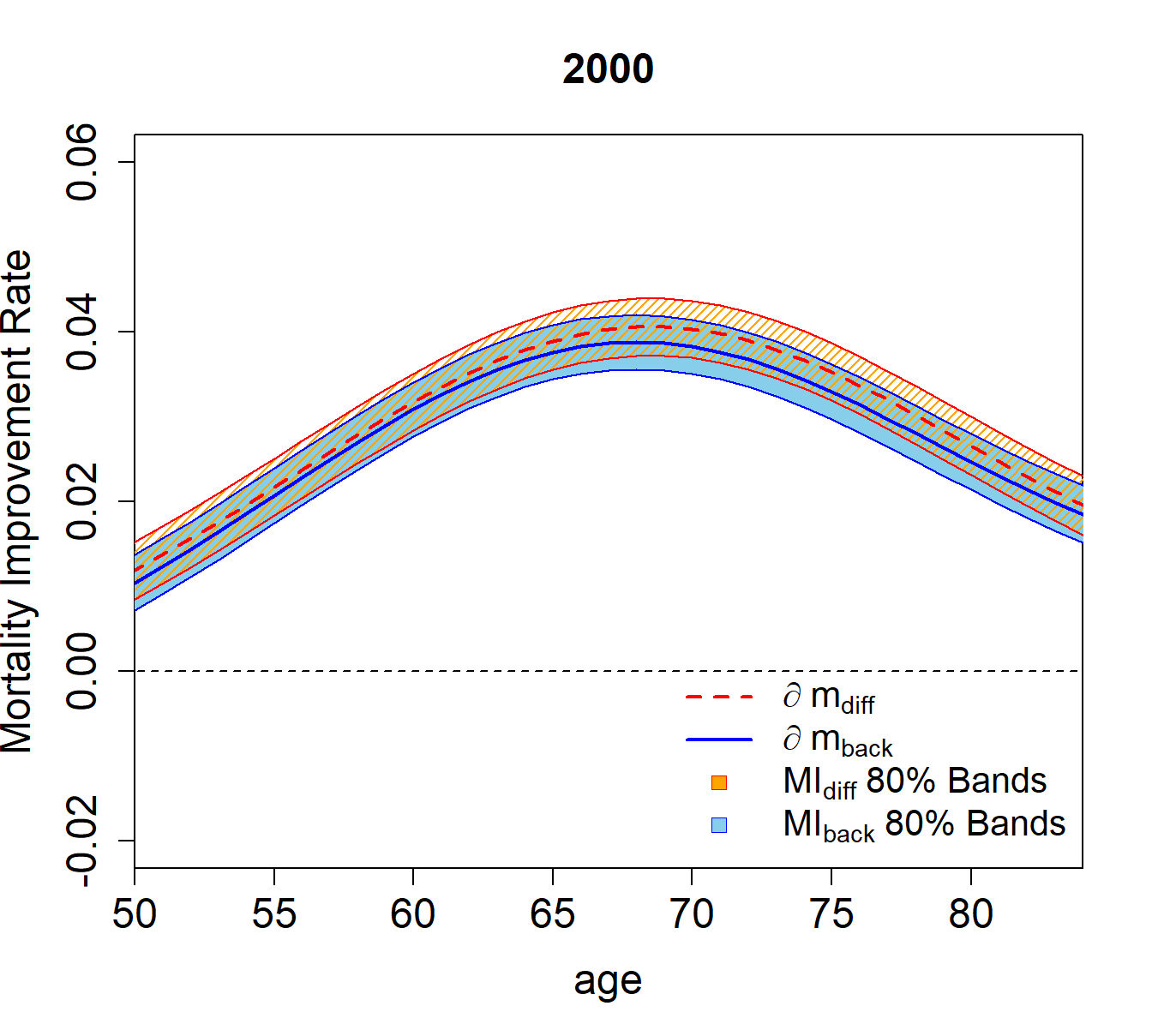}  \includegraphics[width=0.47\textwidth,height=0.20\textheight]{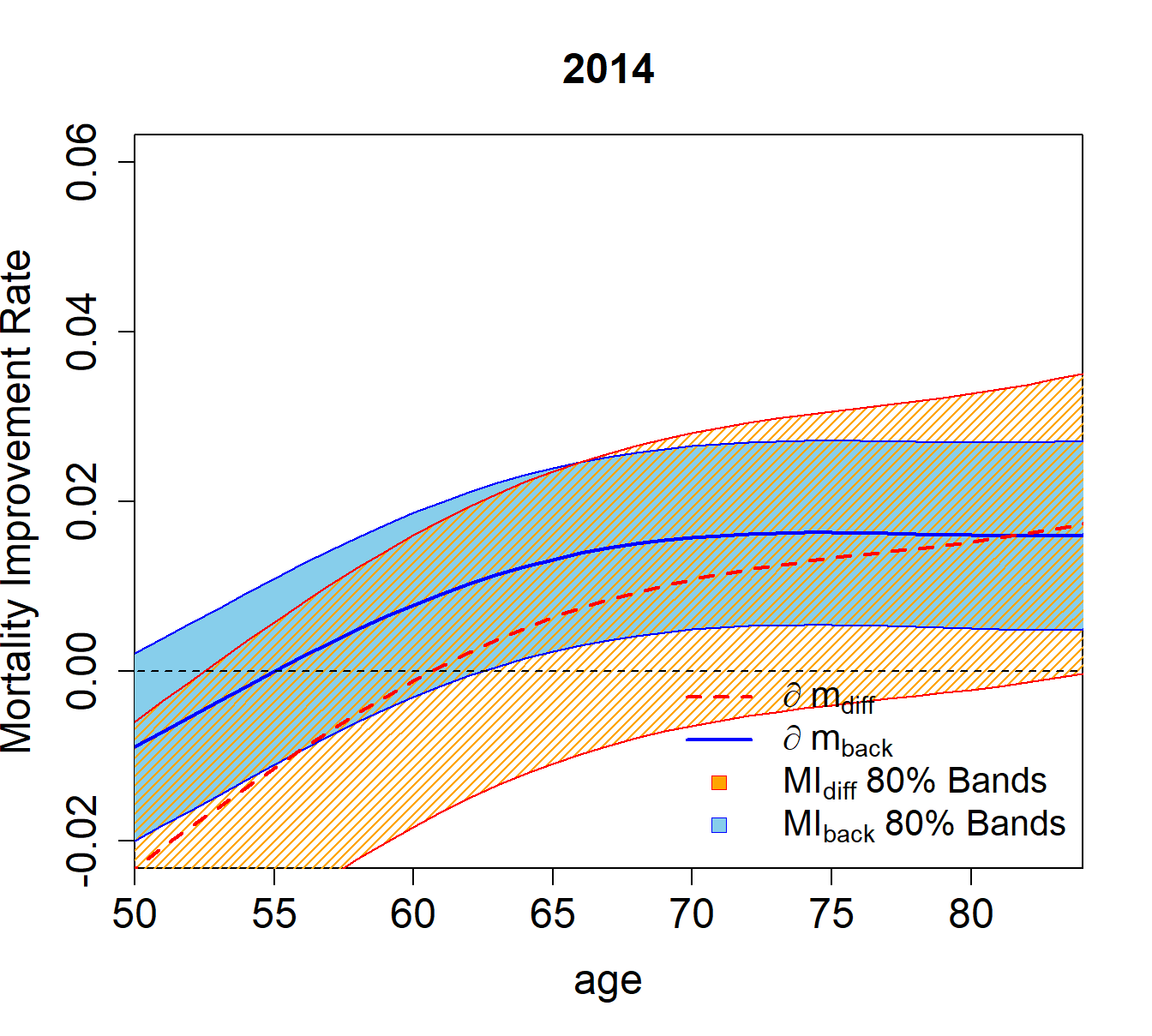}  \\ \hline
  UK Females \\
   \includegraphics[width=0.47\textwidth,height=0.20\textheight]{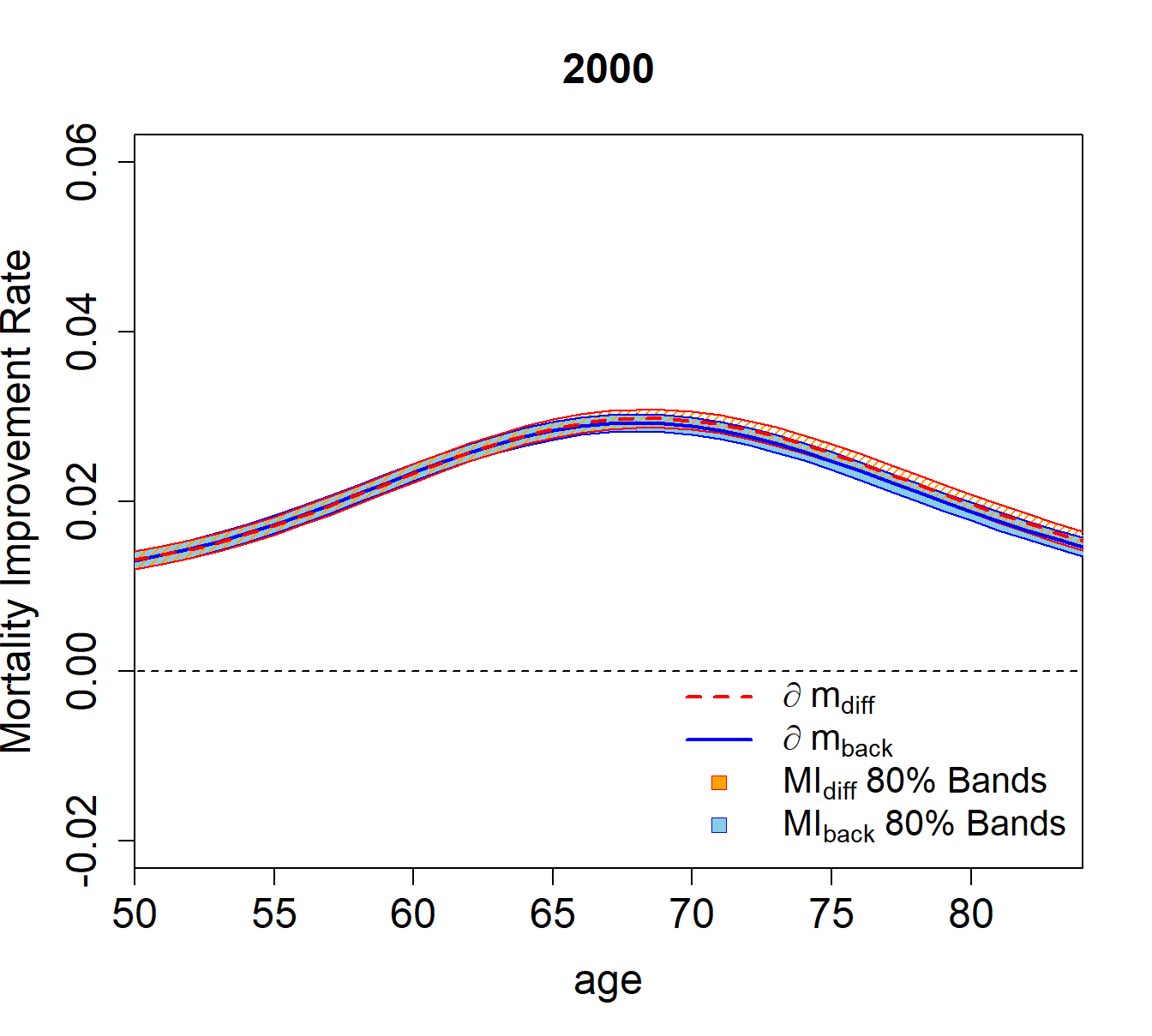}  \includegraphics[width=0.47\textwidth,height=0.20\textheight]{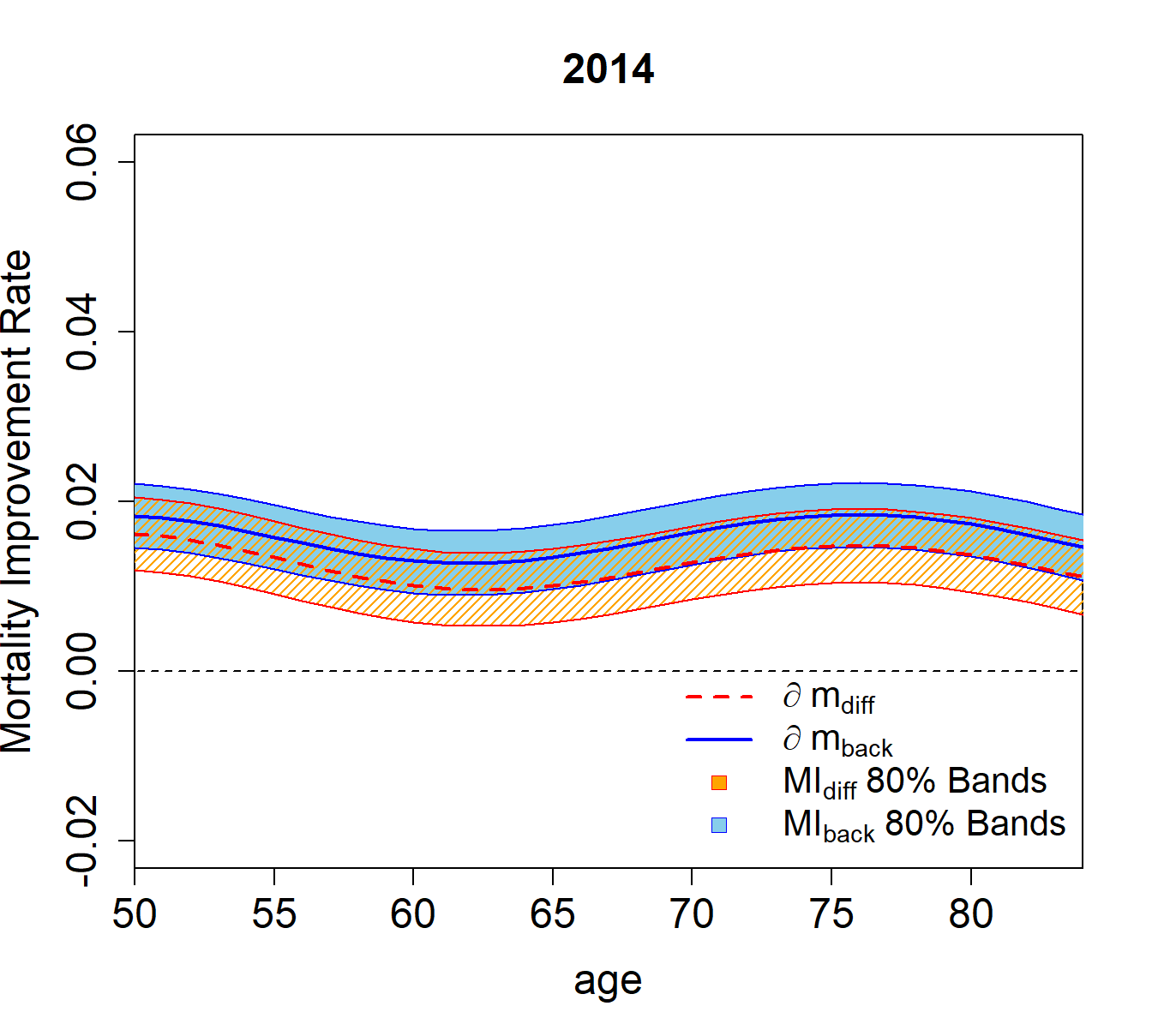} \\ \hline
  \end{tabular}
  \caption{Estimated annualized mortality improvement using the differential GP model (instantaneous improvement) and the YoY improvement from the original GP model. We show the means and 80\% credible bands for $MI_{diff}^{GP}$ and $MI_{back}^{GP}$ for  ages 50--84 and years 2000 \& 2014.  Models used are fit to All Data with $m(x) = \beta_0$. \label{fig:app-improvbandplot}}
\end{figure}

\end{document}